\documentclass[11pt]{article}
\usepackage[perpage]{footmisc}

\usepackage[utf8]{inputenc}
\usepackage{authblk}

\usepackage{amsthm,amsmath,amssymb}
\numberwithin{equation}{section}
\usepackage{amsfonts,mathrsfs}
\usepackage{caption,subcaption}
\usepackage{threeparttable}
\usepackage{booktabs}
\usepackage{multirow}
\usepackage{longtable}
\usepackage[dvipsnames]{xcolor}
\usepackage[linesnumbered,ruled,vlined]{algorithm2e}
\usepackage[letterpaper, margin=1in]{geometry}
\usepackage{xspace,makecell,tikz}
\usepackage{siunitx}
\usepackage{float}
\usepackage[sort,numbers]{natbib}
\usepackage{hyperref}
\usepackage[capitalise]{cleveref}
\usepackage{bbm}
\usepackage{enumerate}
\usepackage[shortlabels]{enumitem}
\usepackage{tocloft}
\usepackage{comment}
\usetikzlibrary{arrows.meta, positioning,calc, shapes.geometric}
\hypersetup{
linktocpage=true,
colorlinks=true,
linkcolor=red,
filecolor=magenta,
urlcolor=cyan,
citecolor=blue,
}
\tikzset{  
    thickline/.style={line width=2pt}, 
}  
\usetikzlibrary{decorations.pathreplacing}
\usepackage[toc,page]{appendix}

\DeclareMathOperator*{\argmax}{arg\,max}

\newcommand{\floor}[1]{\left\lfloor #1 \right\rfloor}

\newcommand{\calD}{\mathcal{D}}
\newcommand{\calF}{\mathcal{F}}

\newcommand{\calT}{\mathcal{T}}

\newcommand{\boldu}{\boldsymbol{u}}
\newcommand{\boldv}{\boldsymbol{v}}
\newcommand{\boldw}{\boldsymbol{w}}

\newcommand{\norm}{f}

\newcommand{\E}{\mathop{\mathbb{E}}}

\newcommand{\R}{\mathbb{R}}

\newcommand{\Rnng}{\mathbb{R}_{\geq 0}}

\renewcommand{\l}{\ell}
\newcommand{\activeelem}{\mathrm{Ac}}

\newcommand{\XOS}{\textnormal{xos}}
\newcommand{\XOSS}{\textnormal{binary-XOS}\xspace}
\newcommand{\xoss}{\textnormal{bxos}}
\newcommand{\sym}{\textnormal{sym}}

\newcommand{\poly}{\mathsf{poly}}

\newcommand{\elt}{\mathsf{elt}}
\newcommand{\polylog}{\mathrm{polylog}\xspace}
\newcommand{\gapfunction}{\Lambda}
\newcommand{\Rdistribution}{\mathbb{P}_R}
\newcommand{\nonadaptivestrategy}{\textsc{AlgPath}\xspace}
\newcommand{\adaptivestrategy}{\textsc{AlgTree}\xspace}
\newcommand{\adaptcost}{\mathsf{Adap}}
\newcommand{\nonadaptcost}{\mathsf{Nadp}}
\newcommand{\alg}{\nonadaptcost}
\newcommand{\adap}{\adaptcost}
\newcommand{\OPT}{\mathsf{OPT}}
\newcommand{\level}{L}
\newcommand{\cc}{\mathcal{A}}
\newcommand{\imp}{\mathsf{Imp}}
\newcommand{\fspecial}{f_{\mathrm{xos}}}
\newcommand{\depth}[1]{\mathsf{depth}(#1)}
\newcommand{\eat}[1]{}

\newtheorem{theorem}{Theorem}[section]
\newtheorem{lemma}[theorem]{Lemma}
\newtheorem{observation}{Observation}[section]

\newtheorem{corollary}[theorem]{Corollary}

\newtheorem{remark}[theorem]{Remark}

\newtheorem{fact}[theorem]{Fact}

\theoremstyle{definition}
\newtheorem{definition}{Definition}[section]

\newtheorem{innercustomcase}{Case}
\newenvironment{mycase}[1]
{\renewcommand\theinnercustomcase{#1}\innercustomcase}
  {\endinnercustomcase}

\newtheorem{innercustomfact}{Fact}
\newenvironment{myfact}[1]
{\renewcommand\theinnercustomfact{#1}\innercustomfact}
{\endinnercustomfact}

 

\SetKwComment{Comment}{\textcolor{Magenta}{$\triangleright$\ }}{}
\SetKw{KwMaintain}{\textbf{Maintain:}}
\SetKw{KwFind}{\textbf{find}}
\SetKw{KwDefine}{\textbf{define}}
\DontPrintSemicolon

\title{Adaptivity Gaps for Stochastic Probing with Subadditive Functions}

\author{Jian Li \quad\qquad Yinchen Liu \quad\qquad Yiran Zhang}

\affil{%
  \small Institute for Interdisciplinary Information Sciences, Tsinghua University, China.\\
  \texttt{lapordge@gmail.com, liuyinch23@mails.tsinghua.edu.cn,
  zhangyir22@mails.tsinghua.edu.cn}
}

\date{}
\begin{document}
\maketitle
\begin{abstract}\makeatletter\phantomsection\def\@currentlabel{(abstract)}\makeatother

In this paper, we study the stochastic probing problem under a {\em general monotone norm} objective. We are given a ground set \(U = [n] = \{1, 2, \dots, n\}\), where each element \(i\) is associated with an independent nonnegative random variable \(X_i\) (with a known distribution). We may probe these elements adaptively, and upon probing an element \(i\), its value \(X_i\) is realized. The sequence of probed elements must satisfy a prefix-closed feasibility constraint \(\mathcal{F}\), such as a matroid, an orienteering constraint, or any other downward-closed constraint. We also have a monotone norm function \(f: \mathbb{R}_{\geq 0}^n \to \mathbb{R}_{\geq 0}\). Let \(P\subseteq U\) be the set of probed elements. Then the reward is \(f(X_P)\), where \(X_P\) is an \(n\)-dimensional vector whose \(i\)-th coordinate equals the realized value of \(X_i\) if \(i\in P\) (i.e., element $i$
is probed), and 0 otherwise. Our objective is to design a probing strategy that maximizes the expected reward \(\mathbb{E}[f(X_P)]\).


We study the {\em adaptivity gap} of the problem, defined as the ratio between the expected reward of an optimal adaptive strategy and that of an optimal non-adaptive strategy. 
A small adaptivity gap allows us to focus on designing non-adaptive strategies, which are typically simpler to represent and analyze.
Establishing tight adaptivity gaps is a central challenge in stochastic combinatorial optimization and has been studied extensively for stochastic probing problems with various objective functions. In this paper, we resolve a central open problem
in this line of research, posed in \cite{GNS17,KMS24}, by proving that the adaptivity gap for stochastic probing with general monotone norms is bounded by $ O(\log^2 n)$.
With a refined analysis, we can further strengthen the bound to
$
O(\log r \log n /\log\log n)
$ 
where \(r\) is the maximum length of a sequence in the feasibility constraint
(\(2 \leq r \leq n\)).
As a by-product, we obtain an asymptotically {\em tight} adaptivity gap $\Theta(\log n/\log\log n)$ for Bernoulli stochastic probing with \XOSS objectives, matching the lower bound in \cite{GNS17}.
We also obtain an \(O(\log^3 n)\) upper bound for Bernoulli stochastic probing with general subadditive objectives. Furthermore, for {\em monotone symmetric norms}, we prove that the adaptivity gap can be bounded by \(O(1)\), 
answering an open question posed in 
\cite{PRS23} and improving upon their 
\(O(\log n)\) upper bound.

\end{abstract}
\newpage
\small \tableofcontents
\newpage

\section{Introduction}
\label{sec:intro}



In this paper, we investigate the \emph{stochastic probing} problem, a fundamental framework in the algorithmic study of optimization under uncertainty. This problem has been widely explored in the presence of various combinatorial constraints and objective functions (see, e.g., \cite{GN13,GNS16,GNS17,fu2018ptas,BSZ19,PRS23,KMS24}). Further discussions can be found in \cref{sec:related}. 

Stochastic probing has numerous applications across different domains, including Bayesian mechanism design \cite{GN13}, online learning \cite{chen2016combinatorial}, influence maximization \cite{GNS17}, robot path planning \cite{GNS16, GNS17} and database management \cite{LPRY08}. In this work, we focus on a general stochastic probing problem characterized by a monotone norm objective function and general combinatorial constraints. 

\begin{definition}[Stochastic Probing]
\label{stoc:probing}
We are given a ground universe \(U = [n] = \{1, 2, \ldots, n\}\) of elements, where each element \(i\) is associated with an independent nonnegative random variable \(X_i \in \mathbb{R}_{\ge 0}\) (with a known distribution that may vary across elements). 
The exact values of the \(X_i\)'s are initially unknown. We are allowed to \emph{probe} the elements adaptively, and upon probing an element, we observe the realized value of the corresponding random variable. 
In addition, we are provided with an objective function \(f: \mathbb{R}_{\ge 0}^U \to \mathbb{R}_{\ge 0}\) and a \emph{prefix-closed} feasibility constraint \(\calF \subseteq U^*\),\footnote{
\(U^*\) denotes the set of all sequences of distinct elements in \(U\) of length at most \(n\). We use $|F|$ to denote the length of the sequence $F$ for $F\in U^*$. $\calF$ is \emph{prefix-closed} means that if a sequence $F\in \calF$, then all its prefixes are also included in $\calF$.
} such that probing is only allowed along sequences in \(\calF\) (which captures combinatorial constraints like matroids, orienteering, or any arbitrary downward-closed constraint). Let \(P \subseteq U\) denote the set of probed items. 
The reward is then given by \(f(X_P)\), where \(X_P \in \mathbb{R}_{\ge 0}^U\) is defined by setting its \(i\)-th coordinate to \(X_i\) if \(i\) is probed (i.e., if \(i \in P\)) and to 0 otherwise. Our objective is to design a probing strategy that maximizes the expected reward \(\E[f(X_P)]\).
\end{definition}

We also examine a special case where each random variable \(X_i\) follows a Bernoulli distribution.

\begin{definition}[Bernoulli Stochastic Probing]
\label{intro:bernoulli}
In the stochastic probing problem, each random variable $X_i$ is 1 with probability $p_i$ and 0 with probability $1-p_i$ for each $i\in U$. If we probe element $i$
and $X_i$ is realized to $1$, we say that element $i$ is {\em active}. The objective function $f$
simplifies to a set function 
\(f: 2^U \to \mathbb{R}_{\ge 0}\).
Let \(P \subseteq U\) denote the set of probed items,
and $\activeelem(P)\subseteq P$ be the set of active elements in $P$.
The reward in this case is $f(\activeelem(P))$.
All other settings are the same as Definition~\ref{stoc:probing},
and our goal is also to find a probing strategy to maximize the expected reward \(\E[f(\activeelem(P))]\).
\end{definition}

In this work, we consider general objective functions, including monotone norms, subadditive functions, fractionally subadditive (XOS) functions, and \XOSS functions (a subclass of XOS functions where weight vectors have 0/1 coordinates, see \cite{bxos}). The detailed definitions and relationships among these function classes are provided in \cref{subsec:mathprel}.

In general, the optimal strategy for stochastic probing is \emph{adaptive}, meaning that elements are probed sequentially, with each decision based on previously observed outcomes. This decision process can be represented as an exponentially large tree, making it conceptually challenging and computationally expensive.

To address these difficulties, researchers often consider \emph{non-adaptive} strategies, which commit in advance to a probing sequence \(P \in \mathcal{F}\) without utilizing observed realizations. The expected reward in this case is \(\E[f(X_P)]\), where \(X_P\) contains the `revealed' values in $P$ with unprobed entries set to zero. Non-adaptive strategies significantly simplify computation and storage by eliminating the need to maintain a large decision tree.

To evaluate the effectiveness of non-adaptive strategies, we define the \emph{adaptivity gap} as the ratio between the expected reward of the optimal adaptive strategy and that of the best non-adaptive strategy.\footnote{It is known that one can restrict attention to deterministic adaptive strategies, which can be modeled as decision trees; see \cref{bernoulli:prel} for further discussion.} A small adaptivity gap justifies the use of non-adaptive strategies, which are typically easier to analyze. Consequently, proving tight adaptivity gaps has been a central challenge in stochastic optimization, with prior studies covering problems such as stochastic knapsack \cite{dean2004approximating}, stochastic covering \cite{goemans2006stochastic,agarwal2019stochastic}, stochastic matching \cite{BGL12}, stochastic orienteering \cite{BN14}, stochastic TSP \cite{jiang2020algorithms}, and the general stochastic probing problems with various objective functions \cite{GNS16,GNS17,PRS23,KMS24}.

We first review prior results on the adaptivity gap in stochastic probing. The Bernoulli stochastic probing problem with general feasibility constraints was introduced in \cite{GN13} and its adaptivity gap was systematically studied by Gupta and Nagarajan~\cite{GNS16}, who show a constant adaptivity gap for prefix-closed constraints when \(f\) is a linear or weighted coverage function. Gupta, Nagarajan, and Singla~\cite{GNS17} then extended these results by establishing constant adaptivity gaps for both monotone and non-monotone submodular functions. In the same paper, they considered the more general {\em subadditive functions},\footnote{Normally, the objective $f$ is always supposed to be \textbf{monotone} and \textbf{normalized} ($f(\emptyset) = 0$ or $f(\mathbf{0}) = 0$) unless stated otherwise, see also \cref{subsec:mathprel}. We usually omit the two conditions for clarity.} and conjectured that the adaptivity gap for all subadditive functions is poly-logarithmic in the size of the ground set (Conjecture 1). Recall that a set function $f:2^U\rightarrow
\mathbb{R}_{\geq 0}$ is subadditive if
$f(X\cup Y)\leq f(X)+f(Y)$ for all $X,Y\in 2^U$.
The class of subadditive functions is very general and includes submodular functions and XOS functions as special cases. They also showed that the adaptivity gap for XOS functions is \(O(\log W)\), where \(W\) denotes the width of the XOS representation. It is known that a subadditive function can be approximated by an XOS function
within a logarithmic factor \cite{Dob07, approxxos2}. Therefore, to resolve
Conjecture 1, it suffices to provide a poly-logarithmic upper bound
for XOS functions (for the Bernoulli case).
However, there are subadditive functions (even
submodular functions) that require exponentially 
large width in XOS representation \cite{BH11,BDF+}. For lower bound, the authors also show that there are instances where the adaptivity gap is at least $\Omega(\log W/\log\log W)$ for $W=\Theta(n)$.

The conjecture has attracted significant attention in recent years. 
Patton et al.~\cite{PRS23} proved the conjecture for symmetric norm objectives and general non-negative 
random variables (not necessarily Bernoulli).
In particular, they showed the adaptivity gap for 
symmetric norms is $O(\log n)$, and posed the open question that whether the true adaptivity gap is sub-logarithmic or even a constant
 (Conjecture 2).
Kesselheim et al.~\cite{KMS24} approached 
the problem from the notion of $p$-supermodularity
for norms. In particular, a norm is $p$-submodular
if its $p$th power is supermodular. They showed that
for $p$-supermodular norms, the adaptivity gap is at most $O(p)$. 
Motivated by the conjecture of \cite{GNS17},
Kesselheim et al.~\cite{KMS24} extended the conjecture to monotone norms,\footnote{
A monotone norm can be also be written as 
the max over linear functions, hence can be seen as
an extension of XOS set functions to the domain $\mathbb R_{\geq 0}^U$
} and conjectured that 
the adaptivity gap for stochastic probing is $\polylog n$ for any monotone norm  (Conjecture 3).

In this paper, we resolve all three conjectures
mentioned above.
For the Bernoulli stochastic probing problem,
we prove the following theorem, thus answering Conjecture 1
\cite{GNS17} affirmatively.

\begin{theorem}[See also \cref{thm:6.8}]
\label{thm:conj1}
The adaptivity gap for Bernoulli stochastic probing with any subadditive objective $f$ is upper bounded by $O(\log^3 n)$.
\end{theorem}
 
For the (general) stochastic probing problem with general monotone norms, 
we have the following theorem, confirming Conjecture 3 \cite{KMS24} affirmatively:
\begin{theorem}[See also \cref{thm:6.6}]
\label{main:conjecture}
The adaptivity gap for stochastic probing with any monotone norm $f$ (not necessarily symmetric) is upper bounded by
$O\left(\frac{\log r\log n}{\log\log n}\right)$, where $2\leq r\leq n$ is an upper bound for the maximal length of a sequence in $\calF$.
\end{theorem}

In view of the lower bound $\Omega\left(\log n/\log\log n\right)$ 
established in \cite{GNS17}, we reduce the gap between the upper and lower bounds to a factor of $O(\log r)$. 
Moreover, 
as a by-product of our proof for \cref{main:conjecture}, we establish that the adaptivity gap for Bernoulli stochastic probing with \XOSS objectives is upper bounded by $O(\log n/\log\log n)$. Notably, in the lower bound instance in \cite{GNS17}, the objective function $f$ is indeed a \XOSS objective. Therefore, our upper bound is asymptotically tight, as it matches the existing lower bound.
\begin{theorem}[See also \cref{thm:tight}]
\label{main:01XOS}
The adaptivity gap for Bernoulli stochastic probing with \XOSS objective is 
$\Theta \left(\frac{\log n}{\log\log n}\right)$.
\end{theorem}

Finally, we improve the result in \cite{PRS23} by reducing the upper bound of the adaptivity gap for monotone symmetric norms from \(O(\log n)\) to \(O(1)\), thereby answering one of the open problems posed in that work
(Conjecture 2).

\begin{theorem}[See also \cref{thm:6.7}]\label{main:sym}
The adaptivity gap for stochastic probing with monotone symmetric norm is $O(1)$.
\end{theorem}

\subsection{Other Related Work}
\label{sec:related}

Stochastic probing is a versatile framework that encompasses many well-studied problems in stochastic combinatorial optimization. For instance, stochastic matching \cite{cik09, ADA11, BGL12}—with applications in kidney exchanges \cite{RSU05, AW12} and online dating—considers a random graph \(G\) where probing an edge \(e\) reveals the outcome of an independent Bernoulli variable \(X_e\), indicating whether \(e \in E(G)\). Here, the objective function \(f\) is the size of the maximum matching, and the feasibility constraints are the valid matchings on the vertex set. Beyond matching, stochastic probing has been applied to Bayesian mechanism design \cite{GN13}, robot path planning \cite{GNS16, GNS17}, orienteering \cite{GM09, GKNR12, BN14}, and stochastic set cover problems in databases \cite{LPRY08, DHK14}.

The study of adaptivity in the Bernoulli stochastic probing problem with arbitrary prefix-closed constraints \(\mathcal{F}\) was initiated by Gupta et al. \cite{GNS17}. They showed that even with a simple cardinality constraint, there exists a monotone function \(f:2^U\to \mathbb{R}_{\ge 0}\) for which the adaptivity gap is at least \(2^{\Omega(\sqrt{n})}\), implying that further structural assumptions are necessary for tighter bounds. They further proved that for non-negative submodular functions \(f:2^U\to \mathbb{R}_{\ge 0}\), the gap is bounded by \(3\) (or \(40\) when \(f\) is non-monotone), with Bradac et al. \cite{BSZ19} later tightening the bound to \(2\) for monotone submodular functions. 

For specific constraints, Asadpour and Nazerzadeh \cite{AN16} established that the adaptivity gap for monotone submodular functions under a single matroid constraint is exactly \(\frac{e}{e-1}\). In addition, building on \cite{GNS16}, Bradac et al. \cite{BSZ19} demonstrated that the adaptivity gap for weighted rank functions of an intersection of \(k\) matroids or a \(k\)-extendible system (see \cite{Mes06,CCPV11}) lies between \(k\) and \(O(k\log k)\).

Recent research has focused on general objectives such as ordered norms \cite{byrka2018constant,chakrabarty2018interpolating} and symmetric norms \cite{chakrabarty2019approximation,chakrabarty2019simpler,ibrahimpur2020approximation,deng2023generalized,abbasi2023parameterized,KMS24,chen2025new}. Approximation algorithms for norm minimization, initiated by Chakrabarty and Swamy \cite{chakrabarty2019approximation}, have been applied to problems including generalized load balancing \cite{deng2023generalized,ayyadevara2023minimizing}, stochastic optimization \cite{ibrahimpur2020approximation}, online algorithms, and stochastic probing \cite{PRS23}, as well as parameterized algorithms \cite{abbasi2023parameterized}. A key technique in these works is to approximate general norms by classical ones (e.g., ordered, \textsf{Top-}\(k\), Orlicz, or super/submodular norms; see \cite{KMS23,KMS24}), which is particularly effective for analyzing the adaptivity gap.

In particular, Patton et al.~\cite{PRS23} showed that any monotone symmetric norm can be approximated within a factor of \(O(\log n)\) by a submodular norm, yielding an adaptivity gap of \(O(\log n)\). Kesselheim et al. \cite{KMS24} introduced \(p\)-supermodular norms and established an \(O(p)\) adaptivity gap for this class, further showing that any monotone symmetric norm can be approximated by an \(O(\log n)\)-supermodular norm, leading again to an \(O(\log n)\) gap.

A summary of these results, along with a comparison of our findings on adaptivity gaps for stochastic probing with prefix-closed constraints, is provided in \cref{tab:adaptivity_gap}.

\renewcommand{\arraystretch}{1.5}  

\begin{table}[ht]
    \centering
    \small
    \begin{threeparttable}
    \begin{tabular}{p{0.15\columnwidth} p{0.4\columnwidth} p{0.35\columnwidth}}
        \toprule
        Variables $X_i$ & Objective Class $f$\tnote{*} & Adaptivity Gap \\
        \midrule
        Bernoulli & Weighted rank of $k$-extendible system  & $[k, O(k\log k)]$~\cite{BSZ19} \\
        Bernoulli & Submodular & $\le 3$~\cite{GNS17} (= $2$ in~\cite{BSZ19}) \\
        Bernoulli & Non-monotone submodular & $\le 40$~\cite{GNS17} \\
        Bernoulli & XOS of width $W$  & $O(\log W)$~\cite{GNS17}\\
        Bernoulli & \XOSS & $\Omega\!\left(\frac{\log n}{\log\log n}\right)$~\cite{GNS17}\\
        Bernoulli & \XOSS & $O\!\left(\frac{\log n}{\log\log n}\right)$~(\cref{main:01XOS})\\
        Bernoulli & XOS & $O\!\left(\frac{\log r\log n}{\log\log n}\right)^\dagger$~(Theorem \ref{thm:strong}) \\
         Bernoulli & Subadditive  & $O(\log^3 n)$~(\cref{thm:conj1}) \\
        \midrule
        General & Symmetric norm & $O(\log n)$~\cite{PRS23,KMS24} \\
        General & Symmetric norm & $O(1)$~(\cref{main:sym}) \\
        General & $p$-supermodular norm & $O(p)$~\cite{KMS24} \\
        General & General norm & $O\!\left(\frac{\log r\log n}{\log\log n}\right)$~(Theorem \ref{main:conjecture}) \\
        \bottomrule
    \end{tabular}
    \begin{tablenotes}
        \footnotesize
        \item[*] All objectives are assumed to be \textbf{monotone} and \textbf{normalized} ($f(\emptyset) = 0$ or $f(\mathbf{0}) = 0$) unless stated otherwise.
        \item[$\dagger$] $2\leq r\leq n$ denotes an upper bound for the maximal length of a sequence in feasibility constraint $\calF$.
    \end{tablenotes}
    \end{threeparttable}
    \caption{Adaptivity gaps for stochastic probing with prefix-closed constraints}
    \label{tab:adaptivity_gap}
\end{table}

\subsection{Organization of the Paper}
The remainder of the paper is organized as follows. In \cref{subsec:overview}, we provide a detailed overview of our proof. 
In \cref{section2}, we present the necessary background and preliminaries. In \cref{section 3}, we prove our main theorem (\cref{thm:weak}), showing that the adaptivity gap for Bernoulli stochastic probing with any XOS objective is upper bounded by \(O(\log r \log n)\), where \(2\leq r\leq n\) is an upper bound for the maximum length of a sequence in \(\calF\). 
In \cref{Section 4}, through more refined analysis, we further improve this bound to \(O\!\Bigl(\log r \cdot \frac{\log n}{\log\log n}\Bigr)\) (see \cref{thm:strong}) and also obtain the tight bound $\Theta\!\Bigl(\frac{\log n}{\log\log n}\Bigr)$ of the adaptivity gap for Bernoulli stochastic probing with \XOSS objective.
In \cref{section5}, we establish that the adaptivity gap for Bernoulli stochastic probing with any symmetric norm objective (defined in \cref{bernoulli:symmetric}) is \(O(1)\) (see \cref{thm:sym}). 
Finally, in \cref{section 6}, we invoke a well-known result that any subadditive objective can be approximated by an XOS function within logarithmic factors to obtain poly-logarithmic adaptivity gap for Bernoulli stochastic probing with any subadditive objective (see \cref{thm:6.8}). In the same section, we present a reduction from the setting of general random variables to the Bernoulli setting, which leads to a poly-logarithmic adaptivity gap for (general) stochastic probing with any monotone norm (see \cref{thm:6.6}) and $O(1)$ gap for (general) stochastic probing with symmetric monotone norm (see \cref{thm:6.7}). 
\subsection{Proof Overview}\label{subsec:overview}

In this section, we provide an overview of the proof for \cref{main:conjecture}, the main result of this paper. For a roadmap of our proof, please refer to \cref{figroad}. For simplicity, we present only the main ideas that establish an \(O(\log^2 n)\) upper bound on the adaptivity gap for any monotone norm objective.

\begin{figure}[htbp]
    \begin{centering}
    \begin{tikzpicture}[
        node distance = 3em and 13em,  
        box/.style = {draw, rectangle, minimum width=3.5cm, minimum height=1.6em, align=center},
        >={Latex[length=2mm,width=2mm]}  
    ]
    
    \node[box] (box1) {Stochastic Probing \\with Random Variables\\ and General Norm};
    \node[box, right=of box1] (box2) {Bernoulli Stochastic Probing\\ with General Norm};
    
    \node[box, below=of box2] (box3) {Bernoulli Stochastic Probing\\ with a \emph{Special} XOS Norm};
    \node[box, below=of box1, yshift= 0.6em] (box4) {$O(\log n\log r)$\\ Adaptivity Gap};
    
    \draw[->] (box1) -- node[above]{\cref{section 6}} (box2);
         \draw[->] (box2) -- node[left, pos=0.48]{\cref{sec3:redcution}} node[right, pos=0.5]{Losing $O(\log r)$} (box3);
    \draw[->] (box3) -- node[above]{\cref{sec3:prepare,sec3:main}}  node[below]{\footnotesize  $O(\log n)$ Gap for \emph{Special} XOS} (box4);
    A\end{tikzpicture}
    \caption{Roadmap for \cref{main:conjecture}.}\label{figroad}
    \end{centering}
    \end{figure}
    
\subsubsection{Step 1: Reduction to Bernoulli Setting}
\label{subsubsec:red2Ber}

We first reduce general random variables to  Bernoulli random variables.
The idea of the reduction is standard
and already appears implicitly in \cite{Sin18,GNS17}. 
Suppose that we aim to prove that the adaptivity gap is bounded by the function $\gapfunction(n)\geq 2$. We denote the expected reward for the optimal (adaptive) strategy by $\OPT$. 
Without loss of generality, we assume $\OPT=1$ by scaling.


We say that an outcome $x_i$ of $X_i$ is $\lambda$-large if $f(x_i)\geq \lambda\cdot\OPT$.\footnote{$f(x_i)$ is a shorthand for $f(0,0,\cdots,0,x_i, 0,0,\cdots,0)$.}
Otherwise, we say it is $\lambda$-small.
Then, for any probed sequence $P \subseteq [n]$, we can decompose the random variable $X_P$
into two parts: $Y_{P,\lambda}$ only contains $\lambda$-large outcomes (with placeholder zero if an outcome is $\lambda$-small), and $Z_{P,\lambda} = X_P -Y_{P,\lambda}$ contains the remaining $\lambda$-small outcomes. We also call a random variable $\lambda$-small if all of its possible outcomes are $\lambda$-small.
An analogue of the work in \cite{GNS17} shows that, if $\E[f(Y_{P,\lambda})]\geq \mathsf{OPT}/2$, then the adaptivity gap is at most $O(1+1/\lambda)$, see also \cref{lm:general:large}. 
Therefore, we can always assume $\E[f(Z_{P,\lambda})]\geq \mathsf{OPT}/2$ using subadditivity of $f$, if we set $\lambda = \gapfunction(n)^{-1}$. Hence, we can replace $X_i$ by the random variables $Z_{i,\lambda}$ with only 
$\lambda$-small outcomes. 
Since $\OPT=1$, by the above procedure, 
we can assume that all outcomes of $X_i$ are $1\geq 2\gapfunction(n)^{-1}$-small. We can also ignore any outcome $x_i$ of $X_i$ such that $f(x_i) \leq 1 / (2n)$
since their total contribution is at most $1/2$.
Now, we round the outcome of $X_i$ to the nearest negative power of $2$. Hence, each random variable $X_i$ has at most $2\log n$ different outcomes,
which implies that each node in the decision tree (corresponding to an adaptive strategy)
has at most $O(\log n)$ children.
We can replace the current decision tree by a binary decision tree, which corresponds to a 
Bernoulli stochastic probing problem with a monotone norm objective. 
The details can be found in \cref{sec6:proof}.
Moreover, it is well known that the restriction of any monotone norm on $2^U$ defines an XOS function, see \cref{subsec:mathprel} for details. Hence, we obtain the following theorem.

\begin{theorem}[See also \cref{thm:geneXOS}]
    Suppose the adaptivity gap for \emph{Bernoulli stochastic probing with XOS objective} is upper bounded by $\gapfunction_{\XOS}(n)$. Then, the adaptivity gap for stochastic probing with any general monotone norm is upper bounded by $O(\gapfunction_{\XOS}(n\log 8n))$.
\end{theorem} 

\subsubsection{A Natural Non-adaptive Strategy}
\label{subsec:intro_adaptive_nonadaptive}
In the following, it suffices to consider the Bernoulli setting.
We first need to define some notations regarding the decision tree for an adaptive strategy, and then introduce the construction of the non-adaptive strategy.

\vspace{0.1cm}
\noindent \textbf{Adaptive Strategies.} 
We denote the optimal adaptive strategy by \adaptivestrategy, which can be represented by a rooted binary tree \(\mathcal{T}\), where each internal node \(v\) represents an element \(i \in U\), denoted as \(\elt(v) = i\). We also say that the node $v$ \textbf{represents} the element $i$. If \(i\) is {\em active} (with probability \(p_i\) given in the definition of Bernoulli setting), the \textsf{Yes} arc is followed; otherwise, the \textsf{No} arc is taken. We require that no element can be represented by more than one node on any root-to-leaf path in the tree, so the length of each root-leaf path is trivially bounded by $n$. Each root-to-leaf path must be feasible, i.e., the element sequence must be contained in $\calF$. The leaves indicate the termination of the probing process but do not represent elements in \(U\).

Each leaf \(\ell\) is associated with a unique path \(P_\ell\), which consists of all nodes visited from the root to \(\ell\). 
For a node $v$, we call an ancestor $u$ of $v$ its \textbf{active ancestor} if the \textsf{Yes} arc from $u$ is taken on the path from root to $v$. 
Further, we call a node of \textbf{depth} $k$ if it has $k$ active ancestors. An illustration is shown in \cref{fig1}. The set of active ancestors of $\l$ on \(P_\ell\) is denoted as \(A_\ell\). Nodes in one single $A_\l$ must have different depth. A node belongs to $A_\l$ if its corresponding element is active when probed in the adaptive strategy \adaptivestrategy.
For any subset $S \subseteq P_\ell$, we denote the
corresponding set of elements as $\elt(S) = \{\elt(v) \in U : v \in S\}$.

The tree induces a probability distribution \(\pi_{\mathcal{T}}\) over its leaves. Given a monotone function \(f\), the expected reward of \adaptivestrategy\ is:  
\[
\adaptcost(\mathcal{T}, f) := \E_{\ell \leftarrow \pi_{\mathcal{T}}} \left[f(\elt(A_\ell))\right].
\]

\noindent \textbf{A Natural Non-adaptive Strategy.}
We consider a natural non-adaptive strategy 
which is derived from \adaptivestrategy (with tree $\mathcal{T}$). The non-adaptive strategy first selects a leaf \(\ell \leftarrow \pi_{\mathcal{T}}\), probes all elements on 
the path \(P_\ell\), and selects the active elements. Clearly, this non-adaptive strategy satisfies the feasibility constraint.
We denote such strategy as \nonadaptivestrategy. The same non-adaptive strategy has been studied in \cite{GNS16, GNS17,BSZ19}.
Note that the realization of the elements during the 
probing process of \nonadaptivestrategy
is independent of that in
\adaptivestrategy (e.g., a node $v\in A_{\ell}$
may be inactive in \nonadaptivestrategy). 
To capture the randomness of the realizations in 
the probing process of \nonadaptivestrategy,
we define $\Rdistribution$, which is the distribution of the random subset $R$ of nodes on $\calT$. 
In particular, the random subset $R$ is obtained as follows:
for each element $e$, with independent probability $p_e$, all nodes $v$ that $\elt(v) = e$ are included in $R$. 
In other words, each node \(v \in P_\ell\) is included in \(R\) with probability \(p_{\elt(v)}\), which means that its corresponding element is active in \nonadaptivestrategy. 
Hence, we can describe the non-adaptive strategy
\nonadaptivestrategy\ as first sampling a leaf \(\ell \leftarrow \pi_{\mathcal{T}}\), then collecting all  elements in $R\cap P_\l$, where $R\sim \Rdistribution$. Therefore, the expected reward of 
\nonadaptivestrategy\ is:
\[
\nonadaptcost(\mathcal{T}, f) := \E_{\ell \leftarrow \pi_{\mathcal{T}}} 
\left[ \E_{R\sim \Rdistribution} \left[f(\elt(R \cap P_\ell))\right] \right].
\]  

By an argument similar to \cref{subsubsec:red2Ber},
we can assume without loss of generality that there are no 
"large" elements (those $e$ with $f(e)\geq \gapfunction(n)^{-1}\adaptcost(\mathcal{T}, f)$), where 
$\gapfunction(n)$ is an upper bound of the adaptivity gap.
Hence, by scaling the function $f$, it suffices to focus on the case where $\adaptcost(\mathcal{T}, f) = \gapfunction(n)$ and $f(e_j)\leq 1, \forall j\in U $, and we aim to show that $\alg(\calT,f)\geq 1$
in such a setting. 

\subsubsection{Step 2: Reduction to a Special XOS Norm}
\label{subsecspecial}
Now, we would like to further simplify the XOS norm 
to a special family of XOS norms.  
We first recall the XOS norm as  
$
f(S)=\max_{i\in [W]}\Bigl\{\sum_{j\in S}w_{j}^{(i)}\Bigr\},
$ where 
$W$ is the width of the XOS norm and $w_{j}^{(i)}\geq 0$
for all $i,j$. Since $f(e_j)\leq 1$, all $w_{j}^{(i)}\leq 1$ for all $i\in [W], j\in U$. 
Next, by rounding each weight $w^{(i)}_j$ 
to a negative power of $2$ and ignoring those less than $1/(2n)$, we are left with an instance with only 
$O(\log n)$ possible weights. By picking the one with the largest contribution
(suppose it is the one with weight $2^{-j_0}$), by losing a factor of at most $O(\log n)$, 
it suffices to deal with the case where all weights are $2^{-j_0}$ or zero.

\eat{
To further expand $f(\elt(A_\l))$, for each $\l$, we need to assign an index $I(\l)\in [W]$ so the maximum in $f(\elt(A_\l))$ is achieved by $\sum_{j\in \elt(A_\l)}w^{(I(\ell))}_{j}$. Then, we can lower bound $f(\elt(R \cap P_\ell))$ as follows:
$$
f(\elt(R \cap P_\ell))=\max
\limits_{i \in [W]}\left\{\sum_{j\in  \elt(R\cap P_\l)}w^{(i)}_{j}\right\} \geq \max
\limits_{\l'}\left\{\sum_{j\in  \elt(R\cap P_\l)}w^{(I(\l'))}_{j}\right\}\geq \max
\limits_{\l'}\left\{\sum_{j\in  \elt(R\cap P_\l)\cap \elt(A_{\l'})}w^{(I(\l'))}_{j}\right\}.
$$
Currently, the weighted version is still hard to deal with. We use the classical technique of rounding to reduce the general weighted to 0/1 weights losing only an $O(\log n)$ factor. 
Intuitively speaking, the process rounds each weight $w^{(I(\ell))}_{\elt(u)},u\in A_\ell$ for each $\l$ to a negative power of $2$ (ignore less than $1/(2n)$ weight). Then, we only pass to consider one of the $O(\log n)$ possible weights that contribute to the maximum in $\E_\l[f(\elt(A_\l))]$, and the non-adaptive strategy can be lower-bounded by \emph{only} the contribution of such weight. Therefore, by losing an $O(\log n)$ factor, we are left to deal with all weights are a fixed negative powers of $2$ or zero.

Indeed, for each $\ell$ and integer $1\leq j\leq \log n+2$, we define $A_{\ell,j}=\left\{u:u\in A_\ell, 2^{-j}< w^{(I(\ell))}_{\elt(u)}\leq 2^{-i+1}\right\}$ and an additional $A_{\ell,0 }$ to be $A_\l\backslash \bigcup_{i=1}^{\lfloor\log n\rfloor+2} A_{\l,i}$. For any node $ u \in A_{\l,0}$, $w^{(I(\ell))}_{\elt(u)} \leq 1/(2n)$. 
}

Through the above manipulations (see the full details
in \cref{lm:reduce}), 
we arrive at a special family of XOS norms as follows: for each $\l$, there is a subset $A_\l' \subseteq A_\l$, and the special XOS norm objective (with respect to the subsets $A_\l'$) is
$$
\fspecial(S) = \max_{\ell} \left\{ | \elt(A_\ell') \cap S | \right\}. 
$$
\eat{
Suppose that we are able to prove when $\E_{\ell \leftarrow \pi_{\mathcal{T}}}\left[|A_\ell'|\right]  =\Omega(h(n))$, 
$$
\E_{\ell \leftarrow \pi_{\calT}} \left[ \E_{R \sim \Rdistribution} \left[f_0(\elt(R\cap P_\ell))\right] \right]\geq \frac{1}{h(n)}\E_{\ell \leftarrow \pi_{\mathcal{T}}}\left[f_0(\elt(A_\l'))\right] = \frac{1}{h(n)}\E_{\ell \leftarrow \pi_{\mathcal{T}}}\left[|A_\ell'|\right],
$$
then plugging in $A_\l' = A_{\l,j_0}$, we already have:
\begin{align*}
    \alg(\mathcal{T}, f)  &\geq 2^{-j_0} \E_{\ell \leftarrow \pi_{\calT}} \left[ \E_{R \sim \Rdistribution} \left[ \max \limits_{\l'}\left\{|\elt(R\cap P_\l)\cap \elt(A_{\l',j_0})|\right\} \right] \right] \\
    & = 2^{-j_0}\E_{\ell \leftarrow \pi_{\calT}} \left[ \E_{R \sim \Rdistribution} \left[f_0(\elt(R\cap P_\ell))\right] \right] \geq \frac{1}{2^{j_0}h(n)}\E_{\ell \leftarrow \pi_{\mathcal{T}}}\left[|A_{\l,j_0}|\right] = \Omega\left( \frac{1}{\log n}\right),
\end{align*}
and thus by scaling $f$ and $h$ by $\log n$, we can conclude that the adaptivity gap is $O(h(n)\log n)$.
}
Let us assume that, for $\fspecial$, we can prove the following statement:
when $\E_{\ell \leftarrow \pi_{\mathcal{T}}}\left[|A_\ell'|\right]  =\Omega(\frac{\gapfunction (n)}{\log n})$, we have that
$$
\E_{\ell \leftarrow \pi_{\calT}} \left[ \E_{R \sim \Rdistribution} \left[\fspecial(\elt(R\cap P_\ell))\right] \right]\geq \frac{\log n}{\gapfunction (n) }\E_{\ell \leftarrow \pi_{\mathcal{T}}}\left[|A_\ell'|\right].
$$
As we argued before, it suffices to study the case where $\adaptcost(\mathcal{T}, f) = \gapfunction(n)$, and we aim to show that $\alg(\calT,f)\geq 1$.
By the construction of $\fspecial$, $f\geq 2^{-j_0}\fspecial$ and $\E_{\ell \leftarrow \pi_{\mathcal{T}}}\left[|A_\ell'|\right]  =\Omega\left(2^{j_0}\frac{\gapfunction (n)}{\log n}\right)$ (by the maximality of the contribution of weight $2^{-j_0}$ to
$\adaptcost(\mathcal{T}, f)$).
Hence, we have the following:
\begin{align*}
    \nonadaptcost(\mathcal{T}, f)  &\geq 
    2^{-j_0} \nonadaptcost(\mathcal{T}, \fspecial) =
     2^{-j_0}\E_{\ell \leftarrow \pi_{\calT}} \left[ \E_{R \sim \Rdistribution} \left[\fspecial(\elt(R\cap P_\ell))\right] \right] \\
     & \geq \frac{\E_{\ell \leftarrow \pi_{\mathcal{T}}}\left[|A'_{\l}|\right]}{2^{j_0}}\cdot \frac{\log n}{\gapfunction(n)}   = \Omega\left( 1\right).
\end{align*}
Thus, we can conclude that the adaptivity gap is $O(\gapfunction(n))$.
We encapsulate the above argument 
(with $\gapfunction(n)=O(\log^2 n)$) as follows:
\begin{lemma}[See also \cref{lm:reduce}]\label{view:lm:reduce}
Suppose that for any decision tree $\calT$ and any choice of subsets $A_\ell' \subseteq A_\ell$ satisfying 
$
\E_{\ell \leftarrow \pi_{\mathcal{T}}}[|A_\ell'|] \geq 20\log n,
$ 
the following condition holds:
    $$\nonadaptcost(\mathcal{T}, \fspecial)=\E_{\ell \leftarrow \pi_{\calT}} \left[ \E_{R \sim \Rdistribution} \left[\fspecial(\elt(R\cap P_\ell))\right] \right]\geq \frac{1}{20\log n}\E_{\ell \leftarrow \pi_{\mathcal{T}}}\left[|A_\ell'|\right].$$
    Then, the adaptivity gap of Bernoulli stochastic probing with XOS objective $f(\cdot)$ is upper bounded by $O(\log^2 n)$.    
\end{lemma}

\subsubsection{Warm Up: A Simplified Setting}
\label{subsec:warmup}
From the previous sections, it suffices to prove the condition of \cref{view:lm:reduce} holds.
To begin with, we consider the following simplified case, 
where the number of leaves of $\calT$ is $O(n)$. 
The simplified case motivates the most technical part of the proof,
presented in \cref{subsec:classify}.

When the number of leaves of $\calT$ is $O(n)$,
we can directly observe the lower bound 
\begin{align}
\label{eq:lowerbound_fxos}
\fspecial(\elt(R\cap P_\ell))\geq |\elt(R\cap P_\ell) \cap \elt(A_\l') | = |R\cap A_\l'|
\end{align}
for any leaf $\l$.
Then, by definition of the distribution $ \pi_{\calT}$, $p_\l$ is 
the probability of reaching the leaf $\l$, hence the product of the probability for each node on $P_\l$ taking the corresponding \textsf{Yes} or \textsf{No} arc. 
Therefore, using AM-GM inequality, we obtain
$$
p_\l = \prod_{u\in A_\l} p_{\elt(u)}\cdot \prod_{u\in P_\l\backslash (A_\l\cup \{ \l \})} (1-p_{\elt(u)})\leq \prod_{u\in A_\l'} p_{\elt(u)}\leq \left(\frac{\sum_{u\in A_\l'} p_{\elt(u)}}{|A_\l'|}\right)^{|A_\l'|}.
$$
Moreover, it is clear that $\E_{R\sim  \Rdistribution}|R\cap A_\l'| = \sum_{u\in A_\l'} p_{\elt(u)}$. Putting together, we have
\begin{align*}
\nonadaptcost(\mathcal{T}, \fspecial)& =
\E_{\ell \leftarrow \pi_{\calT}} \left[ \E_{R \sim \Rdistribution} \left[\fspecial(\elt(R\cap P_\ell))\right] \right] 
\geq \E_{\ell \leftarrow \pi_{\calT}} \left[ \E_{R \sim \Rdistribution} \left[|R\cap A_\l'|\right] \right] \\
&\geq\E_{\ell \leftarrow \pi_{\calT}} \left[|A_\l'| p_\l^{1/|A_\l'|} \right] = \sum_{\l}|A_\l'| p_\l^{1 + 1/|A_\l'|}.
\end{align*}
Now, suppose that $\E_{\ell \leftarrow \pi_{\calT}} |A_\l'|=\sum_{\l}p_\l|A_\l'|\geq 10\log n$.
We only need to consider the “good” leaves $\l$ such that $|A_\l'|\geq \log n$ and $p_\l\geq 1/n^2$, since the contribution of other leaves is at most $ \log n+O(n) / n^2$ (recall the number of leaves is bounded by $O(n)$), and hence is at most half of $\E_{\ell \leftarrow \pi_{\calT}} |A_\l'|$. Now, for a good leaf $\l$, we have $p_\l^{1/|A_\l'|}\geq n^{-2 / \log n } = \Omega(1)$. Therefore, we conclude
$$
\nonadaptcost(\mathcal{T}, \fspecial)
\geq \sum_{\l \text{ good}}|A_\l'| p_\l^{1 + 1/|A_\l'|}\geq \Omega(1) \cdot \sum_{\l \text{ good}}|A_\l'| p_\l \geq \Omega(1) \cdot \E_{\ell \leftarrow \pi_{\calT}} [|A_\l'|].
$$
Therefore, the condition for \cref{view:lm:reduce} indeed holds for this special case (with even a better factor). 

\vspace{0.2cm}
\noindent
{\bf Limitation of the Above Proof:}
For the general setting, the above proof
does not work anymore. 
Indeed, consider the example where $A_\l = A_\l'$ and $\calT$ is a complete binary tree of height $h$, where each node has probability $\lambda$ to be active. 
Simple calculation shows that $\E_{\ell \leftarrow \pi_{\calT}}| A_\l| = \lambda h$ and $\E_{\ell \leftarrow \pi_{\calT}}|R\cap A_\l| = \lambda^2h$. 
If $\lambda$ is extremely small,
$\E_{\ell \leftarrow \pi_{\calT}}|R\cap A_\l| \ll \E_{\ell \leftarrow \pi_{\calT}}| A_\l| $.
For example, if $\lambda h = \log^{k} n$, and $\lambda =\log^{-2k} n$, $k =\omega(1)$, the above proof does not yield a
poly-logarithmic ratio.
Note that in this example, 
the number of nodes is $2^h\gg n$,
the probability $p_{\l}$ of a typical leaf 
is extremely small (much smaller than $1/\poly(n)$).

\vspace{0.2cm}
\noindent
{\bf Remedy:}
We observe in the above example that $R$ typically has very small intersection with $A_\ell$ (when $\lambda$ is very small).
In fact, simply lower bounding $\fspecial(\elt(R\cap P_\ell))$ by $|R\cap A'_\l|$ as in \eqref{eq:lowerbound_fxos} 
is no longer sufficient in this case. 
The remedy is to consider some leaf $\l'$ other than $\l$ itself to make $|\elt(R\cap P_\ell)\cap\elt(A_{\l'}')|$ as large as possible (thus improving \eqref{eq:lowerbound_fxos}).
For example, in \cref{fig:binary_tree_part}, 
$|\elt(R\cap P_\ell)\cap\elt(A_{\l})|=1$ (only $\elt(v_4)$), but we can find another leave $\ell_3$
such that $|\elt(R\cap P_\ell)\cap\elt(A_{\l'}')|=4$
($\elt(v_1),\elt(v_2),\elt(v_3),\elt(v_4)$).

In order to find leaf $\ell'$
such that the intersection $|\elt(R\cap P_\l \cup A_{\l'}')|$ is large, 
we design a greedy algorithm. On input $(\l,R)$, the algorithm starts from leaf $\l$, climbs up along $P_\l$ and tries to select a subset of nodes in $R \cap P_\l$ that is guaranteed to be contained within $A_{\l'}'$ for some other leaf $\l'$.
The main idea of the greedy algorithm is presented below in \cref{subsec:classify}.
Note that we do not care the efficiency of the algorithm, which is only used for 
the analysis. 
We use $B(\l,R)$ to denote the output of the algorithm on input $(\ell, R)$,
which is guaranteed to be a subset of $\elt(R\cap P_\l) \cup \elt(A_{\l'}')$
for some leaf $\ell'$.
Therefore, upon termination of the algorithm, the reward 
$\fspecial(\elt(R\cap P_\ell))$
is at least $B(\l,R)$.
We prove a {\em tail bound} stating that the probability that the algorithm terminates with $|B(\ell,R)|$ being much smaller than $A_\ell$ is exponentially small
(see \eqref{eq:tail}).

\subsubsection{Step 3: The Greedy Algorithm and Proof of \cref{view:lm:reduce}
for $A_\l = A_\l'$}
\label{subsec:classify}
In this section, we sketch the main idea of the proof. 
To avoid confusion, we recall that the word \emph{active ancestor}
refers to the active nodes in \adaptivestrategy, 
while we use $v\in R$ to indicate that $\elt(v)$ is realized \emph{active} in the probing of the nonadaptive strategy \nonadaptivestrategy. We only consider the case $A_\l = A_\l'$ for simplicity: extending to the general case follows essentially the same idea but with more technical details.
Recalling \cref{view:lm:reduce}, in this section, we aim to prove that under condition $\E_{\l}[|A_\l|]\geq 20\log n$, it holds that
\begin{align}
\label{eq:lowerbound_nonalgo_xos}
\E_{\ell \leftarrow \pi_{\calT}} \left[ \E_{R \sim \Rdistribution} \left[\fspecial(\elt(R\cap P_\ell))\right] \right]\geq \frac{1}{20\log n}\E_{\ell \leftarrow \pi_{\mathcal{T}}}\left[|A_\ell|\right],
\end{align}
where $\fspecial(S) = \max_{\ell} \left\{ | \elt(A_\ell) \cap S | \right\}$.

As alluded in \cref{subsec:warmup} above, for each $(\ell,R)$, we want to find a leaf $\ell_0$ to make $|\elt(R\cap P_\ell)\cap \elt(A_{\ell_0})|$ as large as possible.
The intuition is to use a \emph{greedy} algorithm to find this $\ell_0$. We describe the algorithm as follows.
The input of the algorithm is $(\l, R)$.
We iterate from the leaf $\l$ to the root in $P_\ell$, while maintaining a sequence $B=(e_1,\ldots,e_m)$, where $e_i$ are elements in $U$. $B$ is initialized as an empty sequence. At each node $v\in P_\l$, we examine whether the following two conditions hold:
\begin{enumerate}[(1)]
\item $v\in R$. 
\item There is a leaf $\l'$ of $v$ such that $v\in A_{\l'}$, and for each element $e_i$ in the current sequence $B$, there is a descendant of $v$ (not allowed to be $v$ itself) in $A_{\l'}$ that \emph{represents} $e_i$.
\end{enumerate}
If both conditions hold, we append $\elt(v)$ to the end of sequence $B$. 
Otherwise, we do nothing.
Then we continue to consider the parent of $v$. Finally, the algorithm outputs $B$, denoted as $B(\l,R)$. We use $|B(\l,R)|$ to denote the number of elements in the sequence $B(\l,R)$.
An example of the algorithm is illustrated in \cref{fig:binary_tree_part}, where only a portion of the binary tree is drawn for clarity.  
\begin{figure}[t]    
    \centering  \scalebox{0.8}[0.8]{
    \begin{tikzpicture}[
        every node/.style={draw, circle, fill = none, inner sep=1.5pt},
        green node/.style={draw, circle, fill = green, inner sep=2.25pt},
        left edge/.style={draw=blue, line width=1.2pt, ->},
        right edge/.style={draw=red, line width=1.2pt, ->}
      ]

    \node (legend_node) at (8,3.6) {};
    \node[draw=none, fill=none] (legend_no) at (6.8,2.2) {};
    \node[draw=none, fill=none] (legend_yes) at (9.2,2.2) {};
    \draw[left edge] (legend_node) -- node[left, draw=none, fill=none, pos=0.35] {\textsf{No}} (legend_no);
    \draw[right edge] (legend_node) -- node[right, draw=none, fill=none, pos=0.35] {\textsf{Yes}} (legend_yes);

    \node (n4) at (0,-3.6) [label=right:$v_1$, green node]{};
    \node (u2c2) at (-1,-4.4) {}; 
    \node (n6) at (-1,-2.8) [label=left:$w$]{};
    \node (v2) at (0,-2.0) [label=right:$v_2$, green node] {}; 
    \node (u1) at (1,-2.8) {}; 
    \node (u2) at (2,-3.6) [label=right:$v_1'$, green node]{}; 
    \node (u2c) at (3,-4.4) [label=right:$\l_1$] {}; 
    \node (v1) at (1,-4.4) [label=right:$\l$]  {}; 
    \node (n9) at (-1,-1.2) [label=left:$u_1$] {};
    \node (v3) at (0,-0.4) [label=left:$v_3$, green node] {}; 
    \node (v2p) at (1,-1.2) [label=right:$v_1''$, green node]{};; 
    \node (v1pp) at (2,-2.0)[label=right:$v_2'$, green node]  {}; 
    \node (v1ppc1) at (3,-2.8){};
    \node (v1ppcc1) at (4,-3.6) [label=right:$\l_2$] {};
    \node (n10) at (1,0.4) {};
    \node (v4) at (1,3.6) [label=left:$v_4$, green node]{};
    \node (u5) at (0,1.2) [label=right:$u_2$]{};
    \node (u6) at (-1, 2.0) [label=left:$u_3$]{};
    \node (u7) at (0, 2.8) {};
    \node (u8) at (2, 2.8) {};
    \node (u9) at (1, 2.0) {};
    \node (u4) at (3,0.4) {};
    \node (v3p) at (4,-0.4)[label=right:$v_2''$, green node]{};
    \node (n12) at (2,1.2) [label=right:$v_3'$, green node]{};

    \node (n9c) at (3,-1.2)  {};
    \node (n9cc) at (4,-2.0)  {};
    \node (n9ccc) at (5,-2.8)  {};
    \node (v2pc) at (5,-1.2) [label=right:$v_1'''$, green node]  {};
    \node (v1ppc) at (6,-2.0) [label=right:$\l_3$]  {};
    \draw[right edge]  (n4)-- (v1);
    \draw[right edge] (n6) -- (n4);
    \draw[left edge] (v2) -- (n6);
    \draw[right edge] (v2) -- (u1);
    \draw[right edge]  (u1) -- (u2) ;
    \draw[right edge] (n9) -- (v2);
    \draw[left edge] (v3) -- (n9);
    \draw[right edge]  (v3)--(v2p);
    \draw[right edge] (v2p)--(v1pp) ;

    \draw[left edge] (n10) -- (v3);
    \draw[right edge] (u5) -- (n10);
    \draw[right edge] (u6) -- (u5);
    \draw[left edge] (u7) -- (u6);
    \draw[left edge] (v4) -- (u7);
    \draw[left edge] (u8) -- (u9); 
    \draw[right edge] (v4) -- (u8); 

    \draw[right edge] (u9) -- (n12);
    \draw[right edge] (n12) -- (u4);
    \draw[right edge] (u4) -- (v3p);

    \draw[left edge]   (v3p) -- (n9c);
    \draw[right edge] (n9cc) -- (n9ccc);
    \draw[right edge] (n9c) -- (n9cc);
    \draw[right edge] (v3p) -- (v2pc) ;
    \draw[right edge] (v2pc)--(v1ppc);
    \draw[left edge] (n4)--(u2c2);
    \draw[right edge] (u2)--(u2c);
    \draw[right edge] (v1pp)--(v1ppc1);
    \draw[right edge] (v1ppc1)--(v1ppcc1);

    \end{tikzpicture}}
    \caption{\small In the decision tree, red edges indicate \textsf{Yes} arcs, and blue edges indicate \textsf{No} arcs. We suppose nodes labeled with the same letters (but with different number of primes) represent the same element: $\elt(v_1)=\elt(v_1')=\elt(v_1'')=\elt(v_1''')$, $\elt(v_2)=\elt(v_2')=\elt(v_2'')$, etc. $v_4$ is the root of the whole tree. \newline
    Suppose that the specific outcome $R_0$ includes all green nodes in $R_0$, but excludes all white nodes. The algorithm with input $(\l,R_0)$ first visits $\l$ and next discovers that $v_1\in A_\l$ satisfies both conditions. Thus, $B\gets \{\elt(v_1)\}$. After moving up one step without updating, the algorithm detects $v_2\in R_0\cap A_{\l_1}$ has a descendant $v_1'$ such that $v_1,v_1'$ represent the same element. Then $B\gets \{\elt(v_1),\elt(v_2)\}$. One should note that we do not require $v_2 \in A_\l$, so we are considering all nodes in $P_\l$ from the leaf to the root, not just the nodes in $A_\l$. After two steps, $v_3\in R_0\cap A_{\l_2}$ is detected, and $B\gets \{\elt(v_1), \elt(v_2), \elt(v_3)\}$. The same thing happens for $v_4\in R_0\cap A_{\l_3}$, where $B\gets \{\elt(v_1), \elt(v_2), \elt(v_3),\elt(v_4)\}$, and the algorithm terminates with output $B(\l,R_0 ):=B = \{\elt(v_1), \elt(v_2), \elt(v_3),\elt(v_4)\}$.}
    \label{fig:binary_tree_part}
\end{figure}
    
Recall that different nodes in $P_\l$ \emph{represents} distinct elements. Therefore, due to the two conditions in the algorithm,  $|\elt(R\cap P_\ell)\cap \elt(A_{\ell'})|\geq |B|$ always holds throughout the algorithm, where $\l'$ is the descendant leaf of $v$ that satisfies condition (2) and incurs the last update for $B$. 
Thus, upon termination, the reward for the non-adaptive strategy is at least the output $|B(\l,R)|$.
Hence, we just need to prove 
the following lower bound
$$
\E_{\ell \leftarrow \pi_{\calT},R\sim \Rdistribution}[|B(\l,R)|]\overset{?}{\geq} \E_\l|A_\l|/20\log n .
$$

The idea to prove this result is to bound the tail of the probability that $|B(\l,R)|$ is at most $0.1|A_\l|/\log n$ over $\ell \leftarrow \pi_{\calT}$ and $R\sim \Rdistribution$. In particular, for any $0\leq s \leq n$, and any possible sequence $B_0$, we can prove that if $|B_0|\leq 0.1s/\log n$ (here, $|B_0|$ denotes the number of elements in the sequence $B_0$),
\begin{align}
\label{eq:tail}
\Pr_{\ell \leftarrow \pi_{\calT},R\sim \Rdistribution}[B(\l,R) =B_0, |A_\l|=s]\leq 2^{-0.9s}.
\end{align}

Before presenting the proof of \eqref{eq:tail}, we first show how
\eqref{eq:tail} leads to \eqref{eq:lowerbound_nonalgo_xos}.
Notice that for each fixed size $m=|B_0|$, there are at most $n^m$ different $B_0$'s, since elements are in $U$. We consider the total contribution of $(\l,R)$ such that $|A_\ell|>10\log n|B(\l,R)|$ inside $\E_\l[|A_\l|]$ by summing all possible $B_0$ and $|A_\l|$. When $n$ is sufficiently large,
\begin{align}
\label{eq:unionboundoverB}
\E_{\ell \leftarrow \pi_{\calT},R\sim \Rdistribution}[\mathbf{1}_{|A_\l|>10\log n\cdot |B(\l,R)||}|A_\l|]
\leq \sum_{m\geq 0,s>10m\log n}2^{-0.9s}\cdot n^{m}\cdot s 
\leq \sum_{s\geq 1}2^{-0.8s}\cdot s^3=O(1).
\end{align}
Using the condition that
$\E_{\ell \leftarrow \pi_{\calT}}[|A_\l|]\geq 20\log n$,
\begin{align*}
\E_{\ell \leftarrow \pi_{\calT},R\sim \Rdistribution}[|B(\l,R)|] &\geq\frac{1}{10\log n}\E_{\ell \leftarrow \pi_{\calT},R\sim \Rdistribution}[\mathbf{1}_{|A_\l|\leq 10\log n\cdot |B(\l,R)||}|A_\l|]\\ &\geq
\frac{\E_{\ell \leftarrow \pi_{\calT}}[|A_\l|]-O(1) }{ 10\log n}\geq \frac{\E_{\ell \leftarrow \pi_{\calT}}[|A_\l|]}{20\log n},
\end{align*}
which confirms the condition of \cref{view:lm:reduce} as promised.

\vspace{0.1cm}
\noindent
{\bf Proof of Inequality \eqref{eq:tail}:}
Now, the only remaining piece of the proof is to prove the tail probability \eqref{eq:tail}, which is the most technical part of the whole proof. In the following, we fix the algorithm output $B_0$ and only consider the leaves $\l$ that $|A_\l| = s$. Let $S(B_0)$ be the set of all leaves $\l$ such that there exists some $R \sim \Rdistribution$ satisfying $B(\l, R) = B_0$. Note that the set $S$ does not depend on the specific choice of $R$.

In the following, we would like to identify nodes $u$ in the tree 
with the following property (the property is independent of $R$): 
if a leaf $\l $ which is a descendant of $u$ and the outcome $R$ satisfy $B(\l,R) = B_0$, then it is guaranteed that $u$ not in $R$ (this is a consequence of $u$ being an \textbf{impossible} node, a notion formally defined later). 
We collect all such nodes to form a set $T$. We can observe the inclusion of the event (over $\l$ and $R$)
$$
[B(\l,R) = B_0]\subseteq [\l \in S(B_0)] \wedge [(P_\l \cap T)\cap R=\emptyset]
$$
(because the latter event is a consequence of the former).
We will show that if the length of $B_0$ is small enough compared to $s$,
the number of impossible nodes in each $A_\ell$ (for $\l \in S(B_0),|A_\l | = s $) is large, 
hence the probability that
$(P_\l\cap T)\cap R=\emptyset$ is negligible. 
Therefore, by the inclusion of the event, the probability (over $\l,R$) of $B(\l,R) = B_0$ is also small.

Recall that in one single execution with input $(\l,R)$, when the iterator reaches a node $u$, whether $u$ updates $B$ depends on two conditions: (1) $u \in R$; (2) there is a leaf $\l'$ of $u$ such that $u\in A_{\l'}$, and for each element $e_i$ in sequence $B$, there is a (strict) descendant of $u$ in $A_{\l'}$ that \emph{represents} $e_i$. The following observation is crucial to our analysis:
\begin{observation}\label{ob:overview}
    For any descendant leaf $\l $ of $u$ and any outcome $R$ that $B(\l,R) = B_0$, in such execution with input $(\l,R)$, the maintained sequence $B$ must be a \emph{prefix} of $B_0$ at the time the iterator visits $u$ (independent of $\l$ and $R$). Specifically, the \emph{prefix} $B$ contains all elements $e_i$ in $B_0$ excluding those represented by the ancestors of $u$ (including $u$). 
\end{observation}


From the above observation, whether condition (2) is satisfied does not depend on the particular input $R$ or leaf $\l$, as long as $\l$ is a descendant of $u$ and $B(\l,R) = B_0$. We then call node $u$ an \textbf{impossible} node (for fixed $B_0$) if $\elt(u)\notin B_0$ and condition (2) is satisfied. As a direct consequence, in any execution such that $\l$ is a descendant of $u$ and $B(\l,R) = B_0$, $u$ cannot trigger an update (since $\elt(u)\notin B_0$) and hence $u$ must not be in $R$ since it already satisfies condition (2). We conclude that for any impossible node $u$ (for $B_0$), if $B(\l,R) = B_0$ and $\l$ is a descendant of $u$, then $u\notin R$.  We note that the conclusion holds for \emph{any} descendant leaf $\ell$ of $u$, no matter $\ell$ is in $u$'s $\mathsf{Yes}$ or $\mathsf{No}$ subtrees (note that
if $\ell$ is in the $\mathsf{No}$ subtree of $u$, $u\not\in A_\ell$).

An illustration is drawn in \cref{fig:binary_tree_part2}. The main portion of the decision tree is essentially the same as \cref{fig:binary_tree_part}, where nodes of the same letters stand for the same nodes. In addition, some irrelevant parts are abbreviated, while other crucial parts are emphasized.

\begin{figure}[t]    
    \centering
      \scalebox{0.8}[0.8]{
\begin{tikzpicture}[
    every node/.style={draw, circle, fill=none, inner sep=1.5pt},
    green node/.style={draw, circle, fill=green, inner sep=2.25pt},
    black node/.style={draw, circle, fill=black, inner sep=3pt},
    left edge/.style={draw=blue, line width=1.2pt, ->},
    right edge/.style={draw=red, line width=1.2pt, ->},
    tree node/.style={draw, regular polygon, regular polygon sides=3, fill=none, inner sep=2pt, minimum size=4mm},
  ]
    \node (legend_node) at (7,3.6) {};
    \node[draw=none, fill=none] (legend_no) at (5.8,2.2) {};
    \node[draw=none, fill=none] (legend_yes) at (8.2,2.2) {};
    \draw[left edge] (legend_node) -- node[left, draw=none, fill=none, pos=0.35] {\textsf{No}} (legend_no);
    \draw[right edge] (legend_node) -- node[right, draw=none, fill=none, pos=0.35] {\textsf{Yes}} (legend_yes);
  \node (n4) at (0,-3.6) [label=left:$v_1^*$, green node]{};
  \node (u2c2) [tree node] at (-1,-4.4) {}; 
  \node (n6) at (-1,-2.8)[label=left :$v_2^*$, green node] {};
  \node (v2) at (0,-2.0) [label=right:$v_2$, green node] {};
  \node (u1) at (1,-2.8) {};
  \node (u2) at (2,-3.6) [label=right:$v_1'$, green node] {};
  \node (u2c) at (3,-4.4) [label=right:$\l_1$] {};
  \node (v1) at (1,-4.4) [label=right:$\l^*$]  {};
  \node (n9) at (-1,-1.2) [label=left:$u_1$, black node] {};
  \node (n11) at (-2, -2) {};
  \node (subtree) at (-0.5,-2.4) [label=left:$w$] {};
    \node (subtree1) at (0.5,-3.2) [label=left:$v_1$, green node] {};
  \node (subtreeleaf) at (1.5, -4.0)[label=right:$\l$] {};
  \node (v3) at (0,-0.4) [label=left:$v_3$, green node] {};
  \node (v2p) at (1,-1.2)  [label=right:$v_1''$, green node]{};
  \node (v1pp) at (2,-2.0) [label=right:$v_2'$, green node]{};
  \node (v1ppc1) at (3,-2.8){};
  \node (v1ppcc1) at (4,-3.6) [label=right:$\l_2$] {};
  \node (n10) at (1,0.4) {};
  \node (u5) at (0,1.2) [label=right:$u_2$, black node]{};
  \node (u'child) at (-4.0,-2.0){};
    \node (u'child1) at (-3.0,-1.2){};
      \node (u'child2) at (-2.0,-0.4){};
        \node (u'child3) at (-1.0,0.4){};
   \node (u'childtree) [label=right:$\calT_1$, tree node]at (-3.0,-2.8){};
    \node (v4) at (1,3.6) [label=left:$v_4$, green node]{};
\node (u6) at (-1, 2.0) [label=left:$u_3$, black node]{};
    \node (u7) at (0, 2.8) {};
  
     \draw[left edge]  (u7) -- (u6);
   \draw[left edge]  (v4) -- (u7);
  \draw[right edge]  (n4) -- (v1);
  \draw[right edge] (n6) -- (n4);
  \draw[left edge] (v2) -- (subtree);
    \draw[right edge] (subtree) -- (subtree1);
     \draw[right edge] (subtree1) -- (subtreeleaf);
  \draw[right edge] (v2) -- (u1);
  \draw[right edge]  (u1) -- (u2);
  \draw[right edge] (n9) -- (v2);
  \draw[left edge] (v3) -- (n9);
  \draw[right edge]  (v3) -- (v2p);
  \draw[right edge] (v2p) -- (v1pp);
  \draw[left edge] (n10) -- (v3);
  \draw[right edge] (u5) -- (n10);
\draw[right edge] (u6) -- (u5);
\draw[left edge] (n9) -- (n11);
\draw[right edge] (n11) -- (n6);
  \draw[left edge] (n4) -- (u2c2);
  \draw[right edge] (u2) -- (u2c);
  \draw[right edge] (v1pp) -- (v1ppc1);
  \draw[right edge] (v1ppc1) -- (v1ppcc1);
  \draw[left edge] (u5) -- (u'child3);
    \draw[left edge] (u'child3) -- (u'child2);
    \draw[left edge] (u'child3) -- (u'child2);
    \draw[left edge] (u'child2) -- (u'child1);
    \draw[left edge] (u'child1) -- (u'child);
    \draw[right edge] (u'child) -- (u'childtree);
\end{tikzpicture}}
  \caption{\small The decision tree is essentially the same as \cref{fig:binary_tree_part}, where nodes of the same letters stand for the same nodes. The notation $\triangle$ means an abbreviated subtree. Suppose $B_0=\{\elt(v_1),\elt(v_2),\elt(v_3),\elt(v_4)\} $ is the fixed sequence. By the execution $B(\l,R_0) = B_0$ shown in \cref{fig:binary_tree_part}, we have $\l \in S(B_0)$. Now, nodes $u_1,u_2,u_3$ are marked black (they are in $A_\l$ and represent elements $\notin B_0$). Recall that the nodes $v_j$ for $1\leq j\le 4$ trigger updates for $B$ in the execution $B(\l,R_0)=B_0$. Therefore, we can deduce that among $\elt(v_j)$'s
  ($1\leq j\le 4$), only $\elt(v_3), \elt(v_4)$ are represented by nodes above $u_1$, because different nodes in $P_\l$ represent distinct elements. Similarly, only $ \elt(v_4)$ is represented by nodes above $u_2,u_3$.
  \newline
  Now, we focus on an input $(\l^*, R)$ that satisfies $B(\l^*,R)=B_0$, where $\l^*$ is an arbitrary descendant leaf of $u_1$. In such execution, some nodes that \emph{represent} $\elt(v_j)$ for $1\leq j\le 4$ must update $B$. Since only $\elt(v_3), \elt(v_4)$ are represented by nodes above $u_1$, when the iterator approaches $u$ in the execution $B(\l^*,R)$, the current $B$ must be $\{\elt(v_1),\elt(v_2)\}$, confirming \cref{ob:overview}. Now, $u_1$ satisfies condition (2) by taking $\l' = \l_1$ and hence is an impossible node. Similarly, $u_2,u_3$ are impossible nodes by taking $\l' = \l_2$.
  \newline 
  It is straightforward to see $u_1\notin R$, since otherwise $u_1$ would trigger update for $B$. Similarly, for any descendant leaf $\l_0$ of $u_1$ (where $\ell_0$ can be $\ell_1$ or any leaf in the left subtree of $v_1^*$), if $B(\l_0,R) = B_0$ then $u_1\notin R$. Similarly, for any descendant leaf $\l_0$ (including all leaves drawn in the figure, and possibly leaves in the subtree $\calT_1$) of impossible nodes $u_2$ or $u_3$, if $B(\l_0,R) = B_0$, then $u_2,u_3\notin R$.}
  \label{fig:binary_tree_part2}
\end{figure}

Now, we need to show that impossible nodes are not rare. From the example in \cref{fig:binary_tree_part2}, one may guess that for each $\l\in S(B_0)$, any node $u$ in $A_\l$ that $\elt(u)\notin B_0$ is an impossible node. In fact, this is the case. We consider the specific input $R_0$ and $\l$ such that $B(\l,R_0) =B_0 $, whose existence is guaranteed by $\l \in S(B_0)$. When the iterator approaches $u$, we consider the node $v$ below $u$ in $P_\l$ that incurs the latest update ($v\neq u$ because $\elt(u)\notin B_0$). The node $v$ must have a descendant leaf $\l'$ that satisfies condition (2). Note that $\l'$ is also a descendant of $u$ and hence $u\in A_{\l'}$ because $\l$ and $\l'$ share the same active ancestors above $v$. Since condition (2) does not depend on the input $R$ or leaf $\l$, $u$ is indeed an impossible node. Therefore, for each $\l\in S(B_0)$, the number of impossible nodes in $A_\l$ is at least $|A_\l|-|B_0|$, counting the nodes mentioned above.

Now we present the crucial `tail' bound.
\begin{lemma}[See also \cref{liu:induction}]
    \label{intro:liuinduc}
    Let $S$ be a subset of leaves in the tree $\calT$. There is a subset $T$ of nodes in $\cal T$, which does not contain any leaf of $\calT$. There exists an integer $h\in \mathbb N$ such that for each $\l \in S$, $| A_\l\cap T|\geq h $. Then the following inequality holds:
    \[
    \E\limits_{\ell \leftarrow \pi_{\calT}} \left[ \mathbf{1}_{\ell \in S}\Pr_{R \sim \Rdistribution} \left[R\cap P_\ell\cap T=\emptyset\right]\right]\leq 2^{-h}.
    \]
\end{lemma}
The proof of this lemma is simply applying induction on the length of the maximum root-leaf path.
In each induction step, we consider whether the tree root $r_0$ is in $T$, then divide the tree into two parts and use induction hypothesis respectively. 

With this lemma, if we fix $B_0$ and $s>10\log n\cdot |B_0|$, we define $S= S(B_0)\cap \{\l:|A_\l|=s\}$ and put all impossible nodes (for $B_0$) into $T$ (again, we emphasize that the defined $S,T$ do \emph{not} depend on $R$). The above discussion shows that for each $\l \in S$, $|A_\l \cap T|\geq h:=s-|B_0|$.
Then by \cref{intro:liuinduc},
\begin{align*}
\Pr_{\ell \leftarrow \pi_{\calT},R\sim \Rdistribution}[B(\l,R) =B_0,|A_\l|=s]&\leq \E_{\ell \leftarrow \pi_{\calT}} \left[ \mathbf{1}_{\ell \in S}\Pr_{R \sim \Rdistribution} 
\left[\{\textbf{impossible} \text{ nodes in } P_{\l}\}\cap R=\emptyset\right]\right] \\
&\leq \E_{\ell \leftarrow \pi_{\calT}} \left[ \mathbf{1}_{\ell \in S}\Pr_{R \sim \Rdistribution} \left[R\cap P_\ell\cap T=\emptyset\right]\right]\leq 2^{|B_0|-s}\leq 2^{-0.9s}.
\end{align*}

\vspace{0.1cm}
\noindent
{\bf Remark:}
We need to point out that the algorithm and the definition of $B$ here are different from the actual proof in the main part, mainly because the actual proof needs to deal with $A_\l'$, a subset of $A_\l$. 
In particular, the nodes in $A_\l$ for $\l\in S(B_0)$ that represent elements not in $B_0$ are no longer ensured to be impossible nodes, since they may not be in $A_{\l'}'$. 
In the proof of the general setting, we need to 
include additional information,
the depths $d_i$ of the elements $e_i$ and an additional sequence $(y_j)$ (the depth of the last $j\log n$-th node in $A_\l'$,
in the definition of sequence $B$ 
to control the positions of nodes in $A_\l'$.

\section{Preliminaries}
\label{section2}
\subsection{Mathematical Preliminaries and Notations}\label{subsec:mathprel}
This subsection introduces basic mathematical definitions and results. Throughout the paper, all objective and functions (whose domain can be $2^U$ or $\mathbb R_{\geq 0}^U$) are assumed to be \textbf{monotone} ($f(\boldu)\leq f(\boldv)$ if $\boldu \leq \boldv$, or $f(S)\leq f(T)$ if $S\subseteq T$) and \textbf{normalized} ($f(\emptyset) = 0$ or $f(\mathbf{0}) = 0$). We denote the ground set by $U=[n]$. Throughout this paper, $\log$ denotes the logarithm to base $2$. 
\begin{definition}[(Monotone) Norm Objective]
\label{def:norm}
\noindent A function $\norm:\R_{\geq 0}^U\rightarrow\R_{\geq 0}$ is a {\em norm objective}, if the following holds:
\begin{enumerate}[(1)]
\item $\norm(\boldsymbol{0})=0$;
\item $f(\boldu)\leq f(\boldv)$ for any $\boldu\leq \boldv$, that is, $u_i\leq v_i$, $ \forall 1\leq i\leq n $, where $\boldu=(u_i)_{i\in U}$ and $\boldv=(v_i)_{i\in U}\in \Rnng^U$.
\item $\norm(\boldu+\boldv)\leq\norm(\boldu)+\norm(\boldv)$ for all $\boldu,\boldv\in\Rnng^U$;
\item $\norm(\theta\boldv)=\theta\norm(\boldv)$ for all $\boldv\in\Rnng^U,\theta\in\Rnng$.
\end{enumerate}
\end{definition}
\begin{definition}[(Monotone) Symmetric Norm]
 A norm $f:\R_{\geq 0}^U\rightarrow\R_{\geq 0}$ is a {\em symmetric norm} if and only if $f$ remains unchanged under any permutation of its coordinates, i.e., $f(\boldv)=f((v_{\pi(i)})_{i\in U})$, for any permutation $\pi:U\to U$ and $\boldv=(v_i)_{i\in U}\in\Rnng^U$.
\end{definition}
\begin{definition}[XOS (set) Function]
\label{def:xos}
A set function $f : 2^U \to \mathbb R_{\geq 0} $ is XOS (max-of-sums)\footnote{Note that here we implicitly require $f$ to be monotone and normalized.}  if there exist non-negative weight vectors $\boldw^{(1)},\boldw^{(2)},\cdots \boldw^{(W)} \in
\mathbb{R}_{\geq 0}^{U}$, 
such that $f(S) = \max\limits_{i\in [W]}\{\sum_{j \in S} w_j^{(i)}\}$, $\forall S\subseteq U$.
\end{definition}
\begin{definition}[\XOSS (set) Function]
\label{def:xoss}
A set function $f : 2^U \to \mathbb R_{\geq 0} $ is \XOSS if there exist weight vectors $\boldw^{(1)},\boldw^{(2)},\cdots \boldw^{(W)} \in
\{0,1\}^{U}$, 
such that $f(S) = \max\limits_{i\in [W]}\{\sum_{j \in S} w_j^{(i)}\}$, $\forall S\subseteq U$.
\end{definition}
\begin{definition}[Subadditive (set) Objective]
\noindent A function $ f:2^U \to \mathbb R_{\geq 0}$ is a {\em subadditive objective}, if and only if the following holds:
\begin{enumerate}[(1)]
\item $ f(\emptyset)=0$.
\item $f(S)\leq f(T)$ for all $S\subseteq T\subseteq U$.
\item $\norm(S\cup T)\leq\norm(S)+\norm(T)$ for all $S,T\subseteq U$.
\end{enumerate}
\end{definition}
It is clear that a \XOSS\ set function is a special case of an XOS set function, and every XOS function is subadditive. Furthermore, Feige~\cite{XOSfrac} established that a set function is (monotone) fractionally subadditive if and only if it belongs to the XOS class, where a set function \( f:2^U \to \mathbb{R}_{\geq 0} \) is defined as (monotone) fractionally subadditive if it satisfies  
$
f(T) \leq \sum_i \alpha_i f(S_i)
$  
for any collection of sets that \( \mathbf{1}_T \leq \sum_i \alpha_i \mathbf{1}_{S_i} \) and \( \alpha_i \geq 0 \). Here, \( \mathbf{1}_T \) denotes the indicator function of the subset \( T \subseteq U \).

We will also use the well-known Chernoff bound:
\begin{fact}[Chernoff Bound, see \cite{mitzenmacher2005probability}]
\label{lemma:chernoff}
$ X_1, X_2, \dots, X_n $ is a sequence of independent random variables. Each $ X_i$ is a Bernoulli random variable with parameter $p_i\in [0,1]$, i.e.,
$X_i \sim B(p_i)$.
Let $ X = \sum_{i=1}^n X_i $, and the expected value $ \mu = \E[X] = \sum_{i=1}^n p_i $.
Then, for any $ 0 < \epsilon < 1 $,
$$
\Pr\left[X < (1 - \epsilon)\mu\right] \leq \exp\left(-\frac{\epsilon^2 \mu}{2}\right).
$$
\end{fact}

\subsection{Bernoulli Stochastic Probing}
\label{bernoulli:prel}
Recall that we have defined Bernoulli stochastic probing in \cref{intro:bernoulli}.
In this section, we introduce basic definitions and ideas, following the framework of \cite{GNS17}.
\begin{definition}[Bernoulli stochastic probing with XOS ($\XOSS$, subadditive, resp.) Objective]
\label{bernoulli:norm}
In this special case of the stochastic probing problem, each random variable $X_i$ follows a Bernoulli distribution: $X_i = 1$ with probability $p_i$ and $X_i = 0$ with probability $1 - p_i$ for all $i\in U$. 
 The objective function $f$
simplifies to an XOS ($\XOSS$, subadditive, resp.) set function 
\(f: 2^{U} \to \mathbb{R}_{\ge 0}\).
Let \(P \subseteq U\) denote the set of probed items,
and $\activeelem(P)\subseteq P$ be the set of active elements in $P$.
The reward in this case is given by $f(\activeelem(P))$.
All other settings are the same as in Definition~\ref{stoc:probing},
and our goal is also to find a probing strategy to maximize the expected reward \(\E[f(\activeelem(P))]\).
\end{definition}
In the Bernoulli setting, the objective degenerates into a set function: $f(S)$ for $S\subseteq U=[n]$ is defined as $f(\boldv)$ where $\boldv$ is an $n$-dimensional vector where the $i$-th coordinate is 0 if $i\not\in S$ and is $1$ otherwise. We can always assume without loss of generality that $p_i > 0$ for all $i$. If $p_i = 0$, we can remove the corresponding element from the ground set $U$ and modify the norm accordingly by inserting a zero placeholder at the corresponding position. This modification does not affect the reward or any properties of the objective function $f$.

We denote the supremum of the adaptivity gap for \emph{Bernoulli Stochastic Probing with any \textnormal{XOS} Objective} and any \emph{prefix-closed} feasibility constraint $\calF$ with $\max\limits_{F\in \calF }|F|\leq r$ as $\gapfunction_{\XOS}(n, r)$. We also denote the supremum of the adaptivity gap for \emph{Bernoulli Stochastic Probing with any \XOSS Objective} and any \emph{prefix-closed} feasibility constraint $\calF$ as $\gapfunction_{\xoss}(n)$.

Now we formally introduce the decision tree model of an adaptive strategy.

\vspace{0.1cm} 
\noindent \textbf{Adaptive Strategies.}
An adaptive strategy tree, denoted by $\mathcal{T}$, is a binary rooted tree where each internal node $v$ represents an element $i \in U$, denoted by $\elt(v) = i$. Every internal node has two outgoing arcs: the \textsf{Yes} arc, which leads to the next node if the element $i = \elt(v)$ is active (with probability $p_i$) when probed, and the \textsf{No} arc, which leads to a different node if $i$ is not active (with probability $1 - p_i$).
We require that no element can be represented by more than one node on any root-to-leaf path in the tree. Additionally, any root-to-leaf path in $\mathcal{T}$ must satisfy the feasibility constraint $\calF$, meaning it represents a valid sequence of probes for both adaptive and non-adaptive strategies.
Each leaf $\ell$ in $\mathcal{T}$ is associated with a unique root-to-leaf path, denoted by $P_\ell$, where the elements probed along this path are given by $\elt(P_\ell)$. For any subset $S \subseteq P_\ell$, we denote the distinct elements represented by the nodes in $S$ as $\elt(S) = \{\elt(v) \in U : v \in S\}$.
Let $A_\ell$ represent the active nodes along the path $P_\ell$, i.e., the nodes on $P_\ell$ where the \textsf{Yes} arc is taken.
We note that the leaves of the tree do not represent elements in $U$; rather, they simply signal the termination of the probing process. We associate
each leaf with a dummy element (denoted as $0$).
    
The tree $\mathcal{T}$ naturally gives us a probability distribution $\pi_{\mathcal{T}}$ over its leaves: start at the root, at each node $v$, follow the \textsf{Yes} branch with probability $p_{\elt(v)}$ and the \textsf{No} branch otherwise, and finally reach a leaf upon termination.
    
Given a monotone function $f$ and a tree $\calT$, the associated adaptive strategy probes elements following a root-to-leaf path, observes the realized values until we reach a leaf $\ell$, and receives the reward $f(\elt(A_\ell))$. Let $ \adaptcost(\mathcal{T}, f) $ denote the expected reward obtained this way; it can be written compactly as
$$\adaptcost(\mathcal{T}, f) :=\E_{\ell \leftarrow \pi_{\mathcal{T}}}\left[f(\elt(A_\ell))\right].$$

In fact, any deterministic adaptive strategy can be modeled by such a decision tree model. Moreover, it can be easily proven that the reward of any \emph{randomized} adaptive strategy is bounded by the reward of \emph{some} deterministic adaptive strategy. Since the cardinality of a decision tree is finite, the \emph{optimal} adaptive strategy exists. With the central theme to upper bound the gap between \emph{optimal} adaptive strategies and \emph{optimal} non-adaptive ones in mind, we will only consider deterministic adaptive strategies and the decision tree model described above.
We denote the optimal adaptive strategy by \adaptivestrategy.

\vspace{0.1cm}
\noindent \textbf{A Natural Non-adaptive Strategy.} We can also define a \emph{natural} non-adaptive algorithm given the optimal adaptive strategy \adaptivestrategy\
(with tree $\mathcal{T}$): just pick a leaf $\ell \leftarrow \pi_{\mathcal{T}}$ from the distribution given by $\mathcal{T}$, probe all elements on that path, and choose the active elements to get the reward. Clearly, such strategy satisfies feasibility constraint. We denote this non-adaptive strategy as \nonadaptivestrategy. 
We note that the same non-adaptive strategy has been studied in \cite{GNS16, GNS17,BSZ19}.
Since the objective $f$ is always assumed to be monotone throughout the paper, the reward is actually the function value of the set of \emph{all} active elements.
To capture the randomness of the realizations in 
the probing process of \nonadaptivestrategy,
we define $\Rdistribution$, which is a distribution of the random subset $R$ of nodes on $\calT$.
A node $u$ is in the random subset $R$, if 
the element $\elt(u)$ is active in the probing process
of \nonadaptivestrategy.
Specifically, for each element $e$, with independent probability $p_e$, all nodes $v$ such that $\elt(v) = e$ are included in $R$. Otherwise, with probability $1 - p_e$, these nodes are not included. 

We denote the expected reward of \nonadaptivestrategy\ by $\alg(\mathcal{T}, f)$, and we can see that
$$
\alg(\mathcal{T}, f) := \E_{\ell \leftarrow \pi_{\calT}} \left[ \E_{R \sim \Rdistribution} \left[f(\elt(R\cap P_\ell))\right] \right].
$$

Another way to view the distribution $R\cap P_\l$ is to first sample $\ell \leftarrow \pi_{\calT}$, then sample $R$ \emph{only} on the path, with each node $v\in P_\l$ included of independent probability $p_{\elt(v)}$.

\vspace{0.1cm}
\noindent \textbf{Bounding the Adaptivity Gap.}  A common technique 
for bounding the adaptivity gap, employed in \cite{GNS16, GNS17,BSZ19}, 
is to bound the ratio between the \emph{optimal} adaptive strategy \adaptivestrategy\ and its 
corresponding non-adaptive strategy \nonadaptivestrategy. 
Formally, for the optimal tree $\calT^*$, if 
we can show that the following holds for 
the function $\gapfunction(n)$ when $n$ is sufficiently large:
$$
 \gapfunction(n)\cdot\alg(\calT^*, f)\geq \adaptcost(\calT^*,f),
$$
then the true adaptivity gap of the problem is at most $\gapfunction(n)$. 

Throughout the paper, we also need the following definitions for trees.
\begin{definition}[Depth]
\label{def:depth}
We say a node has depth $k$ if there are $k$ \textsf{Yes} arcs on the path from the root to this node. We denote the depth of $u$ by $\depth{u}$.
\end{definition}
\begin{definition}[Active Ancestor]\label{def:activeancestor}
For a node $v$, we call an ancestor $u$ of $v$ its \emph{active ancestor} if the \textsf{Yes} arc from $u$ is taken on the path from root to $v$. Thus, a node of depth $k$ has $k$ active ancestors. Also recall that for a leaf $\l$ of $\calT$, $A_\l$ is exactly the set of active ancestors of $\l$.
\end{definition}

\begin{figure}[!t] 
    \centering 
    \includegraphics[width=0.8\textwidth]{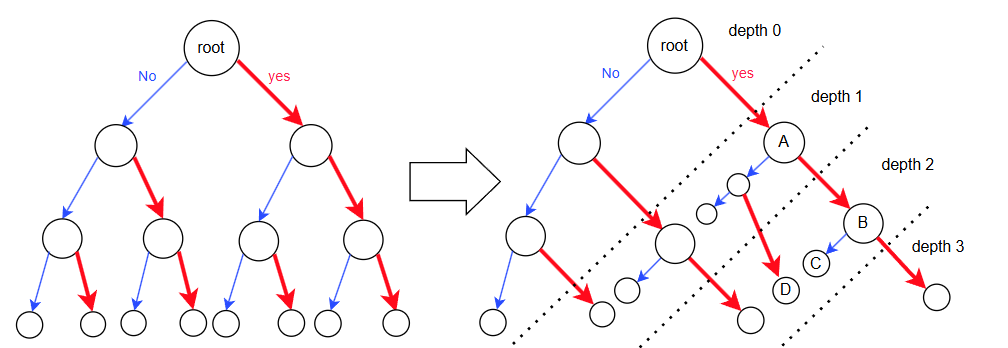} 
    \caption{A binary decision tree of height $3$ is transformed to a layered tree of depth $3$. Note that $A$ is an active ancestor of $B$ and $C$, but not for $D$. The active ancestor set $A_C=\{\textnormal{root},A\}$.}
    \label{fig1} 
\end{figure}

In the following, we will use the definition of depth and active ancestors extensively. It is useful to think of the original tree $\calT$ as being layered by depth, and forms a new layered tree, see \cref{fig1}.

It is clear from definition that different nodes in $A_\l$ have distinct depths. This fact will be frequently used in the algorithm and proofs in \cref{sec3:prepare} and \cref{sec3:main}.

Now, we introduce some tools for the adaptivity gap problem in the Bernoulli setting. As usual, we denote $\mathsf{OPT}=\adaptcost(\calT^*, f)$, the expected reward for \adaptivestrategy.
\begin{definition}[$\lambda$-large (small) elements]
\label{def:bernoullilarge}
For any $\lambda>0$, an element $e$ is $\lambda$-large if $f(e)\geq \lambda\cdot\OPT$ and $\lambda$-small otherwise, where $f(e)$ means the $f$ value when only $e$ is active. 
\end{definition}

The following lemma already appears implicitly in \cite{GNS17}, and the proof is deferred to \cref{AppendixB}.   
\begin{lemma}  
\label{large:case}
For any $\lambda > 0$, suppose that the adaptivity gap in the \emph{Bernoulli setting} is upper bounded by $\gapfunction(n) \geq 1$ for a certain subadditive objective $f$ and a feasibility constraint $\calF$, under the condition that every element is $\lambda$-small with respect to $f$. Then, the adaptivity gap in the \emph{Bernoulli setting} for general instances (with general random variables, the same $f$, and $\calF$) is at most $\max\{2\gapfunction(n),\ 2 + 4/\lambda\}$.
\end{lemma}

\section{Adaptivity Gap for Bernoulli Stochastic Probing}
\label{section 3}
In this section, we analyze {\em Bernoulli stochastic probing with \textnormal{XOS} objective}, as defined in Definition~\ref{bernoulli:norm}.
We follow the notations introduced in Section~\ref{bernoulli:prel}.
Furthermore, recall that $2\leq r\leq n$ denotes an upper bound on the maximal length of a sequence in $\calF$. Intuitively, $r$ serves as an upper bound on the maximum cardinality of elements that a strategy can probe. Consequently, $r$ also bounds the length of any root-to-leaf path in any adaptive strategy tree $\calT$. Formally, for any leaf $\l$, $|A_\l|\leq |P_\l|\leq r$.
Recall that we use $\gapfunction_{\XOS}(n,r)$ to denote the adaptivity gap for 
Bernoulli stochastic probing with XOS objective, as defined in \cref{bernoulli:prel}.
The primary result of this section is the following theorem:
\begin{theorem}
\label{thm:weak}
For $2\leq r\leq n$, the adaptivity gap of {\em Bernoulli Stochastic Probing with any \textnormal{XOS} Objective} under any prefix-closed $\calF$ with $\max\limits_{F\in \calF} |F|\leq r$ is $\gapfunction_{\XOS}(n,r)=O(\log r\log n)$. 
\end{theorem}

\subsection{Reduction to a Special XOS Norm}\label{sec3:redcution}
In this subsection, we present a reduction from the original arbitrary \textnormal{XOS} Objective to a specific norm that is more amenable to analysis. 
Recall the definition of the active ancestor set $A_\ell$ in \cref{def:activeancestor}. 
For each leaf $\ell$, we define a subset $A_\ell' \subseteq A_\ell$
(in the proof of Lemma~\ref{lm:reduce}). 
Using these subsets, we define a new norm $\fspecial$ as follows:
\begin{align}
\label{eq:fxos}
\fspecial (S) = \max_{\ell} \left\{ | \elt(A_\ell') \cap S | \right\},
\end{align}
for any subset $S \subseteq U = [n]$. 

We establish the following lemma.
Recall that $\gapfunction_{\XOS}(n, r)$ and $\gapfunction_{\xoss}(n)$ are defined in \cref{bernoulli:prel} as the adaptivity gaps for corresponding class of functions.

\begin{lemma}
\label{lm:reduce}
Let $\gapfunction(n)\geq 2$ be an increasing function of $n$. Suppose that for any decision tree $\calT$ and any choice of subsets $A_\ell' \subseteq A_\ell$ satisfying 
$$
\E_{\ell \leftarrow \pi_{\mathcal{T}}}[|A_\ell'|] \geq \gapfunction(n),
$$ 
the following condition holds:
$$
\E_{\ell \leftarrow \pi_{\calT}} \left[ \E_{R \sim \Rdistribution} \left[ \fspecial(\elt(R \cap P_\ell)) \right] \right] 
\geq \frac{1}{\gapfunction(n)} \E_{\ell \leftarrow \pi_{\mathcal{T}}} \left[ |A_\ell'| \right].
$$
Then the following two results hold:
\begin{enumerate}[(a)]
\item For $r\geq 2$, $\gapfunction_{\XOS}(n,r)$ is upper bounded by $O(\gapfunction(n) \log r)$.
\item $\gapfunction_{\xoss}(n)$ is upper bounded by $O(\gapfunction(n))$.
\end{enumerate}
\end{lemma}

\begin{proof} 
(a) We first consider $\gapfunction_{\XOS}(n,r)$ and the case where $f$ is an XOS (set) function. From Lemma~\ref{large:case}, we only need to address the case where all elements are $\lambda = 1/(5\gapfunction(n)\log r)$-small, since $1+\lambda^{-1} =O(\gapfunction(n)\log r)$. We further assume that $\mathsf{OPT} = \adap(\calT,f)= 5\gapfunction(n) \log r$ by scaling $f$, and $\calT$ is the optimal decision tree.

Let $f(S)=\max\limits_{i\in [W]}\left\{\sum_{j\in S}w_{j}^{(i)}\right\}$, where $w_{j}^{(i)}\geq 0$ for all $i\in [W],j\in U$. Since all elements are $1/(\gapfunction(n)\log r)$-small, $f(\{j\}) = \max\limits_{i\in [W]}w_{j}^{(i)}< 1$ implies $w_{j}^{(i)}<1$ for all $i,j$.
We note that this XOS function $f$ is the objective function of the stochastic probing problem, 
which is different from $\fspecial (S)$
defined in \cref{eq:fxos}.

For each leaf $\ell$, let $I(\ell)$ be the index $i\in [W]$ such that the linear function $\boldw^{(i)}$ attains the value $f(\elt(A_\l))$ at the set $\elt(A_\l)$, that is, $\mathop{\argmax}\limits_{i\in [W]}\left\{\sum_{j\in \elt(A_\ell)}w_{j}^{(i)}\right\}$ (breaking ties arbitrarily).
Now, for each $\ell$ and integer $1\leq j\leq \log r+1$, we define $A_{\ell,j}=\left\{u:u\in A_\ell, 2^{-j}\leq  w^{(I(\ell))}_{\elt(u)} < 2^{-j+1}\right\}$. We define an additional $A_{\ell,0 }$ to be $A_\l\backslash \bigcup_{j=1}^{\lfloor\log r\rfloor+1} A_{\l,j}$. It is clear that for any node $ u \in A_{\l,0}$, $w^{(I(\ell))}_{\elt(u)} \leq 1/r$.

Therefore, the reward for the adaptive strategy \adaptivestrategy\ is
\begin{align*}
5\gapfunction(n)\log r= \OPT =&\E_{\ell \leftarrow \pi_{\calT}}\left[f(\elt(A_\ell))\right]=\E_{\ell \leftarrow \pi_{\calT}}\left[\sum_{u\in A_\ell}w^{(I(\ell))}_{\elt(u)} \right]
\leq \E_{\ell \leftarrow \pi_{\calT}}\left[\sum_{j=1}^{\lfloor \log r\rfloor+1}\frac{|A_{\ell,j}|}{2^{j-1}}+\frac{|A_{\l,0}|}{r} \right]
\end{align*}
Using the assumption on the maximal length of a sequence in $\calF$, which implies that $|A_{\l,0}|\leq |A
_\l|\leq r$, 
we obtain
\begin{align*}
5\gapfunction(n)\log r\leq &\sum_{j=1}^{\lfloor \log r\rfloor+1}\frac{1}{2^{j-1}}\E_{\ell \leftarrow \pi_{\calT}}\left[|A_{\ell,j}|\right]+ \frac{r}{r} \leq \sum_{j=1}^{\lfloor \log r\rfloor+1}\frac{1}{2^{j-1}}\E_{\ell \leftarrow \pi_{\calT}}\left[|A_{\ell,j}|\right]+ 1.
\end{align*}
Thus, there exists $1\leq j_0\leq \log r+1$ such that $\E_{\ell \leftarrow \pi_{\calT}}\left[|A_{\ell,j_0}|\right]/ 2^{j_0-1}\geq 4\gapfunction(n)\log r/(\log r+1)$ by the averaging principle. Thus,
$$\E_{\ell \leftarrow \pi_{\calT}}\left[|A_{\ell,j_0}|\right]\geq \frac{4\cdot 2^{j_0-1}\gapfunction(n)\log r}{\log r+1}\geq 2^{j_0}\gapfunction(n)\geq \gapfunction(n). $$ 
Consider the $\fspecial$ that is defined by
the subsets $A'_\ell=A_{\ell,j_0}$ for each $\l$.
Then by the lemma condition, 
$$\E_{\ell \leftarrow \pi_{\calT}} \left[ \E_{R \sim \Rdistribution} \left[\fspecial(\elt(R\cap P_\ell))\right] \right]\geq \frac{1}{\gapfunction(n)}\E_{\ell \leftarrow \pi_{\mathcal{T}}}\left[|A_\ell'|\right] = \frac{1}{\gapfunction(n)}\E_{\ell \leftarrow \pi_{\calT}}\left[|A_{\ell,j_0}|\right]\geq 2^{j_0}.$$

Notice that for each set $S\subseteq U$, when $\fspecial(S)=s$ for some nonnegative integer $s$, there is a leaf $\ell$ such that $|S\cap \elt(A_{\ell,j_0})|\geq s$.
Thus, 
$$f(S)\geq \sum_{j\in S}w^{(I(\ell))}_{j}\geq \sum_{j\in S\cap \elt(A_{\ell,j_0})}w^{(I(\ell))}_{j} =\sum_{u\in A_{\ell,j_0},\elt(u)\in S}w^{(I(\ell))}_{\elt(u)} \geq \frac{|S\cap \elt(A_{\ell,j_0})|}{2^{j_0}}=\frac{\fspecial(S)}{  2^{j_0}},
$$
where in the last inequality we use $w^{(I(\ell))}_{\elt(u)}\geq 2^{-j_0}$ when $u\in A_{\l,j_0}$, $1\leq j_0\leq \log r+1$.
Therefore, combining the above two inequalities, the reward for 
the non-adaptive strategy \nonadaptivestrategy is
\begin{align*}
\E_{\ell \leftarrow \pi_{\calT}} \left[ \E_{R \sim \Rdistribution} \left[f(\elt(R\cap P_\ell))\right]\right]&\geq \frac{1}{2^{j_0}}\E_{\ell \leftarrow \pi_{\calT}} \left[ \E_{R \sim \Rdistribution} \left[\fspecial(\elt(R\cap P_\ell))\right]\right]\geq 1.
\end{align*}
Thus, the adaptivity gap is upper bounded by $O(\gapfunction(n)\log r)$.

(b) Next, we consider the case when $f$ is a \XOSS (set) function. From Lemma~\ref{large:case}, we only need to address the case where all elements are $\lambda = 1/\gapfunction(n)$-small, since $1+\lambda^{-1} =O(\gapfunction(n))$. We further assume that $\mathsf{OPT} = \adap(\calT,f)$, and $\calT$ is the optimal decision tree.

Let $f(S)= \max\limits_{i\in [W]}\left\{\sum_{j\in S}w_{j}^{(i)}\right\}$, where $\boldw^{(i)}\in \{0,1\}^U$ for all $i\in [W],j\in U$. 
Since all elements are $\frac{1}{\gapfunction(n)}$-small, $f(\{j\}) = \max\limits_{i\in [W]}w_{j}^{(i)} < \OPT / \gapfunction(n)$ shows that, either $f$ is the zero function (trivial), or $\OPT>\gapfunction(n)$. Similarly, for each leaf $\ell$, let $I(\ell)$ be the index $i\in [W]$ such that the linear function $w^{(i)}$ attains the value $f(\elt(A_\l))$ at the set $\elt(A_\l)$, that is, $\mathop{\argmax}\limits_{i\in [W]}\left\{\sum_{j\in \elt(A_\ell)}w_{j}^{(i)}\right\}$ (breaking ties arbitrarily). Then, we directly set $A_\l'=\{u\in A_\l: w_{\elt(u)}^{(I(\l))} =1\}$. Since different nodes in $A_\l$ represent distinct elements in $U$,
$$
\E_{\ell \leftarrow \pi_{\calT}}\left[|A_\ell'|\right] = \E_{\ell \leftarrow \pi_{\calT}}\left[f(\elt(A_\ell))\right]=\OPT > \gapfunction(n).
$$
Then, by the lemma condition, we obtain
$$
\E_{\ell \leftarrow \pi_{\calT}} \left[ \E_{R \sim \Rdistribution} \left[ \fspecial(\elt(R \cap P_\ell)) \right] \right] 
\geq \frac{1}{\gapfunction(n)} \E_{\ell \leftarrow \pi_{\mathcal{T}}} \left[ |A_\ell'| \right]=\frac{\OPT}{\gapfunction(n)}.
$$
However, we have the following lower bound of the reward
of the path $P_\l$:
$$
f(\elt(R \cap P_\ell))=\max
\limits_{i \in [W]}\left\{\sum_{j\in  \elt(R\cap P_\l)}w^{(i)}_{j}\right\} \geq \max
\limits_{\l'}\left\{\sum_{j\in  \elt(R\cap P_\l)}w^{(I(\l'))}_{j}\right\}= \max
\limits_{\l'}\left\{|\elt(R\cap P_\l)\cap \elt(A_{\l'}')|\right\},
$$
where clearly the last term is exactly $\fspecial(\elt(R\cap P_\l))$. Hence, we obtain $\nonadaptcost(\calT, f)$ as
$$
\E_{\ell \leftarrow \pi_{\calT}} \left[ \E_{R \sim \Rdistribution} \left[f(\elt(R\cap P_\ell))\right]\right]\geq \E_{\ell \leftarrow \pi_{\calT}} \left[ \E_{R \sim \Rdistribution} \left[ \fspecial(\elt(R \cap P_\ell)) \right] \right]\geq \frac{\OPT}{\gapfunction(n)}, 
$$
which shows that $\gapfunction_{\xoss}(n) = O(\gapfunction(n))$.
\end{proof}

The remaining task is to show the premise of 
Lemma~\ref{lm:reduce} holds.
This is highly nontrivial, and we make it a theorem as follows,
and prove it in the next two subsections.

\begin{theorem}
\label{thm:sec3main}
Let $A_\ell'$ be a subset of $A_\ell$ for each leaf $\ell\in \calT$. Assume that
$$\E_{\ell \leftarrow \pi_{\mathcal{T}}}[|A'_\ell|]\geq 200\log n.$$
Then for $n\geq 100$, 
$$\E_{\ell \leftarrow \pi_{\calT}} \left[ \E_{R \sim \Rdistribution} \left[\fspecial(\textnormal{elt}(R\cap P_\ell))\right] \right]\geq \frac{1}{200\log n}\E_{\ell \leftarrow \pi_{\mathcal{T}}}\left[|A_\ell'|\right].$$
\end{theorem}

We can see that taking $\gapfunction(n)=200\log n$ in \cref{lm:reduce}, and \cref{thm:sec3main} completes the proof of \cref{thm:weak}.

\subsection{The Greedy Algorithm}
\label{sec3:prepare}
In this section, we present the \emph{greedy} algorithm on the decision tree $\calT$, which is used for 
proving \cref{thm:sec3main}. 
Recall that $\depth{u}$ denotes the depth of node $u$ 
(formally defined in \cref{def:depth}), $A_\l$ 
denotes the set of \textbf{active ancestors}, defined 
in \cref{def:activeancestor}.
The input of the algorithm is a pair $(\ell,R)$ and the algorithm outputs a sequence $B(\ell,R)$ (the form of a sequence is defined later in Definition~\ref{def:label-seq}).
We say the sequence $B(\ell,R)$ is the {\bf label} of 
input $(\ell,R)$.
We may think the greedy algorithm as a classification
algorithm that classify input $(\ell, R)$ to label $B(\ell, R)$.
As we mentioned before, we need to extend the definition
of the label to include additional information of the depths.
To motivate the need for a new definition, we first briefly outline 
the main ideas to extend the argument in \cref{subsec:classify}
and the limitations of the existing definition of labels.


\subsection{Extending the Augments in \cref{subsec:classify}}

We first briefly explain how we extend the augments in \cref{subsec:classify}
to deal with more general $A_\ell'$ (which is a subset of $A_\ell$).
Similar to \cref{subsec:classify}, we need to design a greedy algorithm 
that takes the input $(\ell,R)$ and outputs a sequence $B(\ell,R)$ such that 
$\fspecial(\elt(R\cap P_\ell)) =|\elt(R\cap P_\ell)\cap \elt(A_{\ell_0}')|\geq |B(\ell,R)|-1,$
where the subtraction of $1$ is only due to technical reasons. 
As we mentioned before, we need to augment the definition of sequence $B(\ell, R)$
with some additional information. 
Hence, the length $|B(\ell, R)|$ is not simply the length of the sequence and will be defined explicitly later.
Thus, we can lower bound
$\E_{\ell \leftarrow \pi_{\calT}} \left[ \E_{R \sim \Rdistribution} \left[\fspecial(\textnormal{elt}(R\cap P_\ell))\right] \right]$ 
by 
$\E_{\ell \leftarrow \pi_{\calT},R \sim \Rdistribution} |B(\ell,R)| -1.$ 
Moreover, for a fixed output label $B(\ell,R)=B_0$ and a non-negative integer $s\geq 0$ such that $|B_0|\leq s / \log n$, our
algorithm ensures the probability 
$\Pr_{\ell \leftarrow \pi_{\calT},R \sim \Rdistribution} [B(\ell,R)=B_0, |A_\ell'| = s]$
is exponentially small similar to \eqref{eq:tail} (see Lemma~\ref{lemma3.11}).
Then, applying a union bound over all possible labels 
similar to \eqref{eq:unionboundoverB}, 
one can easily show that the gap between $\E_{\ell \leftarrow \pi_{\calT}, R \sim \Rdistribution}|B(\ell,R)|$
and $\E_{\ell \leftarrow \pi_{\calT}}[\|A_\ell'|]$
is bounded by $O(\log n)$, thus proving \cref{thm:sec3main}.

Note that we are now working with $A_\ell'$, a subset of $A_\ell$. 
In this setting, we slightly modify the greedy algorithm defined in the beginning of \cref{subsec:classify}. Specifically, at each node $v \in P_\ell$, the algorithm checks whether the following two conditions hold (only the second condition is modified, while the rest of the algorithm remains unchanged):
\begin{enumerate}[(1)]
\item $v \in R$; 
\item there exists a leaf $\ell'$ of $v$ such that $v \in A_{\ell'}'$, and for each element $e_i$ in the current sequence $B$, there is a descendant of $v$ (other than $v$ itself) in $A_{\ell'}'$ that \emph{represents} $e_i$.
\end{enumerate}

To prove the tail probability inequality, we continue to use the same intuition as in \cref{subsec:classify}: fixing a label $B_0$, we would like to identify 
certain nodes $u$ (impossible node) cannot belong to $R$ whenever $u \in P_\ell$ and $B(\ell,R) = B_0$, and show that 
impossible nodes are not rare.
Recall that in \cref{subsec:classify}, for each $\l\in S(B_0)$, the number of impossible nodes in $A_\l$ is at least $|A_\l|-|B_0|$.

\subsubsection{The New Definition of Labels}

However, the original definition of labels no longer guarantees the existence of many impossible nodes in $A_\ell$ for each $\ell$ with some $R_0$ satisfying $B(\ell,R_0) = B_0$. 
In fact, it may happen that for such an $\ell$, there are no impossible nodes in $A_\ell$ at all.
Consider for example, some node $u \notin A_\ell'$.
When $v = u$ triggers an update, the leaf $\ell'$ appearing in condition (2) above may be such that $A_{\ell'}'$ contains no ancestor of $u$. 
Consequently, even though $A_\ell'$ may contain many nodes above $u$, these nodes cannot impose any constraint on the number of impossible nodes in $A_\ell$, because $\ell$ cannot serve as the leaf $\ell'$ in condition (2) that triggers an update (as $u \notin A_\ell'$).

This difficulty arises because the depths of elements in $A_\ell'$ are not sufficiently controlled by the label.
To address this, we augment the definition of labels by introducing a sequence $(y_j)$ to regulate the depths of elements in $A_\ell'$. 
For each $j$, $y_j$ denotes the depth of the $jK$-th element in $A_\ell'$ (counted from bottom to top), where $K$ is a fixed parameter of magnitude
$K = O(\log n)$.
\footnote{
For an input $(\ell, R)$, the greedy algorithm produces a label $B(\ell, R)=B_0$. $B_0$ contains the information of the the depths of elements in $A_\ell'$, which indicates that only those leaves $\ell$ with 
$A_\ell'$ satisfying such depth constraints can possibly be labeled as $B$.
}
Similarly to \cref{subsec:classify}, the algorithm proceeds
from leaf $\ell$ to the root along the path $P_\ell$,
and attempts to append some elements $e_i$ in $R$ to the label $B$. In addition to the set of elements, we introduce in $B$ another sequence $(d_i)$ to denote the depths of the nodes in $P_\ell$ representing element $e_i$. 
We have explained that difficulty in lower-bounding the number of impossible nodes arises from the fact that, during an update, the set $A_{\ell'}'$ corresponding to $\ell'$ may contain far fewer ancestors of $v$ than $A_\ell'$. 
However, with the above depth control, the difference between the numbers of ancestors in $A_\ell'$ and $A_{\ell'}'$ is at most $K$
since the depths of the $jK$-th element in $A_\ell'$ 
and in $A'_{\ell'}$ are the same (dictated by the label).
We also note that we do not control the depth of every element in $A_\ell'$,
since the number of possible labels should be properly bounded
(so that we can apply union bound over labels in proving the tail inequality).
Consequently, the modified algorithm can ensure that the number of impossible nodes in each $A'_\ell$ can be lower-bounded (see \cref{mainlem,defT}).



Now, we formally define the set $\mathfrak{B}$
of possible labels. The greedy algorithm, which we will introduce shortly, classifies each input $(\ell, R)$ to a label $B(\ell, R) \in \mathfrak{B}$.

\begin{definition}
\label{def:label-seq}
We define $\mathfrak{B}$ as the set of sequences of the form $(m;s; d_1,\ldots,d_m;e_1,\ldots,e_m;y_1,\ldots,y_{\lfloor \frac{s}{K} \rfloor})$, satisfying the following conditions:
\begin{itemize}
\item $0\leq s\leq n$, $m\geq 1$;
\item $n\ge d_1>d_2>\ldots>d_m\geq0$, $n\ge y_1>y_2>\ldots> y_{\lfloor \frac{s}{K}\rfloor}\geq0$; 
\item $e_2,\ldots,e_m$ are distinct elements in $U=[n]$, with $e_1=0$ serving as a placeholder.
\end{itemize}
\end{definition}
We refer to such a sequence $B\in\mathfrak{B}$ as a \textbf{label}
and $\mathfrak{B}$ as the set of all possible labels. 
In our context, $m$ is called the \emph{length} of the label (the notation $|B(\l,R)|$ mentioned above denotes the length $m$ of the label $B(\l,R)$, analogous to the notation $|\cdot|$ used in \cref{subsec:classify}). 
$s$ controls the size of $A_\l'$
and
the sequence $(y_j)$ controls the depths of nodes in $A_\l'$.
In other words, 
only leaves $\ell$ with $|A'_\l|=s$ and being consistent with depth information $(y_i)$ 
can possibly be labeled with
$B=(*\, ; s\, ; * \, ; y_1, \ldots, y_{\lfloor \frac{s}{K}\rfloor})$.
For each $j$, $y_j$ indicates the depth of the $jK$th element in $A_\l'$ (from bottom to top).
Moreover, the sequence $(d_i)$ denotes the depths of elements $e_i$.



We also need the notion of a \emph{prefix} of a label in $\mathfrak{B}$.  The \emph{prefix} of length $1\leq a\leq m$ of a label 
$B=(m;s;d_1,\ldots,d_m;e_1,\ldots,e_m;y_1,\ldots,y_{\lfloor\frac{s}{K}\rfloor})$ 
is defined to be 
$$\mathsf{Pref}_a(B)=(a;s;d_1,\ldots,d_a;e_1,\ldots,e_a;y_1,\ldots,y_{\lfloor\frac{s}{K}\rfloor}).$$ 
It is straightforward to verify that $\mathfrak{B}$ is \emph{prefix-closed} (a prefix of a label is also in $\mathfrak{B}$).

\subsubsection{Level sets $L_a(B)$ and Special-Ancestor sets $\cc_u(B)$}

On input $(\l,R)$, the greedy algorithm starts from leaf $\ell$  
progresses upward along $P_\l$ while maintaining a label $B$.
At the leaf $\ell$, the label $B$ is initialized to contain only $m=1$ element, the dummy element $0$ corresponding to a leaf 
(the dummy element is not a real element in $U$), 
the size $s$ of $A'_\ell$ and the sequence $(y_i)$ 
(which controls the depths of 
the $iK$-th node of $A'_\ell$).
Then, at each inner node along $P_\ell$, 
the algorithm attempts to update the label by appending one more element to $B$
if certain condition holds.
To formally describe these conditions in the greedy algorithm,
we also need to introduce the definitions
of level sets and special-ancestor sets (in \cref{lemma3.4}). 
The pseudo-code of the algorithm is presented in \cref{algo2}.

For a given label $B$ of length $m$,
we define $m$ sets of nodes $L_1,L_2,\ldots,L_m$, called the {\em level sets}. 
Intuitively, when $a=1$, $L_1$ is the set of leaves that can potentially produce the label. 
For $a\geq 2$, the $a$-th level set $L_a$ consists of the nodes that can potentially trigger the $a$-th update of the maintained sequence.
We also define the set $\cc_u(B)$, which we call the {\em special-ancestor set} of node $u$ with respect to $B$, as the set of nodes in $A_\l'$ with depths less than $\depth{u}$ for certain leaf $\l$ below $u$.
This set will be particular useful in bounding the number of impossible nodes later.


Now, let us present the formal definitions of level sets and special-ancestor sets, with an example illustrated in the first item of \cref{fig5}. Recall that in a single $A_\l'\subseteq P_\l$,  different nodes have different depths and represent distinct elements. 
\begin{definition}[{\bf Level sets $L_a(B)$ and Special-Ancestor sets $\cc_u(B)$}]
\label{lemma3.4}
\ \newline
Let $B = (m; s; d_1, \ldots, d_m; e_1, \ldots, e_m; y_1, \ldots, y_{\lfloor \frac{s}{K} \rfloor}) \in \mathfrak{B}$ be a label with length $m \geq 1$. Given a positive integer $1 \leq a \leq m$, we define $L_a(B)$ and $\cc_u(B)$ for $u\in L_a(B)$ as follows.
\begin{enumerate}[(a)]
\item When $a=1$, a node $u$ belongs to $L_1(B)$ if and only if:
\begin{itemize}
\item $u$ is a leaf, $\depth{u} = d_1$; \item $|A_u'|=s$;
\item for each $1 \leq i \leq \lfloor \frac{s}{K} \rfloor$, the depth of the $iK$-th node from the bottom of $A_u'$ (i.e., counted upward from the leaf toward the root) is equal to $y_i$.
\end{itemize}We define $\cc_u(B)$ to be the set of nodes in $A_u'$ of depths less than $d_1$.
\item When $a \geq 2$, a node $u$ belongs to $L_a(B)$ if and only if: 
\begin{itemize}
\item $\depth{u} = d_a$, $\elt(u) = e_a$;
\item there exists a leaf $\l \in L_1(B)$ such that the unique node in $A_\l'$ at depth $d_i$ represents $e_i$ for all $2 \leq i \leq a$ (which implies $u \in A_\l'$). Among all such leaves, we select the leftmost one (for tie-breaking), and define $\cc_u(B)$ to be the set of nodes in $A_\l'$ of depths less than $d_a$.
\end{itemize}
\end{enumerate}

\end{definition}

In both cases, any node $v\in \cc_u(B)$, $\depth{v} < d_a=\depth{u}$. Moreover, $v\in \cc_u(B) \subseteq A_\l'$ shows that $v$ is an active ancestor of $\l$, thus
on the path from root to $\l$, the \textsf{Yes} arc from $v$ is taken. At the same time, when $a\geq 2$, $u\in A_\l'$ must be a descendant of $v$ (also true for $a=1$). Therefore, we conclude:
\begin{observation}\label{ob3}
     For any $1\leq a\leq m$, the special ancestor set $\cc_u(B)$ is a subset of active ancestors of $u$.
\end{observation}
 The following observation is immediate from the definition, since membership of a node in $L_a(B)$ depends only on $m,s,y_j$ (for $1\leq j\leq\lfloor\frac{s}{K}\rfloor$) and $d_i,e_i$ (for $i\leq a$); furthermore, the tie-breaking process depends solely on the tree structure.
\begin{observation}\label{ob1}
For any two labels $B_1,B_2\in \mathfrak{B}$ of lengths $m$ and $m^\prime$, respectively, and for any $1\leq a\leq\min\{m,m^\prime\}$, the level sets $L_a(B_1)=L_a(B_2)$, provided that $\mathsf{Pref}_a(B_1)=\mathsf{Pref}_a(B_2)$. Moreover, for any $u\in L_a(B_1)=L_a(B_2)$, the special ancestor sets $\cc_u(B_1)=\cc_u(B_2)$.
\end{observation}
The following observation will be used in the description of \cref{algo2}.
\begin{observation}\label{ob2}
For any two distinct labels $B_1,B_2\in \mathfrak{B}$, both of length $m\geq 1$, if $\mathsf{Pref}_{m-1}(B_1)=\mathsf{Pref}_{m-1}(B_2)$, then 
$L_m(B_1)\cap L_m(B_2)=\emptyset.$
\end{observation}
\begin{proof}
Assume, for contradiction, that $u\in L_m(B_1)\cap L_m(B_2)$. When $m=1$, the contradiction is immediate since any node $u\in L_1(B)$, with $B=(m;s;d_i;e_i;y_j)$, must be a leaf with $s=|A_u'|$, while $y_i$ is equal to the depth of the last $iK$-th node in $A_u'$, and $d_1 =\depth{u}$. Moreover, by definition, $e_1$ is fixed as $0$, hence the label $B$ is uniquely determined by $u$, which gives $B_1=B_2$. 

For $m\geq 2$, by \cref{lemma3.4} (b), if $u\in L_m(B)$ for $B=(m;s;d_i;e_i;y_j)$, then $d_m= \depth{u}, e_m=\elt(u)$. Since $\mathsf{Pref}_{m-1}(B_1)=\mathsf{Pref}_{m-1}(B_2)$, it follows that $B_1=B_2$, contradicting the assumption. Thus, the result is established.
\end{proof}

\subsubsection{Details of the Greedy Algorithm}

We now proceed to present a detailed description of \cref{algo2}.  
The goal of the algorithm is to classify each pair $(\ell,R)$ using appropriate labels from the set $\mathfrak{B}$. Initially, at leaf $\ell$,
the algorithm sets $s$ and $y_j$'s based on $A_\l'$. 
It sets $m=1$ and $d_1=\depth{\l}$.
The iterator $u$ is initialized to the parent of leaf $\ell$.

The core functionality of the algorithm is implemented through a while loop. Within each iteration, the algorithm examines whether there exists a label $B_1 \in \mathfrak{B}$ that satisfies three conditions: it has length $a+1$, its prefix matches the current label (i.e., $B = \mathsf{Pref}_a(B_1)$), and the current node belongs to the corresponding level set (i.e., $u \in L_{a+1}(B_1)$). 
When such a label exists and the additional condition that either $u \in R$ or $u$ is the starting leaf is met, the algorithm identifies the unique label $B_1$ (whose uniqueness is guaranteed by \cref{ob2}). It then updates $B$ to $B_1$, increments $a$ by 1 and advances $u$ to its parent node in the tree. If then $u$ becomes the root of $\mathcal{T}$, then the loop terminates. Upon termination, the algorithm returns the final label $B$.

It should be noted that \cref{algo2} is essentially identical to the algorithm presented in \cref{subsec:classify}, with the interpretation that nodes in the level sets $L_{a+1}(B_1)$ represent those capable of extending the length of the maintained label $B$.

The pseudo-code of the algorithm is as follows:

\begin{algorithm}[H]
\caption{Labeling $(\ell,R)$}\label{algo2}
\KwIn{$(\ell,R)$}
  \KwOut{ $B =(m;s;d_1;\ldots,d_m;e_1,\ldots,e_m;y_1,\ldots,y_{\lfloor \frac{s}{K}\rfloor })\in \mathfrak{B}$}

$s\gets |A_\ell'|$; $d_1\gets \depth{\l}$; $y_i\gets$ the depth of the $iK$-th node from the bottom of $A_\l',1\leq i\leq \lfloor \frac{s}{K}\rfloor$ 

$B\gets (1;s;d_1;0;y_1,\ldots,y_{\lfloor \frac{s}{K}\rfloor })$; $a\gets 1$\Comment*{Initialization of output}

$u\gets \l.parent$

  \While{$u$ is not the root \label{outcond}}{

\If {$\exists B_1\in \mathfrak{B}$ of length $a+1$, s.t. $B = \mathsf{Pref}_{a}(B_1)$, $u\in L_{a+1}(B_1)$ \label{cond}}{

\If  {$u\in R$  {\rm\textbf{or}} $u = \ell$ \label{cond2}}{
\KwFind such unique $B_1$;\Comment*{Uniqueness by \cref{ob2}}
$B\gets B_1$; $a\gets a+1$
}
}
    $u \gets u.\textnormal{parent}$; \Comment*{Climbing up the path $P_{\ell}$. If $u$ is already the root of $\calT$, jump out while loop}
}
\Return{$B$};
\end{algorithm}

The algorithm is deterministic. Therefore, any $(\ell,R)$ is given a unique label, 
which we denote as $B(\ell,R)$. 
For ease of notation, for input $(\ell,R)$,
we use $m(\ell,R)$ to denote the first entry (i.e., length) of the label, $s(\ell,R)$ 
the second entry, etc. 

\subsubsection{Properties of the Output Label $B$}

Before proceeding, we need to examine further the properties of the output label $B$ produced by \cref{algo2}. It is also worth mentioning that $\l$ initializes $d_1=\depth{\ell}$, $e_1 = 0$, each $y_j$ is the depth of the $jK$-th node in $A_{\ell}^\prime$ (counted from $\l$), and $s = |A_{\ell}^\prime|$. These statistics never change after initialization. Moreover, one can observe the \emph{greedy} nature of this algorithm. Specifically, when traversing upward along $P_\ell$, the algorithm always processes lower nodes before higher ones. When conditions \cref{cond} and \cref{cond2} are satisfied, the maintained label $B$ is irrevocably updated. Consequently, a label $B_0$ is fixed, the fact that an outcome $(\ell,R)$ satisfies $B(\ell,R) = B_0$ already imposes specific constraints on $R$. This concept is illustrated in \cref{fig5}. The following lemma formalizes these properties:
\begin{lemma}
\label{reduction:unique}
For each $(\ell,R)$ pair, assume $ B(\ell,R) = B_0=(m;s;d_i;e_i;y_j)\in \mathfrak{B}$. Then the following holds for each $1\leq a\leq m$:
\begin{enumerate}[(a)]
\item $\ell \in L_1(B_0)$. During the execution of the algorithm, we always have $s=|A_\l'|$, and for any $1\leq i\leq \lfloor\frac{s}{K}\rfloor$, $y_i$ denotes the depth of the $iK$-th node from bottom of $A_\l'$.
\item For $a \geq 2$, there exists a unique node $v_a$ of depth $d_a$ in $ P_\ell$, $\elt(v_a) = e_a$, such that $v_a \in R\cap L_a(B_0)$ and the label $B$ is updated from length $(a-1)$ to $a$ at node $v_a$.
\item There does not exist any node $z \in R \cap P_\ell$ satisfying the two conditions:
\begin{enumerate}[(1)]
    \item $z$ is a descendant of $v_{a+1}$, where $v_{a+1}$ is the unique node defined in (b) (this condition is ignored when $a=m$);
    \item \(\exists B_1\in\mathfrak{B}\) of length \(a+1\) such that 
    \(\mathsf{Pref}_a(B_0)=\mathsf{Pref}_a(B_1)\) and \(z\in \level_{a+1}(B_1)\).
\end{enumerate}
\item If a node $z\in P_\l \cap \cc_v(B_0)\text{ for some }v\in \level_a(B_0)$, then the above condition (2) holds; that is, there exists \(B_1\in\mathfrak{B}\) of length \(a+1\) such that 
\(\mathsf{Pref}_a(B_0)=\mathsf{Pref}_a(B_1)\) and \(z\in \level_{a+1}(B_1)\).
\end{enumerate}
\end{lemma}

\begin{proof}
\begin{enumerate}[(a)]

\item For any leaf $\ell$, it follows directly from \cref{lemma3.4} that $\ell\in \level_1(B_1)$.  
It is evident from the algorithm that the length of $B$ never decreases during updates. 
Therefore, $\ell \in \level_1(B_1)=\level_1(B_0)$ by \cref{ob1}, since $B_1 = \mathsf{Pref}_1(B_0)$.
\item 
From \cref{cond}, when $a\ge 2$, the transition of $B$ from length $(a-1)$ to $a$ corresponds to a label $B_1\in \mathfrak{B}$ of length $a$, where $u\in R\cap \level_{a}(B_1)$. Denote this node $u$ as $v_a$. Since the length of $B$ only increases throughout the algorithm, $B_1 = \mathsf{Pref}_a(B_0)$, which by \cref{ob1} implies $\level_a(B_1)=\level_a(B_0)$. According to \cref{lemma3.4}, if $v_a\in \level_{a}(B_1)=\level_a(B_0)$, then $\depth{v_a} = d_a, \elt(v_a)=e_a$, matching the assertion. The uniqueness of $v_a$ follows from the fact that nodes in $P_\ell$ represent distinct elements. Furthermore, even for different outcomes $R_0$ and $R_1$ such that $B(\ell,R_0) = B_0 = B(\ell,R_1)$, the corresponding nodes $v_a$ remain identical.

\item 
 Let us assume, for contradiction, that there exists a node $z$ satisfying the two conditions. Then, assume $B_1 =(a+1;s;d_1,\cdots, d_a,d_{a+1}';e_i;y_j)$. Then, $z\in \level_{a+1}(B_1)$ shows $\depth{z} = d_{a+1}'$ by \cref{lemma3.4}, which should be strictly less than $d_a$ since $B_1\in \mathfrak B$. When examining the execution of \cref{algo2}, we find a contradiction as follows. Between the updates of $B$ from length $a-1$ to $a$ at node $v_a$, and then from length $a$ to $a+1$ at node $v_{a+1}$ (or maintaining length $m$ till termination if $a=m$), no intermediate node should trigger an update. However, the node $z$ would appear in the iterator $u$ after processing $v_a$ ($d_{a+1}' < d_a$) but before reaching $v_{a+1}$ (it is a descendant of $v_{a+1}$). At this point, $B$ equals $\mathsf{Pref}_a(B_0)$. $z\in R$ matches the conditions for updating $B$ in \cref{cond}, forcing an update of $B$, which gives a contradiction. Therefore, no such node $z$ can exist.
 
 \item Now we show that $z\in \cc_v(B_0)\text{ for some }v\in \level_a(B_0)$ suffices to ensure condition (2). We define $d_{a+1}^\prime:= \depth{z}$, $v\in \level_a(B_0)$ and $e_{a+1}^\prime = \elt(z)$. Then we define:
$$B_1 = (a+1;s;d_1\ldots,d_a,d_{a+1}^\prime;e_1,\ldots,e_a,e_{a+1}^\prime;y_1,\ldots,y_{\lfloor \frac{s}{K}\rfloor}).$$
We first verify that $B_1 \in \mathfrak{B}$. By \cref{ob3}, the special ancestor set $\cc_v(B_0)$ only contains nodes of depth less than $d_a$. Thus, $0\leq d_{a+1}^\prime < d_a$. Moreover, from part (b), the nodes $v_i\in P_\ell$ ensure that $\elt(v_i) = e_i$ differ from $\elt(z) = e_{a+1}'$, for $2\leq i\leq a$, as they appear in the same $P_\l$.
Thus, $B_1\in \mathfrak{B}$. 
Given that $z\in \cc_v(B_0)$, there must exist a leaf $\ell_0\in \level_1(B_0)=\level_1(B_1)$ (since $\mathsf{Pref}_a(B_0) = \mathsf{Pref}_a(B_1)$), such that the unique node at depth $d_i$ in $A_{\ell_0}'$ represents $e_i$ for $2\leq i\leq a$, and $z\in \cc_v(B_0)\subseteq A_{\l_0}'$. 
Since $\depth{z} = d_{a+1}^\prime$ and $\elt(z) = e_{a+1}^\prime$, the leaf $\ell_0$ satisfies the conditions in \cref{lemma3.4} that ensures $z\in \level_{a+1}(B_1)$. Hence, condition (2) is satisfied since \(\mathsf{Pref}_a(B_0)=\mathsf{Pref}_a(B_1)\).
\end{enumerate}
\end{proof}

\begin{figure}[t]    
  \centering
   \scalebox{0.8}[0.8]{
\begin{tikzpicture}[
  every node/.style={draw, circle, fill=none, inner sep=1.5pt},
  green node/.style={draw, circle, fill=green, inner sep=2.25pt},
  black node/.style={draw, circle, fill=black, inner sep=3pt},
  left edge/.style={draw=blue, line width=1.2pt, ->},
  right edge/.style={draw=red, line width=1.2pt, ->},
  tree node/.style={draw, regular polygon, regular polygon sides=3, fill=none, inner sep=2pt, minimum size=4mm},
]
  \node (legend_node) at (7,3.6) {};
  \node[draw=none, fill=none] (legend_no) at (5.8,2.2) {};
  \node[draw=none, fill=none] (legend_yes) at (8.2,2.2) {};
  \draw[left edge] (legend_node) -- node[left, draw=none, fill=none, pos=0.35] {\textsf{No}} (legend_no);
  \draw[right edge] (legend_node) -- node[right, draw=none, fill=none, pos=0.35] {\textsf{Yes}} (legend_yes);
\node (n4) at (-2,-3.6) [label=left :$v_2$, green node]{};
\node (u2c2) at (-1,-4.4) {}; 
\node (n6) at (-1,-2.8)[green node] {};
\node (n9) at (-1,-1.2) [label=left:$u$, black node] {};
\node (n11) at (-2, -2) {};
\node (v3) at (0,-0.4)  {};
\node (n10) at (1,0.4) {};
\node (n13) at (2,-0.4)[label=right:$v_2'''$, green node] {};
\node (n12) at (3,-1.2)[label=right:$\ell'''$] {};
\node (u5) at (0,1.2)[label=right:$v_3$, green node] {};
\node (v4) at (1,3.6) [label=left:$v_4$, green node]{};
\node (u6) at (-1, 2.0) {};
\node (u7) at (0, 2.8) {};
\node (u8) at (-2, -5.2) {};
\node (u9) at (-3, -6.0) [label=left:$\ell_0$]{};
\node (u10) at (0, -2.0) {};
\node (u11) at (-0.5, -2.4) [label=right:$v_2'$, green node]  {};
\node (u12) at (0.5, -3.2) [label=right:$\ell'$]  {};
\node (u13) at (1, -1.2) [label=right:$v_2''$, green node]  {};
\node (u14) at (2, -2.0) [label=right:$\ell''$]  {};

   \draw[left edge]  (u7) -- (u6);
 \draw[left edge]  (v4) -- (u7);
\draw[left edge] (n6) -- (n4);
\draw[left edge] (v3) -- (n9);
\draw[left edge] (n10) -- (v3);
\draw[right edge] (u5) -- (n10);
\draw[right edge] (n10) -- (n13);
\draw[right edge] (n13) -- (n12);
\draw[right edge] (u6) -- (u5);
\draw[left edge] (n9) -- (n11);
\draw[right edge] (n11) -- (n6);
\draw[right edge] (n4) -- (u2c2);
\draw[left edge] (u2c2) -- (u8);
\draw[left edge] (u8) -- (u9);
\draw[right edge] (n9) -- (u10);
\draw[left edge] (u10) -- (u11);
\draw[right edge] (u11) -- (u12);
\draw[right edge] (v3) -- (u13);
\draw[right edge] (u13) -- (u14);
  \draw[dashed] ($(u2c2)+(-2,-1.6)$) -- ++(8,6.4)
 node[pos=1, draw=none, fill=none, inner sep=0, yshift=0pt, xshift = 10pt, anchor=south]{$6K$};

\draw[dashed] ($(n6)+(-2.5, -2.0)$)  -- ++(7.5, 6) node[pos=1, draw=none, fill=none, inner sep=0, yshift=-10pt, xshift = 5pt, anchor=south]{$6K-1$};

\draw[dashed] ($(v3)+(-4,-3.2)$) --  ++(7,5.6)
  node[pos=1, draw=none, fill=none, inner sep=0, yshift=-10pt, xshift = 5pt, anchor=south]{$6K-2$};
 
\end{tikzpicture} }
\caption{\small 
In the decision tree, red edges indicate \textsf{Yes} arcs, and blue edges indicate \textsf{No} arcs. Nodes with the same depth are connected using dashed lines, and their depth is labeled on them. 
Nodes higher than $v_4$ and in the $\mathsf{Yes}$ subtree of $v_4$ are not drawn. Let us assume that $K$ is large and $|A_{\ell}'| = 4K$ and the depths of the $K$-th, $2K$-th, $3K$-th, $4K$-th nodes of $A_{\ell}'$ (counted from $\l$) are $4K+10,3K+10,2K+10,K+10$ for all leaves $\ell = \ell_0,\ell',\ell'',\ell'''$. 
Moreover, we assume $v_2 \in A_{\ell_0}'$, $v_2' \in A_{\ell'}'$, $v_2'' \in A_{\ell''}'$, $v_2''' \in A_{\ell'''}'$, $v_3\in A_{\ell'}',A_{\ell''}'$, but $v_3\notin A_{\ell_0}',A_{\ell'''}'$.
Regarding elements, $\elt(v_2)=\elt(v_2')=\elt(v_2'') = \elt(v_2''')$. Consider the label $B_0=(4;4K;6K,6K-1,6K-3,6K-4;0,\elt(v_2),\elt(v_3),\elt(v_4);4K+10,3K+10,2K+10,K+10)$. 
\newline \quad $\bullet$ All leaves $ \ell_0,\ell',\ell'',\ell'''$ are in $\level_1(B_0)$, and all $v_2,v_2',v_2'',v_2'''$ are in $\level_2(B_0)$. $v_3$ is in $\level_3(B_0)$. Now consider $\cc_{v_3}(B_0)$. Note that the satisfying $\ell$ in the definition of $\cc_{v_3}(B_0)$ can only be $\ell'$ or $\ell''$ (it needs $v_3\in A_\l'$), and hence $\cc_{v_3}(B_0)$ is defined to be the nodes of depth less than $6K-3$ in $A_{\ell'}'$ (as $\ell'$ is on the `left' of $\ell''$).
\newline \quad $\bullet$ Suppose that the specific outcome $R_0$ includes all green nodes, but does not include white nodes and the nodes above $v_4$. The algorithm with input $(\ell_0,R_0)$ updates $B$ at $\ell_0,v_2,v_3,v_4$, and hence $B(\ell_0,R_0) = B_0$ (note that $v_3\in \level_3(B_0)$ can trigger an update, even though $v_3 \notin A_{\ell_0}'$).
\newline \quad $\bullet$
Note that $u$ is a descendant of $v_3=v_3(\l_0,B_0)$, and $u\in \cc_{v_2'}(B_0)$ for $v_2'\in \level_2(B_0)$. These facts imply the existence of the label $B_1 =(3;4K;6K,6K-1,6K-2;0,\elt(v_2),\elt(u);4K+10,3K+10,2K+10,K+10)$ that $\mathsf{Pref}_2(B_1) = \mathsf{Pref}_2(B_0)$ by \cref{reduction:unique} (c). This means that $u\in \imp(\l_0, B_0)$. Intuitively $u\notin R_0$, since otherwise, it can update $B$ at $u$ before $v_3$.
}\label{fig5}
\end{figure}

From the above lemma, we can relate $\fspecial(\elt(R\cap P_\ell)), |A_\ell'|$ with the entries of $B(\ell,R)$.
\begin{lemma}\label{lemma3.8}
For any $(\ell,R)$, we have $\fspecial(\textnormal{elt}(R\cap P_\ell))\geq m(\ell,R)-1$, $|A_\ell'| =s(\ell,R) $. Recall that $m(\ell,R),s(\ell,R)$ are the entries $m,s$ in $B(\l,R)$.
\end{lemma}

\begin{proof}
    Assume that the label of $(\l,R)$ is $B_0 = (m;s;d_1,\ldots,d_m;e_1,\ldots,e_m;y_1,\ldots,y_{\lfloor \frac s K \rfloor})$, and $m\geq 2$.
Then, by \cref{reduction:unique} (b), nodes $\depth{v_a} = d_a$ and $v_a\in R\cap P_\l\cap \level_a(B_0)$ for any $2\leq a\leq m$. From \cref{lemma3.4} (b), $v_a\in \level_a(B_0)$ implies $\elt(v_a) = e_a$. In particular, $v_m\in \level_m(B_0)$ shows that there exists a leaf $\l_0\in \level_1(B_0)$ such that the unique node of depth $d_i$ in $ A_{\l_0}^\prime$ represents $e_i$, for any $2\leq i\leq m$. 
This already means that $e_a\in \elt(R\cap P_\l)\cap \elt(A_{\l_0}^\prime)$, $\forall 2\leq a\leq m$. Hence, the conclusion $\fspecial(\elt(R\cap P_\ell))=\max\limits_{\l_1}|\elt(A_{\l_1}^\prime)\cap \elt(R\cap P_\ell)|\geq |\elt(A_{\l_0}^\prime)\cap \elt(R\cap P_\ell)| \geq m-1$ follows. For the second assertion, \cref{reduction:unique} (a) implies that $\l\in \level_1(B_0)$ and $|A_\l^\prime| =  s =s(\l,R)$.
\end{proof}

We now proceed to derive consequences of \cref{reduction:unique}. We first define the subset $S(B_0)$ of leaves as in \cref{subsec:classify}.
\begin{definition}[\bf Generating Set $S(B_0)$ of $B_0$ and the $a$-th Triggering Node $v_a(\ell,B_0)$]\label{defgen} 
\ \newline
For a label $B_0 = (m, s, d_i, e_i, y_j) \in \mathfrak{B}$, we define the generating set of $B_0$, denoted as $S(B_0)$, as the set of leaves $\l$ for which there exists some outcome $R$ such that $B(\l, R) = B_0$, i.e.,
$$
S(B_0) = \left\{ \ell \text{ is a leaf of } \calT : \exists R \text{ such that } B(\ell, R) = B_0 \right\}.
$$
For any leaf $\ell \in S(B_0)$, \cref{reduction:unique} (b) ensures the existence of a unique node $v_a$ of depth $d_a$, $\elt(v_a) = e_a$, such that $v_a \in P_\ell \cap \level_a(B_0)$ for each $2 \leq a \leq m$.
$v_a$ triggers the $a$-th update of the label, and we call it the $a$-th triggering node of $\ell$.
We denote such $v_a$ as $v_a(\ell,B_0)$ to make the  dependency on $\ell,B_0$ explicit.
For notational convenience, we define $v_1(\ell,B_0) := \ell \in \level_1(B_0)$. Note that the definitions of $S(B_0)$ and $v_a(\ell,B_0)$ do \textbf{not} depend on the specific choice of $R$.
\end{definition}

Part (c) of \cref{reduction:unique} tells us that when $B(\ell,R) =B_0$, there are constraints on the active nodes (in \nonadaptivestrategy) along $P_\ell$. In detail, there does not exist a node in $R \cap P_\ell$ that satisfies the following two conditions: it is a descendant of $v_{a+1}(\l,B_0)$; it is in $\level_{a+1}(B_1)$ for some \(B_1\in\mathfrak{B}\) of length \(a+1\) such that \(\mathsf{Pref}_a(B_0)=\mathsf{Pref}_a(B_1)\). Therefore, for each $\ell\in S(B_0)$, we define the subset of $P_\ell$ that cannot be active (i.e. $\not\in R$) as follows (the condition on $v_{m+1}(\ell,B_0)$ when $a= m $ is ignored): 
\begin{align}\label{defH}
     \imp(\ell,B_0):= &
     \bigcup_{a=1}^{m}\{u\in  P_\ell  \cap \level_{a+1}(B_1): B_1\in \mathfrak{B},
 \notag \\
    &\qquad \mathsf{Pref}_a(B_1) = \mathsf{Pref}_a(B_0), u \text{ is a descendant of } v_{a+1}(\ell,B_0)\}.
\end{align}
Notice that $\imp(\ell,B_0)$ is defined with respect to only $\ell$ and $B_0$, and does not depend on $R$, since after fixing $\ell\in S(B_0)$ and $B_0$, the sets $\level_a(B_0)$, $\cc_v(B_0)$ and the nodes $v_a(\ell,B_0)$ are the same for any $R$ that satisfies $B(\ell,R) = B_0$ as mentioned in \cref{reduction:unique} (b). 

As discussed in the beginning of this section, we can view nodes in $\level_a(B_0)$ as nodes that can potentially trigger an update in \cref{subsec:classify}. Therefore, the set $\imp(\ell,B_0)$ defined here corresponds to the notion of \textbf{impossible} nodes defined earlier. In the following lemma, we investigate the properties of $\imp(\ell,B_0)$ and provide a lower bound on the cardinality of $\imp(\ell,B_0)\cap A_\ell$, as we did in \cref{subsec:classify}.
\begin{lemma}
\label{mainlem}
For a fixed label $B_0 =(m;s;d_i;e_i;y_j)\in \mathfrak{B}$, we define $\imp(\ell,B_0)$ for $\ell\in S(B_0)$ as above. Then $\imp(\ell,B_0)$ does not contain any leaf of the original tree $\calT$. Moreover, $|\imp(\ell,B_0)\cap A_\ell|\ge s-mK$.
\end{lemma}
\begin{proof}

First, for any $u\in \imp(\ell,B_0)$ in the $a$-th union, by \cref{lemma3.4}, the depth of $u\in P_\l \cap \level_{a+1}(B_1)$ is the $(a+1)$-th entry of the depth sequence of $B_1$, which is strictly smaller than its $a$-th entry $d_a$. Thus, $u$ is an ancestor of $v_a(\ell,B_0)\neq u$ (they are both in $P_\l$), hence not a leaf of $\calT$. 

Second, we consider the following subset of $\imp(\ell,B_0)$, where $v_a(\ell,B_0)$ is defined for $\ell\in S(B_0)$ in \cref{defgen} (in the following, we abbreviate $v_a(\ell,B_0)$ to $v_a$, 
and the special-ancestor set of $v_a$, i.e., $\cc_{v_a}(B_0)$, to $\cc_{v_a}$, since the leaf and label are fixed):
$$
\imp  = \bigcup_{a=1}^{m}\{u\in \cc_{v_a}:d_{a+1}<\depth{u} < d_{a}\}, $$
where we define $d_{m+1}=-1$ here.


We first prove $\imp\subseteq A_\ell$. By \cref{ob3}, any node $u\in \cc_{v_a} $ is an active ancestor of $v_a\in P_\ell $. In particular, $u\in P_\ell$ and on the path from root to $\l$, the \textsf{Yes} arc from $u$ is taken, and hence $u\in A_\l$, so $\imp\subseteq A_\ell$.

Now, we take $v =v_a\in \level_a(B_0)$ in \cref{reduction:unique} (d).
We still consider a node $u\in \cc_{v_a}$.
Notice that $u\in P_\l\cap \cc_{v_0}$, so \cref{reduction:unique} (d) implies the existence of \(B_1\in\mathfrak{B}\) of length \(a+1\) such that 
\(\mathsf{Pref}_a(B_0)=\mathsf{Pref}_a(B_1)\) and \(u\in \level_{a+1}(B_1)\).
Moreover, $\depth{u} > d_{a+1}$.
So when $a<m$, $u$ is a descendant of $v_{a+1}$.
Thus $u\in\imp(\ell,B_0)$ for any $u\in\imp$.
Therefore $\imp\subseteq \imp(\ell,B_0)$, and thus $\imp\subseteq \imp(\ell,B_0)\cap A_\ell$.

Now we lower bound $|\imp|$.
By definition of $\l_a$, the depth of the last $jK$-th node in $A_{\ell_a}^\prime$ is $y_j$, for $1\leq j\leq \lfloor \frac{s}{K}\rfloor$. 
Denote $y_0 = d_1$. Notice that for some $\l$ with depth $d_1$, $y_i$ denotes the depth of the $iK$-th node from bottom of $A_\l'$.
Thus we have $y_0>y_1>\cdots>y_{\lfloor \frac{s}{K} \rfloor}$.
We consider $[y_{\lfloor \frac{s}{K} \rfloor }, y_0)$.
We find that all $\lfloor \frac{s}{K} \rfloor $ indices from $1\leq j\leq\lfloor \frac{s}{K} \rfloor  $ of $y_j$ divide the interval $[y_{\lfloor \frac{s}{K} \rfloor }, y_0)$ into $\lfloor \frac{s}{K} \rfloor $ intervals $[y_{j},y_{j-1})$. 
However, there are at most $m-1$ indices of $d_i$, $2\leq i\leq m$ in $[y_{\lfloor \frac{s}{K} \rfloor},y_0)$. 
Therefore, there are at least $\lfloor \frac{s}{K} \rfloor - m +1 $ indices $1 \leq j\leq \lfloor \frac{s}{K} \rfloor$, such that the interval $[y_{j},y_{j-1})$ contains no $d_a$, for any $2\leq a\leq m$. 
We define $I(B_0)$ as the union of such intervals $[y_{j},y_{j-1})$. 
See Figure~\ref{fig:implowerbound} for an example.
For each of such intervals, there must be some interval $(d_{a+1},d_{a})$ containing the entire interval $[y_{j},y_{j-1})$, $1\leq a\leq m$, as $(d_{m+1}=-1,d_1) \supseteq [y_{\lfloor \frac{s}{K} \rfloor},d_1)$. Now, the cardinality of nodes in $\cc_{v_a}$ of depth in $[y_{j},y_{j-1})$ is exactly $K$, since the depth of the $jK$-th, $(j-1)K$-th node in $A_{\ell_a}^\prime $ (counted from the bottom $\ell_a$) is $y_{j},y_{j-1}$, and $\cc_{v_a}$ is defined to be all nodes in $A_{\ell_a}^\prime $ with depth less than $d_a\geq y_{j-1}$. 
All such $K$ nodes are entirely contained in the subset $\{u\in \cc_{v_a}: \depth{u}\in  (d_{a+1}, d_a  )\}$. 
This shows that there are $K$ nodes in $\imp$ with depth in $[y_j,y_{j-1})$ as long as $[y_j,y_{j-1})\subseteq I(B_0)$.
Summing up over all intervals in $I(B_0)$ (all intervals $[y_{j},y_{j-1})$ are disjoint), we obtain $ |\imp|\geq K(\lfloor \frac{s}{K} \rfloor - m +1 )\geq s-mK$.

Therefore, $|\imp(\ell,B_0)\cap A_\ell|\geq |\imp|\geq s-mK$.
\end{proof}

\begin{figure}[H]
    \centering
    \begin{tikzpicture}[scale=0.8]
        \def\yfive{1}    
        \def\yfour{3}    
        \def\ythree{5}   
        \def\ytwo{7}     
        \def\yone{9}     
        \def\done{11}    

        \draw[->] (\yfive-0.5,0) -- (\done+1,0) node[right]{};
        \node[above=20pt] at ({(\yfive+\done)/2}, 0) {\textbf{Depth of Nodes in $A_\l'$}};
        \node[below left=0pt and 5pt] at (\yfive-0.5,0) {\small Low};
        \node[below right=0pt and 5pt] at (\done+1,0) {\small High};

        \node[below] at (\yfive,0) {$y_5$};
        \node[below] at (\yfour,0) {$y_4$};
        \node[below] at (\ythree,0) {$y_3$};
        \node[below] at (\ytwo,0) {$y_2$};
        \node[below] at (\yone,0) {$y_1$};
        \node[below] at (\done,0) {$y_0$};

        \draw[thick] (\yfive,0) -- (\yone,0);

        \draw[red, fill=red] (10,0) circle (2pt) node[above] {$d_2$};
        \draw[red, fill=red] (6,0)  circle (2pt) node[above] {$d_3$};
        \draw[red, fill=red] (4,0)  circle (2pt) node[above] {$d_4$};
        \draw[red, fill=red] (11,0) circle (2pt) node[above] {$d_1$};

        \fill[blue!20, opacity=0.5] (\yfive, -0.4) rectangle (\yfour, 0.4);
        \fill[blue!20, opacity=0.5] (\ytwo, -0.4) rectangle (\yone, 0.4);

        \filldraw[black] (\yfive,0.4) circle (2pt);
        \draw[thick] (\yfour,0.4) circle (2pt);
        \node[above] at (\yfive,0.4) {\footnotesize $[$};
        \node[above] at (\yfour,0.4) {\footnotesize $)$};

        \filldraw[black] (\ytwo,0.4) circle (2pt);
        \draw[thick] (\yone,0.4) circle (2pt);
        \node[above] at (\ytwo,0.4) {\footnotesize $[$};
        \node[above] at (\yone,0.4) {\footnotesize $)$};

        \draw[decorate,decoration={brace, mirror, amplitude=5pt}, yshift=-12pt]
            (\yfive,0) -- (\yfour,0) node[midway,above=20pt]{\footnotesize \(K\) nodes};
        \node at ({(\yfour+\yfive)/2},-1) {\small \(I(B_0)\)};

        \draw[decorate,decoration={brace, mirror, amplitude=5pt}, yshift=-12pt]
            (\ytwo,0) -- (\yone,0) node[midway,above=20pt]{\footnotesize \(K\) nodes};
        \node at ({(\ytwo+\yone)/2},-1) {\small \(I(B_0)\)};

        \draw[dashed] (\yfive, -0.5) -- (\yfive, 0.5);
        \draw[dashed] (\yfour, -0.5) -- (\yfour, 0.5);
        \draw[dashed] (\ythree, -0.5) -- (\ythree, 0.5);
        \draw[dashed] (\ytwo, -0.5) -- (\ytwo, 0.5);
        \draw[dashed] (\yone, -0.5) -- (\yone, 0.5);
        \draw[dashed] (\done, -0.5) -- (\done, 0.5);

        \draw[dotted] (\ythree,0) circle (2pt);

    \end{tikzpicture}
    \caption{\small $[y_2,y_1)$ and $[y_5,y_4)$ do not contain any $d_i$, hence contained in $(d_3,d_2)$ and $(-1,d_4)$ respectively. $\cc_{v_2}$ is defined to be nodes in $A_{\ell_2}^\prime $ with depth less than $d_2\geq y_{1}$, hence contain all $K$ nodes at depth $[y_2,y_1)$. Similarly, $A_{\ell_4}^\prime $ contain all $K$ nodes at depth $[y_5,y_4)$.}
    \label{fig:implowerbound}
\end{figure}

\begin{remark}
\label{remark3.7}
In the proof, we have defined $I(B_0)$ as the union of intervals $[y_j,y_{j-1})$ for $1\leq j\leq \lfloor \frac{s}{K} \rfloor $ such that no $d_a$ ($2\leq a\leq m$) is in the interval. 
We can observe that the nodes in $\imp(\ell,B_0)\cap A_\ell$ whose depths are in $I(B_0)$ are at least $|I(B_0)|\geq s-mK$. This fact is not immediately used in this section, but will be used in \rm\cref{Section 4}.
\end{remark}

We define the subset $T(B_0)$ of nodes in $\calT$ as 
$
T(B_0) = \bigcup_{\l \in S(B_0)}\imp(\l,B_0).
$ The following lemma captures the constraints on the nodes $P_\l$ to be in $R$. This result is similar to the one mentioned in \cref{subsec:classify}: if $B(\l,R) =B_0$, then any \textbf{impossible} node along $P_\l$ can not be in $R$.
\begin{lemma}\label{defT}
Fix a label $B_0\in \mathfrak{B}$. For any $\l\in S(B_0),R$ such that $B(\l,R)=B_0$, we have $R\cap P_\l\cap T(B_0) =\emptyset$.
\end{lemma}
\begin{proof}
Assume that $B(\l,R) = B_0$. Also, assume that there is a $u\in T(B_0)$ such that $u\in R\cap P_\l$. By the definition of $T(B_0)$, $u\in \imp(\l',B_0)$ for some $\l'\in S(B_0)$. 
Expanding further \cref{defgen}, there exists $1\leq a\leq m$ such that $u\in P_{\l'}$ is a descendant of $v_{a+1}(\l',B_0)$ (when $a<m$) and $ u\in \level_{a+1}(B_1)$ for some $ B_1\in \mathfrak{B}, \mathsf{Pref}_a(B_1) = \mathsf{Pref}_a(B_0)$. 
As $u\in R\cap P_\l$, we consider \cref{reduction:unique} (c).
The first condition of \cref{reduction:unique} (c) cannot be satisfied for $z=u$ as the second is true.
Hence, $a < m$, and $u$ is not a descendant of $v_{a+1}(\l,B_0)$. Since $u,v_{a+1}(\l,B_0)\in P_\l$, $u$ must be an ancestor of $v_{a+1}(\l,B_0)$ (including $v_{a+1}(\l,B_0)$ itself). 
However, this implies that $v_{a+1}(\l,B_0)$ is a descendant of $v_{a+1}(\l',B_0)$ since $u$ is a descendant of $v_{a+1}(\l',B_0)$. Therefore, $v_{a+1}(\l',B_0)\in P_\l$, which implies $v_{a+1}(\l,B_0) = v_{a+1}(\l',B_0)$, because the two nodes represent the same element $e_{a+1}$ in $P_\l$. This is a contradiction. Then, the conclusion follows.
\end{proof}

\subsection{Proof for \cref{thm:sec3main}}
\label{sec3:main}
In this section, we prove \cref{thm:sec3main}.
Recall that we need to lower bound $\E[\fspecial(\elt(R\cap P_\ell))]$ for the $\fspecial$ defined in the beginning of this section.
We set $K =\lceil 25\log n \rceil + 25$ in this subsection.

First, we need to prove the following lemma:
\begin{lemma}
\label{liu:induction}
Let $S$ be a subset of leaves in the tree $\calT$. There is a subset $T$ of nodes in $\cal T$, not containing any leaf of $\calT$. There exists an integer $h\in \mathbb N$, such that for each $\l \in S$, $| A_\l\cap T|\geq h $. Then the following inequality holds:
$$\E_{\ell \leftarrow \pi_{\calT}} \left[ \mathbf{1}_{\ell \in S}\Pr_{R \sim \Rdistribution} \left[R\cap P_\ell\cap T=\emptyset\right]\right]\leq 2^{-h}.$$
Here, $\mathbf{1}_{E}$ denotes the indicator function of the event $E$.
\end{lemma}

\begin{proof}
The result is trivial for $h = 0$, as the right-hand side is $1$. Therefore, in the following, we may assume that $h\ge 1$.
We use induction on the height (the length of the longest root-leaf path, not the depth) of $\calT $ to prove the desired statement. 

Assume now that the claim holds for trees of smaller height. For the current tree $\calT$, let $\calT_1$ and $\calT_2$ denote the subtrees corresponding to the $\textsf{Yes}$ and $\textsf{No}$ arcs from the root $r_0$, respectively. Let $\lambda$ be the probability that the root is active (if $r_0$ has only one child, we set $\lambda = 0$ or $1$ accordingly). Then, the path distribution $\pi_{\calT}$ assigns probability $\lambda$ to the $\textsf{Yes}$ subtree $\calT_1$ and $1-\lambda$ to the $\textsf{No}$ subtree $\calT_2$, preserving the distributions $\pi_{\calT_1}$ and $\pi_{\calT_2}$.

Now, we treat the following two cases separately. We denote $$E_1:=\E_{\ell \leftarrow \pi_{\calT_1}} \left[ \mathbf{1}_{\ell \in S}\Pr_{R \sim \Rdistribution} \left[R\cap P_\ell\cap T=\emptyset\right]\right], E_2:=\E_{\ell \leftarrow \pi_{\calT_2}} \left[ \mathbf{1}_{\ell \in S}\Pr_{R \sim \Rdistribution} \left[R\cap P_\ell\cap T=\emptyset\right]\right].
$$
\begin{mycase}{1}
$r_0\in T.$
\end{mycase}
In this case, we see that for leaves $\ell$ in $S\cap \calT_1$, there are at least $h-1$ active ancestors of it inside $T\cap \calT_1$. Moreover, we denote the corresponding root-leaf path in $\calT_1$ or $\calT_2$ as $P_{\ell}^\prime=P_{\ell}\backslash\{r_0\}$, then the corresponding relation $ \Pr_{R \sim \Rdistribution} \left[R\cap P_\ell\cap T=\emptyset\right] = (1-\lambda)\Pr_{R \sim \Rdistribution} \left[R\cap P_\ell^\prime\cap T=\emptyset\right]$ holds, as the root $r_0$ has an independent probability of $1-\lambda$ to be active. Therefore, by the induction hypothesis for $\calT_1$,
\begin{align*}
E_1=&(1-\lambda)\E_{\ell \leftarrow \pi_{\calT_1}} \left[ \mathbf{1}_{\ell \in S}\Pr_{R \sim \Rdistribution} \left[R\cap P_\ell'\cap T=\emptyset\right]\right] 
\leq (1-\lambda)2^{-h+1}.
\end{align*}
Similarly, for the $\textsf{No}$ subtree $\calT_2$, each leaf has at least $h$ active ancestors inside $T\cap \calT_2$, since the root is not an active ancestor for leaves in the $\textsf{No}$ subtree.
Hence, 
\begin{align*}
E_2= (1-\lambda)\E_{\ell \leftarrow \pi_{\calT_2}} \left[ \mathbf{1}_{\ell \in S}\Pr_{R \sim \Rdistribution} \left[R\cap P_\ell'\cap T=\emptyset\right]\right] \leq  (1-\lambda)2^{-h}.
\end{align*}
Putting together,
\begin{align*}
\E_{\ell \leftarrow \pi_{\calT}} \left[ \mathbf{1}_{\ell \in S}\Pr_{R \sim \Rdistribution} \left[R\cap P_\ell\cap T=\emptyset\right]\right] &= \lambda E_1 + (1-\lambda)E_2\leq 2^{-h}(2\lambda(1-\lambda)+(1-\lambda)^2)\leq 2^{-h}.
\end{align*}
\begin{mycase}{2}
$r_0\notin T$.
\end{mycase}
In this case, we see that for leaves $\l$ in both subtrees $S\cap \calT_1$ and $S\cap \calT_2$, there are at least $h$ active ancestors of it inside $T\cap \calT_1$ and $T\cap \calT_2$, since $r_0\notin T$. 
Moreover, we denote the corresponding root-leaf path in $\calT_1$ or $\calT_2$ as $P_{\l}^\prime=P_{\l}\backslash\{r_0\}$, then the corresponding  relation $ \Pr_{R \sim \Rdistribution} \left[R\cap P_\ell\cap T=\emptyset\right] = \Pr_{R \sim \Rdistribution} \left[R\cap P_\ell^\prime\cap T=\emptyset\right]$ holds, as the root $r_0\notin T$. 
Therefore, by the induction hypothesis on $\calT_1$ and $\calT_2$, we obtain $E_1,E_2\leq 2^{-h}$. Putting together,
$
\E_{\ell \leftarrow \pi_{\calT}} \left[ \mathbf 1_{\l \in S}\Pr_{R \sim \Rdistribution} \left[R\cap P_\ell\cap T =\emptyset \right]\right] = \lambda E_1 + (1-\lambda)E_2 
\leq 2^{-h}.
$

In conclusion, the result holds for all binary trees $\calT$.
\end{proof}

\begin{lemma}\label{lemma3.11}
For a fixed label $B_0 =(m;s;d_i;e_i;y_j) \in \mathfrak{B}$, the probability that the label of a random leaf and outcome $(\l,R)$ is $B_0$ is at most $2^{mK-s}$, i.e.,
$$
\Pr_{\ell \leftarrow \pi_{\calT}, R \sim \Rdistribution} \left[B(\ell,R) = B_0 \right]\leq 2^{mK-s}. 
$$
\end{lemma}
\begin{proof}
The result is trivial when $s\leq mK$, so we assume $s > mK$. Define $S = S(B_0)$ as in \cref{defgen} and $T = T(B_0)$ as in the discussion preceding \cref{defT}.
Recall that $S(B_0)=\{\ell \text{ is a leaf of } \calT: \exists R \text{ such that } B(\ell,R) = B_0\}$, and $T(B_0)$ is the union of all $\imp(\ell,B_0)$ for $\ell\in S(B_0)$.

It is clear from \cref{mainlem} that $T$ does not contain any leaf of $\calT$. Moreover, for any $\ell \in S$, we have $|\imp(\ell,B_0)\cap  A_\ell| \ge h := s - mK > 0$. Hence, by Lemma~\ref{liu:induction}, 
$$
\E_{\ell \leftarrow \pi_{\calT}} \left[ \mathbf{1}_{\ell \in S}\Pr_{R \sim \Rdistribution} \left[R\cap P_\ell\cap T=\emptyset\right]\right] \leq 2^{-h} = 2^{mK-s}.
$$
By \cref{defT} and the definition of $S$, for any ($\ell,R$), the condition $B(\ell, R) = B_0$ implies that $R \cap P_\ell \cap T = \emptyset$ and $\ell \in S$.
As a result,
$$
\Pr_{\ell \leftarrow \pi_{\calT}, R \sim \Rdistribution} \left[B(\ell, R) = B_0\right]\leq \E_{\ell \leftarrow \pi_{\calT}}\left[ \mathbf{1}_{\ell \in S}\Pr_{R \sim \Rdistribution} \left[R\cap P_\ell\cap T=\emptyset\right]\right] \leq 2^{mK-s}.
$$
\end{proof}

Now we are ready to prove \cref{thm:sec3main} (formally stated in \cref{sec3:redcution}). 
Recall that we aim to prove the following:
$$
\E_{\ell \leftarrow \pi_{\calT}} \left[ \E_{R \sim \Rdistribution} \left[\fspecial(\textnormal{elt}(R\cap P_\ell))\right] \right]\geq \frac{1}{200\log n}\E_{\ell \leftarrow \pi_{\mathcal{T}}}\left[|A_\ell'|\right].
$$
Here, $\fspecial$ is defined as
$$
\fspecial(S) = \max_{\ell} \left\{ | \elt(A_\ell') \cap S | \right\},
$$
where $A_\ell'$ is a subset of the active ancestors of $\ell$, and $\E_{\ell \leftarrow \pi_{\calT}}\left[|A_\ell'|\right] \geq 200\log n$.

\begin{proof}[Proof of \textnormal{\cref{thm:sec3main}}]
Before proceeding, we first give an upper bound on the number of labels in $\mathfrak{B}$ with $s \geq 2mK$. For a fixed $1 \leq s \leq n$, there are at most $s/2K \leq n$ choices for $1 \leq m \leq s/2K$, and for each $m$, there are at most $(n+1)^m \times n^m \times (n+1)^{s/K} \leq (2n)^{2s/K}$ possible combinations of $d_i, e_i, y_j$.

From \cref{lemma3.8}, we can express the term $\adaptcost := \E_{\ell \leftarrow \pi_{\calT}}\left[|A'_\ell|\right]$ as
$
\E_{\ell \leftarrow \pi_{\calT}} \left[ \E_{R \sim \Rdistribution} \left[s(\ell, R) \right] \right].
$
Since for any outcome of $R\cap P_\ell$, the label is unique,
$$
s(\ell, R)  = \sum_{B_0 \in \mathfrak{B}} s(B_0) \cdot \mathbf{1}_{B(\ell,R) = B_0}.
$$
Plugging in this formula and exchanging the order of summation, we obtain
$$
\E_{\ell \leftarrow \pi_{\calT}}\left[|A'_\ell|\right] = \sum_{B_0 \in \mathfrak{B}} s(B_0) \cdot \Pr_{\ell \leftarrow \pi_{\calT}, R \sim \Rdistribution} \left[B(\ell,R) = B_0\right].
$$
We use $(m;s;*)$ to denote a label of the form $(m;s;d_i;e_i;y_j)$.
Now we split the sum over $B_0 = (m; s; *) \in \mathfrak{B}$ into two cases: $s \ge 2mK$ and $s < 2mK$. For the former, we apply \cref{lemma3.11} to bound the probability for each $s \ge 1$, and then sum over all such terms. For the latter, we bound the contribution using the length of the random variable $m(\ell, R)$.

In detail, we have the bound for the case $s \geq 2mK$:
\begin{align*}
\sum_{\substack{B_0=(m;s;*)\in \mathfrak{B}\\1\leq m\leq n,s\geq 2mK}}s\Pr_{\ell \leftarrow \pi_{\calT}, R \sim \Rdistribution} \left[B(\ell,R) =B_0\right]
&\leq \sum_{\substack{B_0=(m;s;*)\in \mathfrak{B}\\1\leq m\leq n,s\geq 2mK}}s\cdot 2^{mK-s}\\
&\leq \sum_{s=2K}^{+\infty}s\cdot 2^{-s/2}\sum_{m=1}^{\lfloor s/2K\rfloor}2^{-K(s/2K-m)}\sum_{B_0=(m;s;*)\in \mathfrak{B}}1\\
&\leq \sum_{s=2K}^{+\infty}s\cdot 2^{-s/2}\cdot 2\cdot (2n)^{2s/K} \leq \sum_{s=2K}^{+\infty}(2n)^{1+2s/K}\cdot 2^{-s/2}.
\end{align*}

We also have the bound for the case $s < 2mK$:
\begin{align*}
\sum_{\substack{B_0=(m;s;*)\in \mathfrak{B}\\1\leq m\leq n, s< 2mK}}s\Pr_{\ell \leftarrow \pi_{\calT}, R \sim \Rdistribution} \left[B(\ell,R) =B_0\right]
&\leq 2K \sum_{{\substack{B_0=(m;s;*)\in \mathfrak{B}\\m, s}}}m\E_{\ell \leftarrow \pi_{\calT}} \left[ \Pr_{R \sim \Rdistribution} \left[B(\ell,R) =B_0\right]\right]\\
&= 2K + 2K\E_{\ell \leftarrow \pi_{\calT}} \left[ \E_{R \sim \Rdistribution} \left[m(\ell, R) - 1 \right]\right],
\end{align*}
where the last equality splits $m$ into $(m - 1) + 1$, and uses the identity $1 = \sum\limits_{{\substack{B_0=(m;s;*)\in \mathfrak{B}\\m, s}}} \mathbf{1}_{B(\ell,R) = B_0}$.

Finally, by \cref{lemma3.8}, we have 
$$
\nonadaptcost:= \E_{\ell \leftarrow \pi_{\calT}} \left[ \E_{R \sim \Rdistribution} \left[\fspecial(\textnormal{elt}(R\cap P_\ell))\right] \right] \geq \E_{\ell \leftarrow \pi_{\calT}} \left[ \E_{R \sim \Rdistribution} \left[m(\ell, R) - 1 \right]\right]. 
$$

Moreover, for $K = \lceil 25\log n \rceil + 25$, when $s \geq 2K$, we have:
$$
(1 + \log n)(1 + 2s/K) - s/2 \leq 1 + \log n + \frac{2s}{25} - \frac{s}{2} \leq \frac{s}{50} + \frac{2s}{25} - \frac{s}{2} \leq -\frac{s}{4}.
$$

Combining all the expressions, we obtain that for $K = \lceil 25\log n \rceil + 25$,
\begin{align*}
\adaptcost &= \E_{\ell \leftarrow \pi_{\calT}}\left[|A'_\ell|\right] = \sum_{\substack{B_0=(m;s;*)\in \mathfrak{B}\\1\leq m\leq n,s\geq 2mK}}s\Pr_{\ell \leftarrow \pi_{\calT}, R \sim \Rdistribution} \left[B(\ell,R) =B_0\right] + \sum_{\substack{B_0=(m;s;*)\in \mathfrak{B}\\1\leq m\leq n, s< 2mK}}s\Pr_{\ell \leftarrow \pi_{\calT}, R \sim \Rdistribution} \left[B(\ell,R) =B_0\right] \\
&\leq \sum_{s=2K}^{+\infty}(2n)^{1+2s/K}\cdot 2^{-s/2} + 2K + 2K\E_{\ell \leftarrow \pi_{\calT}} \left[ \E_{R \sim \Rdistribution} \left[m(\ell,R) - 1 \right]\right] \\
&= \sum_{s=2K}^{+\infty}2^{(1 + \log n)(1 + 2s/K)-s/2} + 2K + 2K\cdot \nonadaptcost \leq \sum_{s=2K}^{+\infty}2^{-s/4} + 2K(1 + \nonadaptcost) \leq 1 + 2K(1 + \nonadaptcost).
\end{align*}

Assuming that $\adaptcost = \E_{\ell \leftarrow \pi_{\calT}}\left[|A'_\ell|\right] \geq 200\log n$, we conclude that
$$
\nonadaptcost \geq \frac{1}{2K}(\adaptcost - 2K - 1) \geq \frac{\adaptcost - 1}{90\log n} - 1 \geq \frac{\adaptcost}{200\log n},
$$
as desired, provided that $n \geq 100$.
\end{proof}
\section{Tight Bounds for Binary-XOS}
\label{Section 4}
In this section, we improve upon the result in \cref{section 3}. In light of \cref{lm:reduce}, we only need to consider the special norm $\fspecial$. Specifically, we define a subset $A_\ell'\subseteq A_\ell$ for each leaf $\ell$. As usual, we define a new norm with respect to $A_\ell'$ as 
 $
\fspecial(S) = \max_{\ell} \left\{ | \elt(A_\ell') \cap S | \right\}$. Our improvement to \cref{thm:sec3main} is summarized in the following theorem:

\begin{theorem}\label{thm2}
Let $A_\ell'$ be a subset of $A_\ell$ for every leaf $\ell\in \calT$, and define
$
K=50\lceil \frac{\log n}{\log\log n}\rceil+50.
$
Assume that
$
\E_{\ell \leftarrow \pi_{\mathcal{T}}}[|A'_\ell|]\geq 6K.
$
Then, when $n$ is sufficiently large,
$$
\E_{\ell \leftarrow \pi_{\calT}} \left[ \E_{R \sim \Rdistribution} \left[\fspecial(\elt(R\cap P_\ell))\right] \right]\geq \frac{1}{6K}\E_{\ell \leftarrow \pi_{\calT}}\left[|A_\ell'|\right].
$$
\end{theorem}

By combining \cref{lm:reduce} and setting $\gapfunction(n) = 6K = O\left(\frac{\log n}{\log\log n}\right)$, we obtain the desired bound:

\begin{theorem}
\label{thm:strong}
For $2\leq r\leq n$, the adaptivity gap of {\em Bernoulli stochastic probing with XOS objective} under any prefix-closed $\calF$ with $\max\limits_{F\in \calF} |F|\leq r$ is
$
\gapfunction_{\XOS}(n,r) = O\left(\log r\,\frac{\log n}{\log\log n}\right).
$
\end{theorem}

Using the same lemma, we obtain an upper bound of $O\left(\frac{\log n}{\log\log n}\right)$ for the \XOSS objective. On the other hand, the lower bound construction in \cite{GNS17} demonstrates that there exists an instance of Bernoulli stochastic probing with \XOSS objective for which the adaptivity gap is $\Omega\left(\frac{\log n}{\log\log n}\right)$. This example shows that our adaptivity gap upper bound for the \XOSS objective is \emph{tight} up to a constant factor. Therefore, we conclude the following:

\begin{theorem}
\label{thm:tight}
The adaptivity gap of {\em Bernoulli stochastic probing with \XOSS objective} is $\Theta\left(\frac{\log n}{\log\log n}\right)$.
\end{theorem}

The proof of \cref{thm2} relies heavily on the results in \cref{sec3:prepare,sec3:main}, using essentially the same notation and algorithms. However, the analysis is significantly more technical and intricate.

\subsection{Technical Inequalities}\label{subsec:ineq}
In this section, we introduce a technical function $g(h,p)$ defined as
\[
g(h, p) = p \cdot \exp(-0.1\, h\, p^{1/h}),
\]
where $h \in \mathbb{N}_+$ and $p \in [0,1]$. For completeness, we define $g(0,p) = p$ for $p \in [0,1]$. This is a natural extension since
\(
\lim_{h \to 0^+} g(h,p) = p \cdot \exp(-0.1 \cdot 0) = p.
\)
We now list some useful properties of $g$. Fix $h \in \mathbb{N}_+$. Since $0.1p^{1/h} \leq 0.1 < 1$ for $p \leq 1$, we have:
\begin{equation}\label{eq2}
\frac{\partial g}{\partial p}(h,p) = \exp(-0.1hp^{1/h})(1 - 0.1p^{1/h}) > 0.
\end{equation}
Furthermore,
\begin{equation}\label{eq5}
\frac{\partial^2 g}{\partial p^2}(h,p) = -0.1\, p^{1/h - 1} \exp(-0.1hp^{1/h})\left(1 + \frac{1}{h} - 0.1p^{1/h}\right) \leq 0.
\end{equation}
These equations show that, for fixed $h$, the function $g$ is increasing and concave on $[0,1]$.

The following result is analogous to \cref{liu:induction}, but with a more careful treatment when the proportion of $S$ is given by $p$. In fact, it is a stronger result (up to a constant in $h$ when setting $x_\ell = 1$), since
\[
g(h,p) = p \exp(-0.1h\, p^{1/h}) = \exp\left(h\left(\log p^{1/h} - 0.1p^{1/h}\right)\right) \leq \exp(-0.1h),
\]
where the inequality uses $\log x \leq 0.1(x-1)$ for $x = p^{1/h} \in (0,1]$. The looseness of \cref{liu:induction} for $p < 2^{-h}$ is evident, as the left-hand side is trivially bounded by $p$. The tightness of \cref{lih:induction} when $p = n^{-O(m)}$ is what allows us to strengthen the main result from \cref{thm:sec3main} to \cref{thm2}.

\begin{lemma}
\label{lih:induction}
Let $S$ be a subset of leaves in the tree $\calT$. Suppose there exists a subset $T$ of nodes in $\calT$, not containing any leaves, and an integer $h \in \mathbb{N}$ such that for every $\ell \in S$, $|A_\ell \cap T| \geq h$. Furthermore, assume that each $\ell \in S$ is assigned a weight $x_\ell \in [0,1]$ such that
\(
\E_{\ell \leftarrow \pi_{\calT}}[x_\ell \mathbf{1}_{\ell \in S}] = p.
\)
Then,
\[
\E_{\ell \leftarrow \pi_{\calT}} \left[x_\ell \mathbf{1}_{\ell \in S} \cdot \Pr_{R \sim \Rdistribution} \left[R \cap P_\ell \cap T = \emptyset\right]\right] \leq g(h,p).
\]
\end{lemma}

The proof follows the same strategy as in \cref{liu:induction}: we analyze separately the cases where the root $r_0\in T$ and $r_0\notin T$. For the former, we use a corresponding technical inequality (see \cref{lemma4.1}); for the latter, the concavity in \cref{eq5} suffices. The complete proof is deferred to \cref{lm:inequality}.

\subsection{Proof for \cref{thm2}}
We now present the improved proof of \cref{thm:sec3main}. Although this proof employs the same \cref{lemma3.4}, \cref{algo2}, and \cref{lemma3.8} as before, we refine the construction of the set $T$, choose a different parameter $K$, and perform a more detailed analysis using conditional probability, which improves the adaptivity gap by a $\log\log n$ factor.

To begin, we introduce a refined set $T'$ for a fixed label $B_0$, which differs from the one defined in the discussion before \cref{defT}. 

Fix a label $B_0=(m;s;d_i;e_i;y_j)\in \mathfrak{B}$. We consider all intervals $I_j(B_0) = [y_j, y_{j-1})$, where $1\leq j\leq \lfloor \frac{s}{K}\rfloor$ and additionally $y_0 := d_1$.
Let $J(B_0)$ be the set of all $j$ such that $d_i \in I_j(B_0)$ for some $2 \leq i \leq m$.
We define $I(B_0) = \bigcup_{j\notin J(B_0), 1 \leq j \leq \lfloor \frac{s}{K} \rfloor} I_j(B_0)$. Therefore, $d_i\notin I(B_0)$, for any $1\leq i\le m$.

Recall the definition of $\imp(\l,B_0)$ in \cref{defH}, $S(B_0) = \{\ell \text{ is a leaf of } \calT: \exists R, B(\l,R) = B_0\}$ in \cref{defgen}, and $T(B_0)$ as the union of all $\imp(\l,B_0)$ for $\l \in S(B_0)$. For each label $B_0 \in \mathfrak{B}$, we define $T^\prime(B_0)$ to be the set of nodes in $T(B_0)$ whose depth lies in $I(B_0)$. Formally,
$$
T^\prime(B_0) = \bigcup_{\l \in S(B_0)} \{u : u \in \imp(\l,B_0), \depth{u} \in I(B_0) \}.
$$
Clearly, $T^\prime(B_0)$ is a subset of $T(B_0)$, and every node in it has depth in $I(B_0)$. Thus, we obtain:

\begin{lemma}\label{defTprime}
Fix a label $B_0 \in \mathfrak{B}$. Define $T^\prime(B_0)$ as above. Then $T^\prime(B_0)$ does not contain any leaf of $\calT$, and for every $\l \in S(B_0)$, there are at least $(s - mK)$ active ancestors in $T^\prime(B_0)$.
\end{lemma}

\begin{proof}
Since $T'(B_0) \subseteq T(B_0)$ and by \cref{mainlem}, $T'(B_0)$ contains no leaves. Furthermore, by \cref{remark3.7}, for any $\l \in S(B_0)$, the nodes in $\imp(\l,B_0)$ whose depths lie in $I(B_0)$ contribute at least $(s - mK)$ active ancestors.
\end{proof}

We now recall some consequences of the assumption that $\l \in S(B_0)$. From \cref{reduction:unique} (a), $\l \in \level_1(B_0)$ implies $s = |A_\l^\prime|$, and $y_j$ is the depth of the $jK$-th node of $A_\l^\prime$ (counted from $\l$). 

By part (b) of the same lemma, for $2 \leq a \leq m$, there exists a unique node $v_a = v_a(\l,B_0)$ of depth $d_a$, representing $e_a$, in $P_\ell \cap \level_a(B_0)$ (see also \cref{defgen}). Moreover, \cref{reduction:unique} (c) states that if $B(\l,R) = B_0$, then $v_a(\l,B_0)$ must be the `lowest' active (i.e., in $R$) node in $P_\l$ and satisfy the condition in \cref{cond} of \cref{algo2}; that is, there exists $B_1 \in \mathfrak{B}$ of length $a$ such that $\mathsf{Pref}_{a-1}(B_1) = \mathsf{Pref}_{a-1}(B_0)$ and $u \in \level_a(B_1)$ (here `lowest' refers to the descendant-ancestor order in the tree and should not be confused with the notion of `depth').

We now define an event on $R$, denoted $D(\l,B_0)$, for each label $B_0 \in \mathfrak{B}$ and each leaf $\l \in S(B_0)$ as follows. The event $D(\l,B_0)$ occurs if:
\begin{itemize}
\item For every $2 \leq a \leq m$, $v_a(\l,B_0) \in R$.
\item For every node $u \in P_\l \cap (T(B_0) \setminus T^\prime(B_0))$, it holds that $u \notin R$.
\end{itemize}

The following lemma summarizes key properties of the event $D(\l,B_0)$ for fixed $\l$ and $B_0$. In particular, part (c) asserts that if we fix $\l$ and group the labels with respect to certain technical conditions, then any outcome $R$ cannot satisfy two distinct $D$ events in the same group simultaneously. This reflects the fact that $v_a(\l,B_0)$ is the unique `lowest' node meeting the condition in \cref{cond} of \cref{algo2}, as mentioned above.

\begin{lemma}\label{event}
Fix a label $B_0$ and a leaf $\l\in S(B_0)$. Define the event $D(\l,B_0)$ as above. Then:
\begin{enumerate}[(a)]
\item The following event inclusion holds:
\[
[B(R\cap P_\l)=B_0] \subseteq D(\l,B_0) \cap [R \cap P_\l \cap T^\prime(B_0) = \emptyset].
\]
\item The event $D(\l,B_0)$ is independent of the event $[R\cap P_\l \cap T^\prime(B_0)=\emptyset]$.
\item Let $|A_\l'| = s$, and fix a subset $J\subseteq [\lfloor \frac{s}{K} \rfloor]$. Let $\mathfrak{B}^\prime$ denote the set of all labels $B^\prime$ such that $J(B') = J$ and $\l \in S(B')$. Then,
\[
\sum_{B^\prime \in \mathfrak{B}^\prime} \Pr[D(\l,B^\prime)] \leq 1.
\]
\end{enumerate}
\end{lemma}

\begin{proof}
Since we assume $\l\in S(B_0)$, the nodes $v_a(\l,B_0)$ (for $2\leq a\leq m$, where $m$ is the length of $B_0$) are uniquely determined by $B_0$. We work within a single path $P_\l$, so any two nodes in $P_\l$ are in a descendant-ancestor relation, and two nodes represent the same element if and only if they are identical. For clarity, we abbreviate $v_a(\l,B_0)$ to $v_a$, though these nodes may differ with respect to different labels.

\begin{enumerate}[(a)]
\item This inclusion follows directly from part (b) of \cref{reduction:unique} and from \cref{defT}:
\[
[B(R\cap P_\l)=B_0] \subseteq \left([R \cap P_\l \cap T(B_0) = \emptyset] \wedge \bigwedge_{2\leq a\leq m}[v_a \in R] \right)
= D(\l,B_0) \cap [R \cap P_\l \cap T^\prime(B_0) = \emptyset].
\]

\item It is clear that $D(\l,B_0)$ or $[R \cap P_\l \cap T^\prime(B_0)=\emptyset]$ are just conditions on certain nodes in $P_\l$ being in $R$ or not. The nodes involved in $D(\l,B_0)$ are either $v_a(\l,B_0)$, which is of depth $d_a\notin I(B_0)$, or in $T(B_0) \setminus T^\prime(B_0)$, hence disjoint from $T'(B_0)$. Therefore, nodes involved in $D(\l,B_0)$ and $[R \cap P_\l \cap T^\prime(B_0)=\emptyset]$ are disjoint. Since different nodes in $P_\l$ represents distinct elements, the two events are independent.

\item We show that the events $D(\l,B^\prime)$ for all $B^\prime \in \mathfrak{B}^\prime$ are mutually disjoint—that is, any outcome $R$ can satisfy at most one such event.

Suppose, for contradiction, that there exist two labels $B^1, B^2 \in \mathfrak{B}^\prime$ such that $R \in D(\l,B^1) \cap D(\l,B^2)$. From \cref{reduction:unique}(a), $\l \in S(B^1) \cap S(B^2)$ implies $\l \in \level_1(B^1) \cap \level_1(B^2)$. Thus, in both labels, $s = |A_\l^\prime|$, and $y_j$ is the depth of the $jK$-th node of $A_\l^\prime$ (counted from $\l$). We write $B^1 = (m^1; s; d_i^1; e_i^1; y_j)$ and $B^2 = (m^2; s; d_i^2; e_i^2; y_j)$.

Let $a\geq 2$ be the first index such that $e_a^1 \neq e_a^2$ (note that $e_1^1 = e_1^2 = 0$, and we allow $a = \min\{m^1,m^2\} + 1$). By \cref{defgen} and the assumption that $\l \in S(B^1)\cap S(B^2)$, assume without loss of generality that the corresponding node $v_a^1 := v_a^1(\l)\in \level_a(B^1)$ is an ancestor of $v_a^2 := v_a^2(\l)\in \level_a(B^2)$, and $v_a^1 \neq v_a^2$ (if either node does not exist due to $a= \min\{m^1,m^2\} + 1$, we define the non-existent one to be $v_a^1$, a virtual node).

Since $e_i^1 = e_i^2$ for all $2 \leq i < a$, it follows that $v_i^1 = v_i^2$, and hence $d_i^1 = d_i^2$. Consequently, $\mathsf{Pref}_{a-1}(B^1) = \mathsf{Pref}_{a-1}(B^2)$. Furthermore, since $B^1,B^2 \in \mathfrak{B}^\prime$, we have $J(B^1) = J(B^2) = J$, so $I(B^1) = I(B^2)$, as all $y_j$ values are equal.

Because $d_a^2 \notin I(B^2) = I(B^1)$, the node $v_a^2$ at depth $d_a^2$ is not contained in $T^\prime(B^1)$. On the other hand, the label $\mathsf{Pref}_a(B^2)$ (of length $a$) satisfies $\mathsf{Pref}_{a-1}(\mathsf{Pref}_a(B^2)) = \mathsf{Pref}_{a-1}(B^1)$ and $v_a^2 \in P_\l \cap \level_a(B^2)$. Then $v_a^2$ satisfies all conditions in the $(a-1)$-th union in \cref{defH} for $B^1$, since $v_a^2$ is a descendant of $v_a^1$ (this still holds even when $a = \min\{m^1,m^2\} + 1$).

Thus, $v_a^2 \in \imp(\l, B^1) \subseteq T(B^1)$. Since $v_a^2 \notin T^\prime(B^1)$, it follows that $v_a^2 \in P_\l \cap (T(B^1) \setminus T^\prime(B^1))$, and therefore $v_a^2 \notin R$ because $D(\l,B^1)$ occurs. But this contradicts the assumption that $D(\l,B^2)$ occurs, which would imply $v_a^2 \in R$.

\end{enumerate}
\end{proof}

We now proceed with the proof of the main theorem \cref{thm2}. Intuitively, in \cref{liu:induction} we "consume" the condition $R\cap P_\l \cap T(B_0)=\emptyset$, yet the set $T(B_0)\cap A_\l$ can be much larger than $(s-mK)$. In this proof, we only "consume" the condition $R\cap P_\l\cap T'(B_0) = \emptyset$ and instead employ \cref{lih:induction} in place of \cref{liu:induction}. By leveraging the concavity and monotonicity of $g(h,\cdot)$, we can combine conditional probabilities to obtain the desired bound.

\begin{proof}[Proof of \textnormal{\cref{thm2}}]
First, using \cref{lemma3.8}, we can express
$$
\adaptcost:=\E_{\ell \leftarrow \pi_{\calT}}\left[|A'_\ell|\right] = \sum_{B_0=(m;s;*)\in \mathfrak{B}}s\cdot \E_{\ell \leftarrow \pi_{\calT}} \left[ \Pr_{R \sim \Rdistribution} \left[B(\l, R) = B_0 \right]\right],
$$
where $(m;s;*)$ denotes a sequence of the form $(m;s;d_i;e_i;y_j)$.

We split the sum over $B_0=(m;s;*)\in \mathfrak{B}$ into two cases: $s\ge 2mK$ and $s<2mK$. We bound the latter case using \cref{lemma3.8}:
\begin{align}
\sum_{\substack{B_0=(m;s;*)\in \mathfrak{B}\\1\leq m\leq n, s< 2mK}}s\Pr_{\ell \leftarrow \pi_{\calT}, R \sim \Rdistribution} \left[B(\l, R) = B_0 \right]
&\leq 2K \sum_{B_0=(m;s;*)\in \mathfrak{B}}m\E_{\ell \leftarrow \pi_{\calT}} \left[ \Pr_{R \sim \Rdistribution} \left[B(\l, R) = B_0 \right]\right] \nonumber\\
&= 2K+2K\E_{\ell \leftarrow \pi_{\calT}} \left[ \E_{R \sim \Rdistribution} \left[m(\l, R) -1 \right]\right] \nonumber\\
&\leq 2K+ 2K\E_{\ell \leftarrow \pi_{\calT}} \left[ \E_{R \sim \Rdistribution} \left[\fspecial(\elt(R\cap P_\ell))\right] \right]. \label{eq10}
\end{align}

We denote the target quantity
\(
\nonadaptcost := \E_{\ell \leftarrow \pi_{\calT}} \left[ \E_{R \sim \Rdistribution} \left[\fspecial(\elt(R\cap P_\ell))\right] \right].
\)

Now, we aim to bound the remaining term where $s \geq 2mK$.

For each fixed label $B_0 = (m;s;*)$ with $s \geq 2mK$, we define $T^\prime(B_0)$ and $S(B_0)$ as in the discussion preceding \cref{defTprime}. Moreover, for each $\l \in S(B_0)$, we define $x_\l \in [0,1]$ to be the probability $\Pr_{R\sim \Rdistribution}[D(\l,B_0)]$ that the event $D(\l,B_0)$ occurs.

From \cref{event} (a) and the definition of $S(B_0)$, the inclusion of events hold:
\[
[B(R\cap P_\l)=B_0] \subseteq D(\l,B_0)\cap [R\cap P_\l \cap T^\prime = \emptyset] \cap [\l \in S(B_0)].
\]
Taking expectations, we obtain:
\[
\E_{\ell \leftarrow \pi_{\calT}}\left[ \Pr_{R \sim \Rdistribution} \left[B(\l, R) = B_0 \right]\right] \leq \E_{\ell \leftarrow \pi_{\calT}}\left[ \mathbf 1_{\l\in S(B_0)} \cdot \Pr_{R \sim \Rdistribution} \left[D(\l,B_0)\cap [R\cap P_\l \cap T^\prime = \emptyset] \right]\right].
\]
Furthermore, from \cref{event} (b), for fixed $B_0$ and $\l \in S(B_0)$, the event $D(\l,B_0)$ is independent of $[R\cap P_\l \cap T^\prime = \emptyset]$. Therefore,
\[
\E_{\ell \leftarrow \pi_{\calT}}\left[ \Pr_{R \sim \Rdistribution} \left[B(\l, R) = B_0 \right]\right] 
= 
\E_{\ell \leftarrow \pi_{\calT}}\left[ x_\l \cdot \mathbf 1_{\l\in S(B_0)} \cdot \Pr_{R \sim \Rdistribution} \left[ R\cap P_\l \cap T^\prime = \emptyset \right]\right].
\]
This is exactly the form required by Lemma~\ref{lih:induction}. Hence, using \cref{defTprime}, we denote
\(
p_{B_0} := \E_{\ell \leftarrow \pi_{\calT}}\left[ x_\l \cdot \mathbf 1_{\l\in S(B_0)}\right]
\)
with $h := K\floor{\frac{s}{2K}} \leq s - mK$ (since $s \geq 2mK$), taking $T = T'(B_0)$, $S = S(B_0)$, we apply Lemma~\ref{lih:induction} to get:
\[
\E_{\ell \leftarrow \pi_{\calT}}\left[ \Pr_{R \sim \Rdistribution} \left[B(\l, R) = B_0 \right]\right]
\leq g\left(K\floor{\frac{s}{2K}},p_{B_0}\right).
\]
Let $s_0 := \floor{\frac{s}{2K}}$. Now, we use the fact that the function $g(h,p)$ is concave for fixed $h\geq 1$ and $p\in [0,1]$, as shown via Jensen's inequality (see \cref{eq5}). We then sum over all labels with a fixed $s_0$ (i.e., we sum over all $s$ such that $2Ks_0 \leq s < 2K(s_0+1)$ and $m$ satisfying $1 \leq m \leq s_0$), and define
$$
q_{s_0} = \sum_{s=2Ks_0}^{2K(s_0+1)-1}\sum\limits_{\substack{B_0 = (m;s;*)\in \mathfrak{B}\\ 1\leq m\leq s_0}}[p_{B_0}] 
= \sum_{s=2Ks_0}^{2K(s_0+1)-1}\sum\limits_{\substack{B_0 = (m;s;*)\in \mathfrak{B}\\ 1\leq m\leq s_0}} \E_{\ell \leftarrow \pi_{\calT}}[x_\l \mathbf 1_{\l\in S(B_0)}].
$$

Furthermore, we denote by $r_{s_0}$ the number of labels satisfying $B_0 = (m;s;*)\in \mathfrak{B}$ with $1\leq m\leq s_0$ and $2Ks_0\leq s<2K(s_0+1)$. We can bound $r_{s_0}$ as follows.

There are at most $s_0$ possibilities for $m$ and $2K$ possibilities for $s$. For each $m, s$, there are at most $(n+1)^m \times n^m \times (n+1)^{s/K} \leq (2n)^{2s/K}$ possibilities for $d_i, e_i, y_j$. Therefore,
$$
r_s\leq 2Ks_0(2n)^{2s/K}\leq (2n)^{2s/K+1}\leq (2n)^{4s_0+5}.
$$

Applying Jensen’s inequality, we obtain:
\begin{equation}\label{eq7}
\sum_{s=2Ks_0}^{2K(s_0+1)-1}\sum_{\substack{B_0 = (m;s;*)\in \mathfrak{B}\\1\leq m\leq s_0}}\E_{\ell \leftarrow \pi_{\calT}} \left[ \Pr_{R \sim \Rdistribution} \left[B(\l, R) = B_0 \right]\right] 
\leq
\sum_{s=2Ks_0}^{2K(s_0+1)-1}\sum_{\substack{B_0 = (m;s;*)\in \mathfrak{B}\\1\leq m\leq s_0}} g(s_0K,p_{B_0}) \leq r_{s_0} g\left(s_0K,\frac{q_{s_0}}{r_{s_0}}\right).
\end{equation}

Note that the quantity $r_{s_0} g\left(s_0K,\frac{q_{s_0}}{r_{s_0}}\right) = q_{s_0}\exp\left(-0.1s_0K\left(\frac{q_{s_0}}{r_{s_0}}\right)^{1/(s_0K)}\right)$ is an increasing function of $r_{s_0}$ for fixed $q_{s_0}$. Hence,
\begin{equation}\label{eq11}
r_{s_0} g\left(s_0K,\frac{q_{s_0}}{r_{s_0}}\right)\leq (2n)^{4s_0+5} g\left(s_0K,\frac{q_{s_0}}{(2n)^{4s_0+5}}\right) =  q_{s_0} \exp\left(-0.1s_0K\left(\frac{q_{s_0}}{(2n)^{4s_0+5}}\right)^{\frac{1}{s_0K}}\right).
\end{equation}

To bound $q_{s_0}$, we expand its definition:
\begin{equation*}
q_{s_0} = \sum\limits_{\substack{B_0 = (m;s;*)\in \mathfrak{B}\\ 1\leq m\leq s_0\\2s_0K\leq s<2(s_0+1)K}} \E_{\ell \leftarrow \pi_{\calT}}[\mathbf 1_{\l\in S(B_0)}\Pr_{R\sim \Rdistribution}[D(\l,B_0)]] 
\leq \E_{\ell \leftarrow \pi_{\calT}} \sum\limits_{\substack{B_0 = (*;s;*)\in \mathfrak{B}\\2s_0K\leq s<2(s_0+1)K}}[\mathbf 1_{\l\in S(B_0)}\Pr_{R\sim \Rdistribution}[D(\l,B_0)]],
\end{equation*}
where the last inequality follows by exchanging summation and relaxing the range of $m$.

For fixed $\ell$, note that $\l\in S(B_0)$ implies $s = |A_\l'|$ by \cref{reduction:unique} (a). Thus, if $\l$ contributes to the above sum, then $2s_0K\leq |A_\l'| <2(s_0+1)K$.

Now we group all labels $B_0 = (*; s ;*)$ and $\l \in S(B_0)$ with respect to $J(B_0)\subseteq \left[\lfloor \frac{s}{K} \rfloor\right]$. For a fixed group, \cref{event} (c) ensures that the sum of $\Pr[D(\l,B_0)]$ over all such labels $B_0$ is at most $1$.

Therefore,
\begin{align*}
q_{s_0} &=\E_{\ell \leftarrow \pi_{\calT}} \sum\limits_{\substack{B_0 = (m;s;*)\in \mathfrak{B}\\2s_0K\leq s<2(s_0+1)K}}[\mathbf 1_{\l\in S(B_0)}\Pr_{R\sim \Rdistribution}[D(\l,B_0)]]\\
&\leq \max\limits_{\substack{\l: |A_\l'|=s\\ 2s_0K\leq s<2(s_0+1)K}}\sum_{J\subseteq \left[\lfloor \frac{s}{K} \rfloor\right]} \sum\limits_{B_0:J(B_0) = J,\l\in S(B_0)} \Pr[D(\l,B_0)]\\
&\leq \max\limits_{2s_0K\leq s<2(s_0+1)K}\sum_{J\subseteq \left[\lfloor \frac{s}{K} \rfloor\right]} 1 \leq 2^{2s_0+1}.
\end{align*}

Combining \cref{eq7}, \cref{eq11}, the monotonicity of $g(s_0K, p)$ in $p$ (from \cref{lemma4.1}), and the bound $q_{s_0} \leq 2^{2s_0+1}$, we get (for $s_0\geq 1, n>100$):

\begin{align*}
\sum_{s=2Ks_0}^{2K(s_0+1)-1}\sum_{\substack{B_0 = (m;s;*)\in \mathfrak{B}\\1\leq m\leq s_0}}\E_{\ell \leftarrow \pi_{\calT}} \left[ \Pr_{R \sim \Rdistribution} \left[B(\l, R) = B_0 \right]\right] 
&\leq q_{s_0} \exp\left(-0.1s_0K\left(\frac{q_{s_0}}{(2n)^{4s_0+5}}\right)^\frac{1}{s_0K}\right)\\
&\leq  2^{2s_0+1}\exp\left(-0.1s_0K n^{-10/K}\right).
\end{align*}

Therefore, summing over all $s_0$ (when $n$ is large enough and $s\geq 2K$):
$$
\sum\limits_{\substack{B_0 = (m;s;*)\in \mathfrak{B}\\1\leq m\leq r, s\geq 2mK}}s\E_{\ell \leftarrow \pi_{\calT}} \left[ \Pr_{R \sim \Rdistribution} \left[B(\l, R)=B_0\right]\right]
\leq \sum_{s_0=1}^{+\infty} 2K(s_0+1)\cdot 2^{2s_0+1}\exp(-0.1s_0K n^{-10/K}).
$$

Since $K = 50 \lceil\frac{\log n}{\log\log n} \rceil+ 50$, we have $n^{10/K} =2^{\frac{10\log n}{K}}\leq 2^{\frac{\log\log n}{5}} = \log^{1/5} n \leq K/40$ for sufficiently large $n$. Hence,
\begin{align*}
\sum_{s_0=1}^{+\infty} 2K(s_0+1)\cdot 2^{2s_0+1}\exp(-0.1s_0K n^{-10/K})
&\leq \sum_{s_0=1}^{+\infty} 2K(s_0+1)\cdot 2^{2s_0+1}\exp(-4s_0)\\
&\leq 2K\sum_{s_0=1}^{+\infty} (s_0+1)\cdot 2^{2s_0+1-5s_0}\leq 2K.
\end{align*}

Combining this with \cref{eq10} and the partition strategy of $s\geq 2mK$ and $s<2mK$, we conclude:
$$
\adaptcost\leq 2K+2K\cdot \nonadaptcost+2K \Longleftrightarrow \nonadaptcost\geq \frac{\adaptcost}{2K}-2.
$$

Thus, when $\adaptcost = \E_{\ell \leftarrow \pi_{\calT}}\left[|A'_\ell|\right] \geq 6K$, it follows that $\nonadaptcost\geq \frac{\adaptcost}{2K}-2\geq \frac{\adaptcost}{6K}$. The result follows.
\end{proof}

\section{Adaptivity Gap for Symmetric Norm}
\label{section5}
We first introduce the Bernoulli setting for monotone symmetric norms.  

\begin{definition}[Bernoulli Stochastic Probing with Symmetric Norm Objective]
    \label{bernoulli:symmetric}
    In this special case of the Stochastic Probing problem, each random variable $X_i$ takes value $c_i \in \R_{\geq 0}$ with probability $p_i$, and $0$ with probability $1 - p_i$, for all $i\in U$. Also, the objective function $f$ is a \emph{monotone symmetric norm}. 
    Other settings are the same as Definition~\ref{stoc:probing}.
\end{definition}

It is important to emphasize that 
the definition here is slightly different from Definition~\ref{bernoulli:norm} (in which the Bernoulli random variable is 0/1).
In Definition~\ref{bernoulli:symmetric}, the random variable may take values $c_i \in \mathbb{R}_{\geq 0}$, which is crucial for our analysis. The reason that we do not absorb $c_i$ into the objective function $f$, as doing so would destroy the symmetry property of $f$.

Without loss of generality, we may assume that $p_i, c_i > 0$ for all $i$. 
We denote by $f(S)$ (for $S\subseteq U=[n]$) the value $f(\boldv)$, where the $n$-dimensional vector $\boldv$ has its $i$-th coordinate equal to $c_i$ if $i\in S$ and $0$ otherwise, as we did in \cref{bernoulli:norm}.

Let $\gapfunction_{\sym}(n)$ denote the supremum of the adaptivity gap for {\em Bernoulli stochastic probing with any symmetric norm objective} and any {\em prefix-closed} feasibility constraint $\calF$.
We show in this section that the adaptivity gap for this problem is bounded by a constant, namely, $\gapfunction_{\sym}(n) = O(1)$ (with an explicit upper bound of 2050).

In the remainder of this section, we adopt the binary decision tree model and use the notation for \textbf{active ancestors} $A_\l\subseteq P_\l$ and the random variable $R\sim \Rdistribution$ introduced in \cref{bernoulli:prel}. For simplicity, we write $f(A_\l)$ instead of $f(\elt(A_\l))$, as our analysis here avoids the complicated method of changing the maximizer leaf employed in \cref{section 3} and \cref{Section 4}.

In view of Lemma~\ref{large:case} (whose proof in \cref{AppendixB} applies to this setting as well), by setting $\lambda = 4^{-4}$ and possibly incurring a factor of 2 loss in the adaptivity gap, we may assume that all elements are $4^{-4}$-small. By scaling $f$ by a constant, we normalize the optimal reward (the optimal binary decision tree denoted by $\calT_0$) to 
$$
\mathsf{OPT} = \adaptcost(\calT_0,f) = \E_{\ell \leftarrow \pi_{\mathcal{T}}}\left[f(A_\ell)\right] = 1.
$$

Furthermore, we assume that $f(1,0,0,\cdots,0) = 1$. Indeed, if $f(1,0,0,\cdots,0)=\frac{1}{a}>0$, we can simply divide each $c_i$ by $a$ and define a new function $f_1(\boldv)=f(a\boldv)$ to replace $f$. This transformation preserves the $\lambda$-smallness, the adaptivity gap, and the value of $\mathsf{OPT}$. Consequently, $f(x,0,0,\cdots,0) = xf(1,0,0,\cdots,0) = x$, which we denote by $f(x)$. Now, for any internal node $u\in \calT_0$, we round each $c_{\elt(u)}=f(c_\elt(u)) \leq 4^{-4}$ to the largest negative power of $4$ strictly smaller than $c_\elt(u)$. After rounding, each $c_\elt(u)\leq 4^{-5}$. Using linearity of the norm objective, the final reward $\OPT_1$ is still at least $1/4$.

Next, we modify the optimal decision tree $\calT_0$ slightly into another decision tree $\calT$
with some additional requirements, 
as in the following lemma.
\begin{lemma}
There exists a decision tree $\calT$ satisfying the following requirements:
\begin{enumerate}
    \item Each root-leaf path is a subsequence of some root-leaf path in $\calT_0$. Thus, the corresponding element sequence satisfies the feasibility constraint $\calF$;
    \item For any leaf $\l \in \calT$, $f(A_\l)\leq 2$;
    \item $\E_{\l\leftarrow \pi_\calT}[f(A_\l)]\geq 1/8$.
\end{enumerate}
\end{lemma}
\begin{proof}
    In the original adaptive strategy, the algorithm starts at the root of $\calT_0$, follows the decision tree based on the probing outcomes, and eventually terminates at a leaf $\l$, obtaining the set $A_\l$. In our modified strategy, the algorithm still follows $\calT_0$, but before probing any node $u$, it checks whether $A_u$ (the set of active ancestors of $u$) satisfies $f(A_u) \geq 1$. If it does, the algorithm terminates at node $u$. Looking into the decision tree model, the new decision tree $\calT$ is obtained by replacing, for each node $u$ in $\calT_0$ where $f(A_u) \geq 1$ but $f(A_v) < 1$ ($v$ is the parent of $u$), the entire subtree $\calT_u$ rooted at $u$ with a single leaf $u$. Clearly, no termination signs in the new strategy have an ancestor-descendant relationship in $\calT_0$.

We now argue that the reward for the new decision tree is at least $ 1/8$. To see this, consider any execution of the new strategy terminating at a node $u$ in the original tree $\calT_0$. By the subsequence property (see the formal definition in \cref{subsequence}; the precise statement is not critical for understanding), the expected additional reward after $u$ is at most $1$; otherwise, a better adaptive strategy could be constructed by  first following the tree to $u$ and then applying the adaptive strategy at the subtree $\calT_u$. Formally, if $\pi_{\calT_u}$ denotes the original distribution of leaves $\pi_\calT$ conditioned on $\l \in\calT_u$, then
$$
\E_{\l\gets \pi_{\calT_u}} [f(A_\l\backslash A_u)] \leq \OPT_1 = 1.
$$
Hence, when $f(A_u) \geq 1$, by subadditivity we have
$$
\E_{\l\gets \pi_{\calT_u}} [f(A_\l)] \leq \E_{\l\gets \pi_{\calT_u}} [f(A_\l\backslash A_u)] + f(A_u) \leq 1 + f(A_u) \leq 2f(A_u).
$$
When $f(A_u) < 1$, then $u$ must be a leaf in the original tree, and we have
$$
\E_{\l\gets \pi_{\calT_u}} [f(A_\l)] = f(A_u) \leq 2f(A_u).
$$
Now, we take expectation with respect to the leaf distribution $u\gets \pi_\calT$ (note that now $u$ is a leaf in $\pi_\calT$), we obtain
$$
\OPT_1 = \E_{u\gets \pi_\calT} \E_{\l\gets \pi_{\calT_u}} [f(A_\l)] \leq 2\, \E_{u\gets \pi_\calT}[f(A_u)],
$$
hence $\E_{u\gets \pi_\calT}[f(A_u)] \geq \OPT_1/2\geq 1/8$.

Moreover, in the new tree $\calT$, we have $f(A_u) \leq 2$ for every leaf $u$. To see this, let $v$ be the parent of $u$. Then,
$$
f(A_u) \leq f(A_v) + f(u) \leq 1 + c_{\elt(u)} \leq 1 + 4^{-5} \leq 2,
$$
where we have used the fact that after rounding, each $c_{\elt(u)}\leq 4^{-5}$ for any internal node $u\in \calT \subseteq \calT_0$.
\end{proof}
By the above lemma, it suffices to lower bound the non-adaptive reward $\nonadaptcost(\calT,f)$ for the modified decision tree $\calT$, using the natural non-adaptive strategy  \nonadaptivestrategy discussed in \cref{bernoulli:prel}.
 
For each internal node $u \in \calT$, if $c_{\elt(u)}=4^{-k}$, we place $u$ into a set $Q_k$, and refer to $u$ as being in class $k$. Note that no leaf is included in any $Q_k$, as leaves do not represent any elements. Recall that after rounding, $c_{\elt(u)}\leq 4^{-5}$ for each internal node $u\in\calT$. Therefore, in the current tree, only classes $Q_k$ with $k\geq 5$ are nonempty.

For each leaf $\ell$, let $D_\ell \subseteq A_\l$ be the union of the sets $A_\l\cap Q_k$ for all $k\geq 5$ satisfying the condition $|A_\l\cap Q_k|\geq  2^k$. 
Thus, since each node is in exactly one $Q_k$, we have the following relation:
$$
D_\l = \bigcup_{k:|A_\l \cap Q_k|\geq 2^{k}} A_\l \cap Q_k, 
A_\ell\backslash D_\ell= \bigcup_{\substack{k\geq 5\\ |A_\l \cap Q_k| <  2^{k}}} A_\l \cap Q_k .
$$
Thus, the contribution from $A_\ell \backslash D_\ell$ to the optimal reward is small, leaving a remaining contribution that is at least a constant: 
\begin{align}\label{eq:sec5}
\E_{\ell\leftarrow\pi_{\calT}}[f(A_\ell\cap D_\ell)]&\geq \E_{\ell\leftarrow\pi_{\calT}}[f(A_\ell)-f(A_\ell\backslash D_\ell)]\nonumber\\
&= \E_{\ell\leftarrow\pi_{\calT}}[f(A_\ell)]-\E_{\ell\leftarrow\pi_{\calT}}[f(A_\ell\backslash D_\ell)]\nonumber\\
&\overset{(\star)}{\geq} \E_{\ell\leftarrow\pi_{\calT}}[f(A_\ell)]-\E_{\ell\leftarrow\pi_{\calT}}\left[\sum_{k=5}^{\infty}2^k f(4^{-k})\right]\nonumber\\
&= 1/8-1/16=1/16.
\end{align}
Note that the inequality marked $(\star)$ uses the fact that $f$ is subadditive and symmetric. This bound implies that for each $\l$, we need only consider the classes in which a sufficiently large number of nodes in $A_\l$ appear.

Now, we introduce a technical definition. 
\begin{definition}
For each $k\geq 5$, we call a leaf $\l$ a \textbf{$k$-bad leaf} if $|A_\l\cap Q_k|\geq   2^k$, but the following holds:
$$
\sum_{e\in \elt(P_\l \cap Q_k)} p_e   = \E_{R\sim \Rdistribution}[|R\cap P_\l\cap Q_k|] \leq \frac{1}{8}|A_\l\cap Q_k|.
$$
\end{definition}
Intuitively, a \textbf{$k$-bad leaf} is one for which the non-adaptive algorithm is unlikely to achieve a comparable reward from class $Q_k$. However, we are able to show that the proportion of \textbf{$k$-bad leaves} (with respect to the distribution $\pi_\calT$) is small. The following lemma, whose proof is analogous to those of \cref{liu:induction} and \cref{lih:induction} (by induction on the height of the tree), establishes this bound.

\begin{lemma}
Let $S$ be a subset of leaves in the tree $\calT$, and let $T \subseteq \calT$ be a subset of nodes that does not include any leaf. 
Suppose that for every $\ell \in S$, the number of active ancestors of $\ell$ in $T$ is at least $h$, i.e., 
$|A_\ell \cap T| \geq h$.  
Furthermore, assume there exists a real number $q \in \mathbb{R}$ such that for every $\ell \in S$, the following inequality holds:
$$
|A_\l \cap T|\geq 8\sum_{e\in \elt(P_\l \cap T)} p_e + q.
$$
Then, the following bound holds:
$$
\Pr_{\ell \leftarrow \pi_{\calT}}[\l \in S]\leq e^{-h-q}.
$$
\end{lemma}
\begin{proof}
We prove the claim by induction on the height of $\calT$.

When the height of $\calT$ is $1$, $h = 0, q\leq 0$, since $T$ does not contain any leaf. Hence, the base case is trivial. Assume now that the claim holds for trees of smaller height. For the current tree $\calT$, let $\calT_1$ and $\calT_2$ denote the subtrees corresponding to the $\textsf{Yes}$ and $\textsf{No}$ arcs from the root $r_0$, respectively. Let $\lambda$ be the probability that the root is active (if $r_0$ has only one child, we set $\lambda = 0$ or $1$ accordingly). Then, the path distribution $\pi_{\calT}$ assigns probability $\lambda$ to the $\textsf{Yes}$ subtree $\calT_1$ and $1-\lambda$ to the $\textsf{No}$ subtree $\calT_2$.
Conditioning on the $\textsf{Yes}$ subtree ($\textsf{No}$ subtree resp.), $\pi_{\calT}$ has the same distribution of leaves as in $\pi_{\calT_1}$
($\pi_{\calT_2}$ resp.). 
Denote
$
\Pr_{\ell \leftarrow \pi_{\calT_1}}[\ell \in S] = u \quad \text{and} \quad \Pr_{\ell \leftarrow \pi_{\calT_2}}[\ell \in S] = v.
$
Then, $\Pr_{\ell \leftarrow \pi_{\calT}}[\ell \in S] = \lambda u +(1-\lambda)v$.

We now consider two cases.
\begin{mycase}{1}
$r_0 \notin T$.
\end{mycase}
In this case, the two subtrees inherit essentially the same conditions $(h,q)$ as the original tree because $r_0\notin T$. Thus, by the induction hypothesis, $u,v\leq e^{-h-q}$, so that $\Pr_{\ell \leftarrow \pi_{\calT}}[\ell \in S] = \lambda u +(1-\lambda)v\leq e^{-h-q}$.
\begin{mycase}{2}
$r_0 \in T$.
\end{mycase}
In this case, for leaves $\ell$ in $S\cap \calT_1$, there are at least $h-1$ active ancestors in $T\cap \calT_1$. We denote $P_{\ell}^\prime = P_{\ell}\backslash\{r_0\}$ and $A_{\ell}'  = A_\ell \backslash\{r_0\}$ for $\ell\in S\cap \calT_1$. Then, for $\ell\in S\cap \calT_1$,
$$
|A_\ell' \cap T|\geq 8 \sum_{e\in \elt(P_\ell' \cap T)} p_e + 8\lambda -1+ q.
$$
Applying the induction hypothesis on $\calT_1$, we deduce $u\leq e^{-h+2-8\lambda-q}$. Similarly, for the \textsf{No} subtree, every leaf has at least $h$ active ancestors in $T\cap \calT_2$, and
$$
|A_\ell' \cap T|\geq 8 \sum_{e\in \elt(P_\ell' \cap T)} p_e + 8\lambda + q.
$$
By the induction hypothesis on $\calT_2$, we obtain $v\leq e^{-h-8\lambda-q}$. Combining these estimates and applying \cref{fact2} to the inequality marked $(\star)$, we have
$$
\Pr_{\ell \leftarrow \pi_{\calT}}[\ell \in S] = \lambda u +(1-\lambda)v \leq e^{-h-q-8\lambda} \cdot(\lambda e^{2}+(1-\lambda)) \leq e^{-h-q-8\lambda} \cdot e^{(e^2-1)\lambda} \overset{(\star)}{\leq} e^{-h-q},
$$
which completes the proof.
\end{proof}

Thus, for each $k\geq 5$, if we take the set of \textbf{$k$-bad leaves} as $S$ and the nodes in class $k$ as $T=Q_k$, then by choosing $h = 2^k$ and $q = 0$, the proportion of \textbf{$k$-bad leaves} is at most $e^{-2^k}$. For each $\ell$, let
$$
G_\ell = \bigcup_{\ell \text{ is a } k\textbf{-bad leaf}} (A_\ell \cap Q_k).
$$
Note that for any leaf $\ell$, since $f$ is monotone, $f(G_\ell)\leq f(A_\ell)\leq 2$. Therefore,
\begin{equation}\label{eq:bad}
    \E_{\ell \leftarrow \pi_{\calT}}[f(G_\ell)]\leq 2\Pr_{\ell \leftarrow \pi_{\calT}}[G_\ell \neq \emptyset] 
\leq 2\sum_{k=5}^{+\infty}\Pr_{\ell \leftarrow \pi_{\calT}}[\ell \text{ is a } k\textbf{-bad leaf}] \leq2\sum_{k=5}^{+\infty}2^{-2^{k}}\leq  2\sum_{k=5}^{+\infty}2^{-27-k}\leq 2^{-30}.
\end{equation}

We now state the following lemma, which shows that if $\ell$ is not a \textbf{$k$-bad leaf}, then we can always obtain comparable non-adaptive reward.
\begin{lemma}\label{sec5lem:chernoff}
For any fixed leaf $\ell$, we have 
$$\E_{R\sim \Rdistribution}[f(R\cap P_\ell\cap D_\ell)]\geq \frac{1}{64}f((A_\ell\cap D_\ell)\backslash G_\ell).$$
\end{lemma}
\begin{proof}
For each $k\geq 5$ such that $b_{\ell,k} := |A_\ell\cap Q_k|\geq 2^k$ and $\ell$ is not a \textbf{$k$-bad leaf}, we include such $k$ in a set $K_{\ell} \subseteq \mathbb{N}$. For each $k\in K_\ell$, consider the probability over $R$ that $|R\cap P_\ell\cap Q_k|$ is at least $b_{\ell,k}/32$. 
Observe that $|R\cap P_\ell\cap Q_k|$ is the sum of $t_{\ell,k}=|P_\ell\cap Q_k|\geq b_{\ell,k}$ Bernoulli random variables $Y_1,\cdots,Y_{t_{\ell,k}}$, each corresponding to an element in $\elt(P_\ell\cap Q_k)$. Since $\E[\sum_{i=1}^{t_{\ell,k}} Y_i] =\E[|R\cap P_\ell\cap Q_k|] \geq b_{\ell,k}/8$ (as $\ell$ is not a \textbf{$k$-bad leaf}), the Chernoff bound (see \cref{lemma:chernoff}) with $\epsilon = 3/4$ implies that 
$$\Pr\left[\sum_{i=1}^{t_{\ell,k}} Y_i\leq \frac{b_{\ell,k}}{32}\right]\leq \exp\left(-\frac{t_{\ell,k}}{16}\right)\leq \exp\left(-\frac{b_{\ell,k}}{16}\right)\leq \exp\left(-2^{k-4}\right).$$
By the union bound, the probability that there exists such a $k$ is at most
$\sum_{k=5}^{\infty}\exp\left(-2^{k-4}\right)\leq \frac{1}{2}.$
Therefore, we have 
$$
\Pr_R\Bigl[|R\cap P_\ell\cap Q_k|\geq \frac{b_{\ell,k}}{32},
\,\,\forall k\in K_\ell\Bigr]\geq \frac{1}{2}.
$$
Next, we observe that
$$
(A_\ell\cap D_\ell)\backslash G_\ell = \bigcup_{k\in K_\ell}(A_\ell \cap Q_k).
$$
Since this set is a subset of $P_\ell$ and each node represents a distinct element, its corresponding vector has exactly $b_{\ell,k}$ copies of $4^{-k}$ for each $k\in K_\ell$. By the symmetry and subadditivity of $f$, the $f$-value of this vector is at most $32$ times the $f$-value of a vector with $\lceil b_{\ell,k}/32 \rceil$ copies of $4^{-k}$. We have shown that, with probability at least $1/2$, $|R\cap P_\ell\cap Q_k|\geq \lceil b_{\ell,k}/32 \rceil$ for every $k\in K_\ell$. For such $R$, it follows that
$$f(R\cap P_\ell\cap D_\ell)\geq \frac{1}{32}f((A_\ell\cap D_\ell)\backslash G_\ell).$$
Averaging over all $R$ (and using the fact that $f\geq 0$), the lemma follows.
\end{proof}

Finally, we are in a position to lower bound $\alg(\mathcal{T}, f)$ by a constant.
\begin{theorem}
\label{thm:sym}
The adaptivity gap for {\em Bernoulli Stochastic Probing with Symmetric Norm Objective} is $O(1)$.
\end{theorem}

\begin{proof}
By \cref{sec5lem:chernoff}, \cref{eq:sec5} and \cref{eq:bad}, we have
\begin{align*}
\alg(\mathcal{T}, f) &= \E_{\ell \leftarrow \pi_{\calT}} \left[ \E_{R \sim \Rdistribution} \left[f(R\cap P_\ell)\right] \right] \\
&\geq \E_{\ell \leftarrow \pi_{\calT}} \left[ \E_{R \sim \Rdistribution} \left[f(R\cap P_\ell\cap D_\ell)\right] \right] \\
&\geq \frac{1}{64}\E_{\ell \leftarrow \pi_{\calT}} \left[f((A_\ell\cap D_\ell)\backslash G_\ell) \right] \\
&\geq \frac{1}{64}\E_{\ell \leftarrow \pi_{\calT}} \left[f(A_\ell\cap D_\ell) \right]-\frac{1}{64}\E_{\ell \leftarrow \pi_{\calT}}[f(G_\ell)] \\
&\geq \frac{1}{1024}-\frac{1}{64\cdot 2^{30}}\geq \frac{1}{1025}.
\end{align*}
This completes the proof, as the original $\OPT = 1$.
\end{proof}

\section{Generalization to non-Bernoulli Case and Other Objectives}
\label{section 6}
In this section, we generalize our results on Bernoulli Stochastic Probing to general random variables based on the previous results for Bernoulli variables. 

We state the following three theorems, where $n$ denotes the cardinality of the ground set, and $2\leq r\leq n$ denotes an upper bound for the maximal length of a sequence in $\calF$ (the same as in \cref{sec3:redcution}).
Recall the adaptivity gap functions $\gapfunction_{\XOS}(n,r)$ and $\gapfunction_{\sym}(n)$ for the two Bernoulli settings defined in \cref{bernoulli:norm} and \cref{bernoulli:symmetric}. 
\begin{theorem}[\cite{GNS17}]
\label{thm:genesub}
The adaptivity gap for {\em Bernoulli stochastic probing with any subadditive objective} under any prefix-closed $\calF$ with $\max\limits_{F\in \calF}|F|\leq r$ is $O(\gapfunction_{\XOS}(n,r)\log n)$.
\end{theorem}

\begin{theorem}
\label{thm:geneXOS}
The adaptivity gap for {\em stochastic probing with any monotone norm} under any prefix-closed $\calF$ with $\max\limits_{F\in \calF}|F|\leq r$ is $O(\gapfunction_{\XOS}(n\log 8r,r\log 8r))$.
\end{theorem}

\begin{theorem}
\label{thm:genesym}
The adaptivity gap for {\em stochastic probing with  monotone symmetric norm} is $O(\gapfunction_{\sym}(n\log 8n))$.
\end{theorem}

The first theorem is a direct consequence of the relationship between (monotone, normalized) subadditive objectives and XOS functions, and already appear informally in \cite{GNS17}. The proofs of the other two theorems are very similar to the classical methods in \cite{Sin18} but in slightly different settings. All three proofs are deferred to \cref{sec6:proof}.

Combining the three theorems with the upper bounds of adaptivity gap for Bernoulli settings: \cref{thm:strong} and \cref{thm:sym}, we obtain the promised results for general random variables, subadditive objectives and monotone symmetric norms.
\begin{corollary}
\label{thm:6.8}
For $2\leq r\leq n$, the adaptivity gap for {\em Bernoulli stochastic probing with any subadditive objective} under any prefix-closed $\calF$ with $\max\limits_{F\in \calF}|F|\leq r$ is upper bounded by $$
O\left(\log r\frac{\log^2 n}{\log\log n}\right) = O(\log^3 n).
$$
\end{corollary}

\begin{corollary}
\label{thm:6.6}
For $2\leq r\leq n$, the adaptivity gap for {\em stochastic probing with any monotone norm} under any prefix-closed $\calF$ with $\max\limits_{F\in \calF}|F|\leq r$ is upper bounded by
$$
O\left(\log r\frac{\log n}{\log\log n}\right).
$$
\end{corollary}

\begin{corollary}
\label{thm:6.7}
The adaptivity gap for {\em stochastic probing with  monotone symmetric norm} is $O(1)$.
\end{corollary}

\section{Conclusion and Future Directions}

In this paper, we investigated the stochastic probing problem under a general monotone norm and subadditive objectives. First, we resolved a central open problem posed in \cite{GNS17,KMS24} by establishing that the adaptivity gap for stochastic probing with general monotone norms is bounded by \(O(\log^2 n)\), which can be further refined to \(O\!\Bigl(\log r \cdot \frac{\log n}{\log\log n}\Bigr)\) where \(r\) is the maximum number of elements that can be probed under the feasibility constraint. As a by-product, we obtain an asymptotically tight adaptivity gap $\Theta(\frac{\log n}{\log\log n})$ for Bernoulli stochastic probing with \XOSS objectives. We also derived an \(O(\log^3 n)\) upper bound for Bernoulli stochastic probing with general subadditive objectives. Second, for the special case of monotone symmetric norms, we prove that the adaptivity gap is \(O(1)\) (in fact, bounded by 2050), answering an open question raised in \cite{PRS23} and improving upon their \(O(\log n)\) bound.

Several intriguing open questions remain. Notably, while our upper bound for general monotone norms shows an \(O(\log r)\) gap compared to the best known lower bound of \(\Omega\Bigl(\frac{\log n}{\log\log n}\Bigr)\), bridging this gap likely requires deeper understanding of the non-binary weight vectors in the XOS representation of the objective function $f$, representing an important direction for future research. In addition, refining the constant for monotone symmetric norms to obtain a tight bound remains an appealing challenge. Finally, extending our techniques to more general stochastic optimization settings and further exploring algorithmic applications of our insight in stochastic combinatorial optimization remain interesting further directions.

\section*{Acknowledgments}
We thank Yichuan Wang for valuable feedback on an earlier draft of this manuscript. 
We also thank Sahil Singla for detailed and insightful discussions, 
which simplified the presentation of the overview.
\newpage
\crefalias{section}{appendix}
\begin{appendices}

\section{Basic Lemmas for Stochastic Probing}\label{AppendixB}
In this section, we introduce several key tools for analyzing the general \emph{stochastic probing} problem, as defined in \cref{stoc:probing}. The lemmas and proofs here closely resemble those in \cite{GNS17}. However, unlike \cite{GNS17}, which focuses on the case where all random variables $X_i$ follow Bernoulli distributions, we extend the analysis to handle multi-valued random variables and present the arguments using our own notation. These results play an important role in the proof of \cref{large:case} and are also applied in \cref{sec6:proof} (i.e., the arguments given in \cref{section 6}).

Throughout this section, we only assume that the objective function $f$ satisfies the standard properties of being normalized, monotone, and subadditive, as described in conditions (1), (2), and (3) of \cref{def:norm}. We also assume that the feasibility constraint is \emph{prefix-closed}. In our model, the random variables $X_i$ are allowed to take on multiple values.  Let $\OPT$ denote the expected reward achieved by the optimal adaptive strategy.

\begin{definition}[$\lambda$-large (small) outcomes and variables]
For any $\lambda>0$, we call an outcome $x_i$ of $X_i$ \emph{$\lambda$-large} if and only if $f(0,0,\cdots,0,x_i,0,0,\cdots,0)\geq \lambda\cdot\OPT$, and \emph{$\lambda$-small} otherwise. If every possible outcome of $X_i$ is $\lambda$-large (or $\lambda$-small), then we say that $X_i$ is a $\lambda$-large (or $\lambda$-small) variable.
\end{definition}

\begin{definition}[$\lambda$-large decomposition]\label{decomp}
For any $\lambda>0$ and any coordinate $i\in[n]$, define the random variable $Y_{i,\lambda}$ by
$$
Y_{i,\lambda} = 
\begin{cases}
X_i, & \text{if } X_i \text{ is a } \lambda\text{-large outcome}, \\
0,   & \text{otherwise}.
\end{cases}
$$
Similarly, define $Z_{i,\lambda}$ by replacing ``large'' with ``small'' in the above definition.
\end{definition}

We begin by proving a subsequence property, which is analogous to the subtree property in \cite[Assumption 5.3]{GNS17}. In adaptive strategies, the decision of which element to probe next can depend on the outcomes observed so far. Suppose that during the execution of an adaptive strategy we have already probed a sequence $P_1\in\calF$ (recall that $\calF$ is prefix-closed) and obtained the outcome $X_{P_1}$ (that is, the $n$-dimensional vector whose $i$th coordinate equals $X_i$ if $i\in P_1$ and $0$ otherwise). Then, the remaining probing process produces a distribution of final outcomes, which we denote by $\calD_{P_1}$. For any $X_P\gets\calD_{P_1}$, the difference $X_P-X_{P_1}$ represents the outcomes obtained after probing $P_1$.

\begin{lemma}[Subsequence Property]
\label{subsequence}
For any adaptive strategy and any sequence $P_1\in\calF$ probed during its execution, we have
$$
\E_{X_P\leftarrow\calD_{P_1}}[f(X_P-X_{P_1})]\leq \OPT.
$$
\end{lemma}
\begin{proof}
Consider an alternative adaptive strategy that first probes the sequence $P_1\in\calF$ and receives $X_{P_1}$, and then follows the original adaptive strategy. Since the overall strategy still satisfies the feasibility constraint, its expected reward is at least $\E_{X_P\leftarrow\calD_{P_1}}[f(X_P)]$. By the monotonicity of $f$, we have $f(X_P)\geq f(X_P-X_{P_1})$, so that the expected reward is at least $\E_{X_P\leftarrow\calD_{P_1}}[f(X_P-X_{P_1})]$. Since $\OPT$ is the reward of the optimal adaptive strategy, the claim follows.
\end{proof}

For any probed sequence $P\subseteq U = [n]$, the $\lambda$-large decomposition of $X_P$ is given by
$
X_P = Y_{P,\lambda}+Z_{P,\lambda},
$
where $Y_{P,\lambda}$ and $Z_{P,\lambda}$ consist of the outcomes $Y_{i,\lambda}$ and $Z_{i,\lambda}$, respectively, for each coordinate $i\in P$ (with coordinates for $i\notin P$ set to $0$).

\begin{lemma}
\label{lm:general:large}
Assume that the adaptive strategy is optimal. If $\E[f(Y_{P,\lambda})]\geq \OPT/2$, then there exists a non-adaptive algorithm whose expected reward is at least $\frac{\OPT}{2(1+1/\lambda)}$.
\end{lemma}
\begin{proof}
We begin by constructing an intermediate strategy derived from the optimal adaptive strategy. This new strategy stops probing as soon as it encounters a $\lambda$-large outcome. Let $\mathsf{OPT}^\prime$ denote the expected reward from the (at most one) $\lambda$-large outcome under this strategy. Let $\OPT_0$ be the expected contribution from any additional $\lambda$-large outcomes that would be probed in the original optimal strategy \emph{after} the first one is found.

By construction, every $\lambda$-large outcome seen in any execution of the optimal strategy is either the first such outcome (counted in $\mathsf{OPT}^\prime$), or a later one (counted in $\OPT_0$). Since the objective $f$ is subadditive, we have:
$
\mathsf{OPT}^\prime + \OPT_0 \geq \E[f(Y_{P,\lambda})] \geq \OPT/2.
$

Next, consider one execution of this intermediate strategy. There are two cases:

\begin{mycase}{1}
    The strategy encounters a $\lambda$-large outcome and terminates. By definition of $\lambda$-large, this outcome has $f$-value at least $\lambda \cdot \OPT$. Now imagine continuing to probe as in the original optimal strategy. By the subsequence property (\cref{subsequence}), the expected total reward of these further probes is at most $\OPT$. Thus, the expected value contributed by any additional $\lambda$-large outcomes is also at most $\OPT$ by monotonicity.
\end{mycase} 
\begin{mycase}{2}
     The strategy finds no $\lambda$-large outcome and continues to the end. In this case, the expected contribution by further $\lambda$-large outcomes is zero.
\end{mycase}

Thus, in either case, the intermediate strategy achieves at least a $\lambda$ fraction of the expected value contributed by later $\lambda$-large outcomes in the original strategy. Taking expectation over all executions gives:
$
\lambda \cdot \OPT_0 \leq \mathsf{OPT}^\prime.
$
Combining with
$
\mathsf{OPT}^\prime + \OPT_0 \geq \OPT/2,
$
we derive
$
\mathsf{OPT}^\prime \geq \frac{\OPT}{2(1 + 1/\lambda)}.
$

Now, we construct a \emph{non-adaptive} strategy that matches the reward of the intermediate adaptive strategy. This strategy simulates the behavior of the original optimal strategy. Whenever the adaptive strategy probes an element, the non-adaptive strategy samples an element by itself from the same known distribution \emph{conditioned} on the outcome being $\lambda$-small. It then probes the same element as the adaptive strategy does upon receiving this (unreal) conditioned outcome sampled by the non-adaptive strategy itself.

This strategy is clearly feasible and non-adaptive. We show, by induction on the height of the decision tree for the intermediate strategy, that it achieves at least the same expected reward as the intermediate strategy.

The base case is trivial when the decision tree is of height 1 (the outcome of the probed element has same distribution in both strategies). For the general case, suppose the root element gives a $\lambda$-large outcome with probability $p$, and a $\lambda$-small outcome with probability $1 - p$. The intermediate strategy achieves expected reward $A$ and terminates with probability $p$, and with probability $1 - p$, continues probing based on the received $\lambda$-small outcome, say distributed with probabilities $\{q_i\}$ (so that $\sum q_i = 1 - p$).

For the non-adaptive strategy, it occurs with  probability $p$ (independent of the strategy) that the first probe gives a $\lambda$-large outcome. Under this condition, by monotonicity, the expected reward is at least the $f$ value of only the $\lambda$-large outcome for the first probe, which is exactly $A$. 

There is also another $1-p$ probability (independent of the strategy) that the first probe gives a $\lambda$-small outcome. Hence, conditioning on such event, the probability that the strategy chooses the $i$-th element to probe is exactly $q_i/ (\sum q_i)\times (1-p)=q_i$, the same as the probability of the intermediate adaptive strategy chooses the $i$-th element to probe. Hence, we use induction on the sub-execution of both strategies. Together with the $p$ probability discussed above, we conclude that the reward of the non-adaptive strategy is at least that of the intermediate strategy.

Hence, the non-adaptive strategy achieves expected reward at least $\mathsf{OPT}^\prime\geq \frac{\OPT}{2(1 + 1/\lambda)},
$
as desired.
\end{proof}

We now present the proof of \cref{large:case}.

\begin{proof}
Note that the Bernoulli settings for XOS, subadditive, and \XOSS objectives described in \cref{bernoulli:norm}, as well as the Bernoulli setting for symmetric norm objectives in \cref{bernoulli:symmetric}, can both be understood as cases where the random variables $X_i$ take values in the set $\{0, c_i\}$ for some fixed constant $c_i$.

By applying \cref{lm:general:large}, either the adaptivity gap is at most $2(1+1/\lambda)$, or we are in the case where
$$
\E[f(Z_{P,\lambda})]\geq \E[f(X_P)]-\E[f(Y_{P,\lambda})]\geq \OPT-\OPT/2 = \OPT/2,
$$
which follows from the subadditivity of $f$.

In this second case, we analyze the adaptivity gap under the modified random variables $Z_{i,\lambda}$. Since for each coordinate $X_P \geq Z_{P,\lambda}$, the best non-adaptive strategy on $Z_{P,\lambda}$ cannot outperform the best non-adaptive strategy on $X_P$. Therefore, the adaptivity gap for the $Z_{i,\lambda}$ variables is at most twice the original gap. 

Moreover, since the optimal reward is at least $\E[f(Z_{P,\lambda})] \geq \OPT/2$, the random variables $Z_{P,\lambda}$ can be considered $2\lambda$-small. The final result then follows by substituting $\lambda$ with $2\lambda$.
\end{proof}

\section{Missing Proof of \cref{lih:induction}}
\label{lm:inequality}
This section provides the whole proof for \cref{lih:induction}. Recall the function $g(h, p) = p\cdot \exp(-0.1 hp^{\frac 1 h})$ where $h\in\mathbb N_+$, and $p\in [0,1]$. Additionally, we define $g(0,p) = p$, where $p\in [0,1]$. \cref{eq2} and \cref{eq5} shows that $g$ is increasing and concave on $[0,1]$ when fixing $h \in \mathbb N$.

We first prove the following technical inequality lemma.
\begin{lemma}\label{lemma4.1}
For any reals $p,q,\lambda\in [0,1]$, positive integer $h\ge 1$,  the following inequality holds:
\begin{equation}\label{B.1}
g(h, \lambda p + (1-\lambda )q)\ge (1-\lambda)[\lambda g(h-1, p) + (1-\lambda) g(h, q)].
\end{equation}
\end{lemma}

\begin{proof}
The idea is that, first we use concavity of function $g(h,\cdot)$ as well as the slackness in the coefficient (say, the coefficient $(1-\lambda)$ outside the squared brackets in \cref{B.1}) to rule out most of the non-tight case. The remaining possibility indicates that $\lambda$ is relatively small, and $p$ is larger than a constant multiple of $q$. In such case, we directly use the derivative of $g(h,\cdot)$ to approximate $g(h, \lambda p + (1-\lambda )q)-g(h,q)$. In either case, we need to upper bound $g(h-1,p)$ by $g(h,\cdot)$ using the so-called Young's inequality bellow:
\begin{myfact}{1}[Young's Inequality, see \cite{young1912}]\label{fact1}
If $a,b\ge 0$ are non-negative real numbers, and $x,y>1$ are real numbers such that $\frac 1 x+\frac  1 y = 1$, then 
$
ab\leq a^x/x+b^y/y.
$
\end{myfact}
The inequality can be equivalently stated in another form, by substituting $x = \frac{u+v}{u}$, $y =\frac{u+v}{v}$, and $a\gets a^{u+v}$, $b\gets b^{u+v}$: if $a,b\ge 0$, and $u,v>0$, then 
$
(u+v)(ab)^{\frac{1}{u+v}}\leq ua^{\frac{1}{u}}+ vb^{\frac{1}{v}}.
$

We may also use the following two fundamental inequalities related to the natural logarithm $e$.
\begin{myfact}{2}\label{fact2}
$
e^x\ge x+1, \forall x\in\mathbb R.
$
$(1+\frac{1}{x})^x\leq e,\forall x\in\mathbb R_+.
$
\end{myfact}
Now we treat the following two cases respectively, as mentioned ahead. 
\begin{mycase}{1} 
$(0.2-1.2\lambda)p\leq (1-\lambda)q$.
\end{mycase}
\noindent By Young's inequality (\cref{fact1}), $(h-1)p^{\frac{1}{h-1}}+1 \ge hp^{\frac{1}{h}}$ for $h\ge 2$. Then, for such $h\ge 2$
$$
g(h-1, p)= p\exp(-0.1(h-1)p^{\frac{1}{h-1}})\leq p\exp(-0.1hp^{\frac{1}{h}} + 0.1)\leq 1.2\exp(-0.1hp^{\frac{1}{h}}) = 1.2g(h,p).
$$
Note that the above also holds for $h =1$, as the first inequality $0\leq 0.1-0.1p$ still holds for $p\in [0,1]$.

Now we utilize the property that $g(h,\cdot)$ is a concave function, and $g(h,0)= 0 $ for $h\in\mathbb N_+$. 
Then the following holds for any real $a,b\ge 0$ satisfying $a+b\leq 1$:
$$
ag(h,p)+bg(h,q) = ag(h,p)+bg(h,q)+(1-a-b)g(h,0)\leq g(h,ap+bq+(1-a-b)\cdot 0) = g(h,ap+bq).
$$
Taking $a = 1.2\lambda(1-\lambda)$, $b = (1-\lambda )^2$, $a+b=(1-\lambda)(1+0.2\lambda)\leq (1-\lambda)(1+\lambda)\leq 1$ in the above inequality,
\begin{align*}
(1-\lambda)[\lambda g(h-1, p) + (1-\lambda) g(h, q)] &\leq 1.2\lambda(1-\lambda)g(h,p)+(1-\lambda )^2g(h,q)\\
& \leq g(h, 1.2\lambda(1-\lambda)p+(1-\lambda )^2q )\\
& = g(h, \lambda p+(1-\lambda )q+\lambda[(0.2-1.2\lambda)p-(1-\lambda)q ])\\
& \leq g(h, \lambda p+(1-\lambda )q ),
\end{align*}
where the last inequality uses the monotonicity of $g(h,\cdot)$ and the condition given in this case.
\begin{mycase}{2}
$(0.2-1.2\lambda)p > (1-\lambda)q$.
\end{mycase}
\noindent Note that the condition already implies $0.2-1.2\lambda > 0$, or $\lambda < \frac 1 6$, and $p>\frac{1-\lambda}{0.2-1.2\lambda}q \ge  5q\ge 0 $. We see that $\lambda p +(1-\lambda)q \ge (1+4\lambda)q\ge q $, as well as $\lambda p +(1-\lambda)q < (0.2 -0.2\lambda)p \leq 0.2p$. Putting together,
\begin{equation}\label{bound}
    q \leq \lambda p +(1-\lambda)q < 0.2p.
\end{equation}

\noindent Since $g(h,\cdot)$ is concave when we fix $h$, the partial derivative $ \frac{\partial g}{\partial p}(h, \cdot)$ with respect to the second variable is monotonically decreasing. In the following, we use $g^\prime(\cdot)$ to denote the partial derivative $\frac{\partial g}{\partial p}(h, \cdot)$. Therefore, 
\begin{align*}
    g(h,\lambda p +(1-\lambda)q)-g(h,q) =\int_{q}^{\lambda p + (1-\lambda)q}g^\prime(h,x)dx \ge \lambda ( p - q) g^\prime(h,\lambda p + (1-\lambda)q).
\end{align*}
Thus, by dividing $\lambda$ in the target inequality \cref{B.1}, we only need to prove that
\begin{align*}
    ( p - q) g^\prime(h,\lambda p + (1-\lambda)q) + (2-\lambda)g(h,q) \ge (1-\lambda)g(h-1,p).
\end{align*}
For clarity, we denote $\lambda p +(1-\lambda)q = r$. Using the derivative formula in \cref{eq2}, we have
$$
g^\prime(h,r)=\exp(-0.1hr^{\frac 1 h})(1- 0.1r^{\frac 1 h})\geq \exp(-0.1hr^{\frac 1 h})\exp(-0.2r^{\frac 1 h})=\exp(-0.1(h+2)r^{\frac 1 h}),
$$
where the first inequality uses a small inequality that $1-x\geq \exp(-2x)$, when $x=0.1r^{\frac 1 h}\in [0,0.1] $. One can check that it indeed holds since $e^{2x}\ge 1+2x\ge \frac{1}{1-x}$, for $x\leq \frac{1}{2}$. Therefore, the target inequality can be relaxed further to be
\begin{equation}\label{eq3}
    (p-q)\exp(-0.1(h+2)r^{\frac 1 h})+ (2-\lambda)g(h,q) \ge (1-\lambda)g(h-1,p).
\end{equation}
Now we bound the right-hand side. Similar as in the previous case, we use Young's inequality (\cref{fact1}) and obtain for $h\ge 2$ (using the implicit fact that $p>0$ mentioned in the first line in the current case),    
\begin{align*}
(h-1)p^{\frac{1}{h-1}}+e^2 \frac{r}{p}
\ge (h-1)p^{\frac{1}{h-1}}+ \left(\frac{h+2}{h}\right)^{h}\frac{r}{p}
\ge h {p^{\frac{1}{h}}\times \frac{h+2}{h} \left(\frac{r}{p}\right)^{\frac{1}{h}}}
= (h+2 )r^{\frac 1 h},
\end{align*}
where the first inequality is given by the second inequality in \cref{fact2} together with the substitution $x = \frac{h}{2}$.

In the following, we still need another small inequality that $\exp(0.1e^2x) \leq  1+x,\forall x\leq 0.2$. One can also check it indeed holds since $\exp(-0.1e^2x)\ge\exp(-0.8x)\ge 1-0.8x\ge \frac{1}{1+x}$, for $x\leq 0.2$. Therefore, by \cref{bound}, taking $x = \frac{r}{p}=\frac{\lambda p +(1-\lambda )q}{p}<0.2$ in the above inequality, we obtain for $h\ge 2$:
\begin{align*}
g(h-1,p) = p\exp(-0.1(h-1)p^{\frac{1}{h-1}})
&\leq p\exp \left(-0.1(h+2)r^{\frac{1}{h}}+0.1e^2\frac{r}{p}\right)\\
&\leq p\exp \left(-0.1(h+2)r^{\frac{1}{h}}\right)\left(1+\frac{r}{p}\right)\\
& = ((1+\lambda)p+(1-\lambda)q)\exp \left(-0.1(h+2)r^{\frac{1}{h}}\right).
\end{align*}
Note that the above also holds for $h =1$, as the first inequality $0\leq -0.3r+0.1e^2\frac{r}{p} = (0.1e^2-0.3p)\frac{r}{p}$ still holds for $p\in [0,1]$. Thus, in either case, the right-hand side of \cref{eq3} can be upper bounded as: 
$$
(1-\lambda) g(h-1,p) \leq ((1-\lambda^2)p+(1-\lambda)^2 q)\exp \left(-0.1(h+2)r^{\frac{1}{h}}\right)\leq (p+(1-\lambda) q)\exp \left(-0.1(h+2)r^{\frac{1}{h}}\right).
$$
Plugging in the upper bound above in \cref{eq3} and dividing $2-\lambda$ in both sides, we are left to prove 
$$ g(h,q)\ge q\exp \left(-0.1(h+2)r^{\frac{1}{h}}\right).$$ By expressing $g(h,q)$ by definition and using the monotonicity of $\exp(\cdot)$, such inequality is equivalent to $hq^{\frac{1}{h}}\leq (h+2)r^{\frac{1}{h}}$. But this is obvious, as we've already mentioned in \cref {bound} that $r = \lambda p +(1-\lambda)q \ge q$, Thus, $hq^{\frac{1}{h}}\leq hr^{\frac{1}{h}}\leq (h+2)r^{\frac{1}{h}}$, as desired.
\end{proof}
Now we prove \cref{lih:induction}. Recall the setting as follows: $S$ is a subset of leaves in the tree $\calT$. There is a subset $T$ of nodes in $\cal T$, not containing any leaf of $\calT$. There exists an integer $h\in \mathbb N$, such that for each $\l \in S$, $| A_\l\cap T|\geq h $. Moreover, each $\l\in S$ is assigned a real weight $x_\l \in [0,1]$, we denote $\E_{\ell \leftarrow \pi_{\calT}}[x_\l \mathbf 1_{\ell\in S}] = p$. We need to prove the following inequality:
$$\E_{\ell \leftarrow \pi_{\calT}} \left[x_\l \mathbf 1_{\l \in S}  \cdot \Pr_{R \sim \Rdistribution} \left[R\cap P_\ell\cap T=\emptyset\right]\right]\leq g(h,p).$$
\begin{proof}[Proof of \textnormal{\cref{lih:induction}}]
First, the result is trivially true for $h = 0$, since
\[
\E_{\ell \leftarrow \pi_{\calT}} \left[ x_\ell \mathbf{1}_{\ell \in S} \cdot \Pr_{R \sim \Rdistribution} \left[ R \cap P_\ell \cap T = \emptyset \right] \right]
\leq \E_{\ell \leftarrow \pi_{\calT}}[x_\ell \mathbf{1}_{\ell \in S}] = p = g(0,p).
\]
Therefore, we may assume $h \geq 1$ for the rest of the proof. We proceed by induction on the height of $\calT$.

When the height of $\calT$ is $1$, then $h$ can only be $0$, since $T$ does not contain any leaf. Hence, the base case is already covered. Assume the statement holds for trees of height less than that of $\calT$. Let the root of $\calT$ be $r_0$. Let $\calT_1$ and $\calT_2$ be the subtrees rooted at the \textsf{Yes} and \textsf{No} children of $r_0$, respectively. Denote the probability for the root to be active is $\lambda\in [0,1]$ (note that we can also handle the case where $r_0$ has only one child by setting $\lambda = 0$ or $1$).  It is clear that the path distribution $\pi_{\calT}$ is $\lambda$ probability to $\textsf{Yes}$ subtree $\calT_1$ and $1-\lambda$ to $\textsf{No}$ subtree $\calT_2$, then preserves the distribution given by $\pi_{\calT_1}$ and $\pi_{\calT_2}$. Denote $\E_{\ell \leftarrow \pi_{\calT_1}} \left[ x_\l \mathbf 1_{\l \in S} \right] = u$, and $\E_{\ell \leftarrow \pi_{\calT_2}} \left[ x_\l \mathbf 1_{\l \in S} \right] = v$. Then $\E_{\ell \leftarrow \pi_{\calT}} \left[ x_\l \mathbf 1_{\l \in S} \right] = p = \lambda u +(1-\lambda)v$.

Now we treat the following two cases respectively, as in \cref{liu:induction}.

\noindent Let $E_1:=\E_{\ell \leftarrow \pi_{\calT_1}} \left[ x_\l\mathbf{1}_{\ell \in S}  \cdot \Pr_{R \sim \Rdistribution} \left[R\cap P_\ell\cap T=\emptyset\right]\right]$, $E_2:=\E_{\ell \leftarrow \pi_{\calT_2}} \left[ x_\l\mathbf{1}_{\ell \in S}  \cdot \Pr_{R \sim \Rdistribution} \left[R\cap P_\ell\cap T=\emptyset\right]\right]$.

\begin{mycase}{1}
$r_0\in T.$
\end{mycase}
In this case, for leaves $\l$ in $S\cap \calT_1$, there are at least $h-1$ active ancestors of it inside $T\cap \calT_1$. Moreover, we denote the corresponding root-leaf path in $\calT_1$ or $\calT_2$ as $P_{\l}^\prime=P_{\l}\backslash\{r_0\}$, then the corresponding relation $ \Pr_{R \sim \Rdistribution} \left[R\cap P_\ell\cap T=\emptyset\right] = (1-\lambda)\Pr_{R \sim \Rdistribution} \left[R\cap P_\ell^\prime\cap T=\emptyset\right]$ holds, as the root $r_0$ has an independent probability of $1-\lambda$ to be active. Therefore, by the induction hypothesis on $\calT_1$, we obtain:
\begin{align*}
E_1 = (1-\lambda)\E_{\ell \leftarrow \pi_{\calT_1}} \left[ x_\l \mathbf 1_{\l \in S} \cdot \Pr_{R \sim \Rdistribution} \left[R\cap P_\ell^\prime\cap T=\emptyset\right]\right] \leq (1-\lambda)g(h-1,u).
\end{align*}
Similarly, for the \textsf{No} subtree, each leaf has at least $h$ active ancestors inside $T\cap \calT_2$, since the root is not an active ancestor for leaves in the \textsf{No} subtree. Hence, 
\begin{align*}
E_2= (1-\lambda)\E_{\ell \leftarrow \pi_{\calT_2}} \left[ x_\l \mathbf 1_{\l \in S} \cdot \Pr_{R \sim \Rdistribution} \left[R\cap P_\ell\cap T=\emptyset\right]\right] \leq (1-\lambda)g(h,v).
\end{align*}
Putting together,
$$
\E_{\ell \leftarrow \pi_{\calT}} \left[ x_\l \mathbf 1_{\l \in S} \cdot \Pr_{R \sim \Rdistribution} \left[R\cap P_\ell\cap T=\emptyset\right]\right] 
= \lambda E_1 + (1-\lambda)E_2\leq (1-\lambda)[\lambda g(h-1,u)+(1-\lambda)g(h,v)].
$$
Since $h\ge 1$, and $u,v,\lambda\in [0,1]$ being expectation of a random variable in $[0,1]$, we use \cref{lemma4.1} and concludes:
$$
 (1-\lambda)[\lambda g(h-1,u)+(1-\lambda)g(h,v)] \leq g(h,\lambda u +(1-\lambda)v)=g(h,p).
$$
\begin{mycase}{2}
$r_0\notin T$.
\end{mycase}
In this case, for leaves $\l$ in both subtrees $S\cap \calT_1$ and $S\cap \calT_2$, there are at least $h$ active ancestors of it inside $T\cap \calT_1$ and $T\cap \calT_2$, since $r_0\notin T$. Moreover, we denote the corresponding root-leaf path in $\calT_1$ or $\calT_2$ as $P_{\l}^\prime=P_{\l}\backslash\{r_0\}$, then the corresponding  relation $ \Pr_{R \sim \Rdistribution} \left[R\cap P_\ell\cap T=\emptyset\right] = \Pr_{R \sim \Rdistribution} \left[R\cap P_\ell^\prime\cap T=\emptyset\right]$ holds, as the root $r_0\notin T$. Therefore, by the induction hypothesis on $\calT_1$ and $\calT_2$, we obtain $E_1\leq g(h,u)$ and $E_2\leq g(h,v)$. Putting together,
\begin{align*}
\E_{\ell \leftarrow \pi_{\calT}} \left[ x_\l \mathbf 1_{\l \in S} \cdot \Pr_{R \sim \Rdistribution} \left[R\cap P_\ell\cap T=\emptyset\right]\right] &= \lambda E_1 + (1-\lambda)E_2 
\leq \lambda g(h,u)+(1-\lambda)g(h,v)\\
&\leq  g(h,\lambda u+(1-\lambda)v) = g(h,p),
\end{align*}
where the last inequality uses the concavity of $g(h,\cdot)$ as in \cref{eq5}.
\end{proof}
\section{Missing Proof for \cref{section 6}}
\label{sec6:proof}
Recall that $n$ represents the size of the ground set, and $r$ is an upper bound on the length of a sequence in $\calF$. We first present the proof of \cref{thm:genesub}, the idea has already appeared implicitly in \cite{GNS17}.
\begin{proof}[Proof of \textnormal{\cref{thm:genesub}}]
We consider the Bernoulli setting with a subadditive objective function $f:2^{U}\to \mathbb{R}_{\geq 0}$. It is a known result that any monotone and normalized ($f(\emptyset) = 0$) subadditive set function can be approximated within a factor of $\log |U| = O(\log n)$ by an XOS function \cite{approxxos,approxxos2}. Consequently, the adaptivity gap 
for \emph{Bernoulli Stochastic Probing with Subadditive Objective} is $O(\gapfunction_{\XOS}(n,r) \log n)$.
\end{proof}
For the remaining two theorems, both settings require that the objective $f$ is a monotone norm. At the same time, we consider general random variables. We define $f(x_i) = f(0,0,\dots,0,x_i,0,\dots,0)$ and likewise treat $f(X_i)$ as the corresponding random variable. Since $f$ is a norm, we can assume that $f(0,0,\dots,0,1,0,\dots,0)> 0$, since otherwise $f(x_i)=0$ for any $x_i$. Then, we can remove the element from $U$ and modify the norm accordingly by inserting a zero placeholder at the corresponding position. 

We first establish the following lemma.
\begin{lemma}
\label{general:assumption}
Suppose the adaptivity gap for general random variables $X_i$ and a monotone norm objective $f$ under a feasibility constraint $\calF$ is $\gapfunction\geq 1$, provided that each $X_i$ takes value either $0$ or in a geometric progression of ratio \(2\) and length at most $\log 4r$. Then, the adaptivity gap for general random variables with the same $f$ and $\calF$ is at most $8\gapfunction$.
\end{lemma}
\begin{proof}
By appropriately scaling the objective $f$, we may assume without loss of generality that the optimal adaptive reward is $\OPT = 1$ in both cases. As explained in \cref{subsubsec:red2Ber}, we apply the $1$-large decomposition introduced in \cref{decomp}: $X_P = Y_{P,1} + Z_{P,1}$. 
Under the optimal adaptive strategy. If $\E[f(Y_{P,1})]\geq 1/2$, Lemma~\ref{lm:general:large} implies an adaptivity gap of at most $4$.
Otherwise, we consider the $1$-small random variables $Z_{P,1}$. Since the original optimal strategy ensures $\E[f(Z_{P,1})] \geq 1/2$, and the non-adaptive strategy cannot exceed this value due to monotonicity, the adaptivity gap for the original variables is at most twice that for $Z_{P,1}$. Reusing the notation, we assume that the current random variables $X_i$ satisfy $f(X_i)\leq 1$.

We can further simplify $X_i$ by setting $H_i = X_i$ if $f(X_i) \leq 1/(4r)$ and $H_i = 0$ otherwise. This gives $\E[f(H_P)] \leq r \cdot 1/(4r) \leq 1/4$. Thus, replacing $X_i$ with $H_i$ reduces the expected adaptive reward by at most $1/4$, and the non-adaptive reward does not increase. So, the adaptivity gap increases by at most a factor of $2$. Now, all $f(X_i)$ are either zero or within $[1/(4r), 1]$, and the optimal reward is at least $1/4$. 

Furthermore, for any possible outcome $x_i$ of $X_i$, we scale $x_i$ by a suitable constant $\theta\in[1/2,1]$ to make $f(\theta x_i) =\theta f(x_i)$ the maximum negative power of 2 strictly below the original $f(x_i)$. Now, each $f(X_i)$ can take at most $\log 8r$ possible values, namely $0$ and negative powers of $2$ contained in the interval $[1/(4r), 1/2]$. Due to the homogeneity of the norm $f$, it follows that $X_i$ itself assumes $0$ or values in a geometric progression of ratio \(2\) and length at most $\log 4r$ (note that we are using $f(0,0,\dots,0,1,0,\dots,0)> 0$). Again, the adaptive reward decreases by at most a factor of $2$ by monotonicity, and the non-adaptive reward does not increase. Thus, the adaptivity gap increases by another factor of $2$.

Combining the three reductions, the adaptivity gap increases by at most a factor of $8$.
\end{proof}

By incurring only a constant factor loss in the approximation, we may now assume that all $X_i$ satisfy the condition of Lemma~\ref{general:assumption}. We can represent the optimal adaptive strategy using a decision tree $\calT$, where each node corresponds to an element with at most $\log 8r$ children, each representing an outcome of $X_i$.

For each $i$, let $X_i$ take values $0 = c_{i,0} < c_{i,1} < \cdots < c_{i,k_i}$ ($ c_{i,j},1\leq j\leq k_i$ forms a geometric progression of ratio $2$) with probabilities $p_{i,0}, \dots, p_{i,k_i}$ (summing to 1, and $p_{i,0}$ can be zero). Following the approach in \cite{Sin18}, we introduce $k_i$ new elements $e_{i,1}, \dots, e_{i,k_i}$. For $1 \leq j \leq k_i$, we associate $e_{i,j}$ with a Bernoulli random variable $R_{i,j}$ with activation probability $\frac{p_{i,j}}{\sum_{h=0}^j p_{i,h}}$. It takes value in $0,1$ in the proof of \cref{thm:geneXOS}, and $c_{i,j}$ in the proof of \cref{thm:genesym}.

We construct a new decision tree $\calT'$ using these Bernoulli variables.
At each node $u$ with $\elt(u) = i$ and subtrees $T_0, \dots, T_{k_i}$, we replace $u$ with $k_i$ nodes that represent elements $e_{i,k_i}, \dots, e_{i,1}$ respectively. For each $j$, the node represents $e_{i,j}$ proceeds to $T_j$ if active or to the node represents $e_{i,j-1}$ otherwise. If none of $e_{i,j}$ is active, the path leads to $T_0$. Notice that the replacement is from root to leaf, so the replacement procedure finally terminates. An example is illustrated in \cref{fig2}.
\begin{figure}[ht]
\centering\scalebox{0.8}[0.8]{
\begin{tikzpicture}
    \tikzstyle{node}=[circle, draw, fill=white, inner sep=0pt, minimum size=10mm]
    \tikzstyle{label}=[draw=none, fill=none]

    \node[node] (i) at (-1,-1.5) {$i$};
    \node[node] (T0) at (-4,-3.5) {$T_0$};
    \node[node] (T1) at (-2,-3.5) {$T_1$};
    \node[node] (TJ) at (0,-3.5) {$\cdots$}; 
    \node[node] (Tki) at (2,-3.5) {$T_{k_i}$};

    \draw[blue, thick, ->] (i) -- node[left] {$c_{i,0}$} (T0);
    \draw[red, ultra thick, ->] (i) -- node[left] {$c_{i,1}$} (T1);
    \draw[red, ultra thick, ->] (i) -- (TJ);
    \draw[red, ultra thick, ->] (i) -- node[right] {$c_{i,k_i}$} (Tki);

    \draw[->, thick] (3,-2.5) -- (5,-2.5); 

    \node[node] (root) at (8,0) {$e_{i, k_i}$}; 
    \node[node] (Tki_right) at (6.5,-2) {$T_{k_i}$};
    \node[node] (e_ik1) at (9.5,-2) {$e_{i, k_i-1}$};

    \draw[red, ultra thick, ->] (root) -- node[left] {yes} (Tki_right);
    \draw[blue, thick, ->] (root) -- node[right] {no} (e_ik1);

    \node[node] (Ti) at (8,-4) {$T_{k_i-1}$};
    \node[label] (dots_right_middle) at (10.5,-4) {$\dots$};
    \node[node] (e_i1) at (11,-5) {$e_{i,1}$};

    \draw[red, ultra thick, ->] (e_ik1) -- node[left] {yes} (Ti);
    \draw[blue, thick, ->] (e_ik1) -- node[right] {no} (dots_right_middle);
    \draw[blue, thick, ->] (dots_right_middle) -- (e_i1);

    \node[node] (T1_right) at (10,-7) {$T_1$};
    \node[node] (T0_right) at (12,-7) {$T_0$};

    \draw[red, ultra thick, ->] (e_i1) -- node[left] {yes} (T1_right);
    \draw[blue, thick, ->] (e_i1) -- node[right] {no} (T0_right);

\end{tikzpicture}
}
\caption{Illustration of How to Change the Tree}
\label{fig2}
\end{figure}

We can define a Bernoulli stochastic probing instance according to this decision tree. The elements are all $e_{i,j}$, with Bernoulli random variables $R_{i,j}$. In addition, we define the new feasibility constraint $\calF'$ to contain all prefixes of element sequences that correspond to a root-leaf path for $\calT'$.
The new functions $f$ are defined in the proof sections of the two theorems, as they differ slightly in their definitions.

The following lemma demonstrates that the leaf distributions according to the two trees $\calT$ and $\calT'$ are essentially the same:
\begin{lemma}
\label{lm:treeequal}
For each $1 \leq i \leq n$ and $1 \leq j \leq k_i$, the probability that $e_{i,j}$ is active while all $e_{i,j'}$ with $j' > j$ are inactive is exactly $p_{i,j}$.
\end{lemma}
\begin{proof}
This follows from direct computation of probability:
\begin{align*}
&\frac{p_{i,j}}{p_{i,0}+p_{i,1}+\cdots+p_{i,j}}\prod_{j'=j+1}^{k_i}\left(1-\frac{p_{i,j'}}{p_{i,0}+p_{i,1}+\cdots+p_{i,j'}}\right)\\
=&\frac{p_{i,j}}{p_{i,0}+p_{i,1}+\cdots+p_{i,j}}\prod_{j'=j+1}^{k_i}\frac{p_{i,0}+p_{i,1}+\cdots+p_{i,j'-1}}{p_{i,0}+p_{i,1}+\cdots+p_{i,j'}}\\
=&\frac{p_{i,j}}{p_{i,0}+p_{i,1}+\cdots+p_{i,k_i}}=p_{i,j}.
\end{align*}
\end{proof}

Note that we have defined a new tree $\calT'$, a new feasibility constraint $\calF'$, new elements $e_{i,j}$ and their associated Bernoulli random variables $R_{i,j}$. The current number of elements $e_{i,j}$ is at most $n\log 8r$, and the maximal length of a sequence in the feasibility constraint $\calF'$ is at most the height of the new decision tree, which is clearly bounded by $r\log 8r$. Let $U'$ be the set of all $e_{i,j}$. 

Now, based on the original objective $f$, we should define a new objective $f'$ in the new setting. We show that the adaptivity gap for the new instances is at least that of the original instance.

First, we show the construction for the monotone norm and prove Theorem~\ref{thm:geneXOS}.
\begin{proof}[Proof of \textnormal{Theorem~\ref{thm:geneXOS}}]
We now define an objective
\(
  f'\colon \mathbb{R}_{\ge 0}^{U'} \to \mathbb{R}_{\ge 0}:
  \)
  $$
  f'\bigl((x_{ij})_{i\in U,\,j\in[k_i]}\bigr)
  \;=\;
  f\Bigl(\bigl(\max_{j\in[k_i]}c_{i,j}x_{i,j}\bigr)_{i\in U}\Bigr).
$$
Clearly, \(f'\) is a monotone norm on \(\mathbb{R}_{\ge0}^{U'}\). Thus, its restriction to a set function (as \cref{bernoulli:norm} does) is fractionally subadditive by Jensen’s inequality and therefore can be written as an XOS (set) function, see \cite{XOSfrac}.

Now, we first analyze the optimal reward for the new problem. Clearly, a valid adaptive strategy is to probe elements according to the probed result, following the decision tree $\calT'$.
Given a path $P'$ for $\calT'$ and all the realized $R_{ij}$ along the path, the reward $f'(R_{P'})$ is equal to the value $f(X_{P})$, where $P$ is a root-leaf path in the original tree $\calT$ that corresponds to $P'$ under the transformation. By Lemma~\ref{lm:treeequal}, the leaf distribution of $\calT$ and $\calT'$ is the same. Thus, we conclude that the expected reward following $\calT'$ with objective $f'$ is the same as the original optimal reward. Therefore, the \emph{optimal} adaptive reward is at least the original.

For the non-adaptive strategy in the new Bernoulli instance, it is clear that the feasibility constraint $\calF'$ requires the algorithm to probe a prefix of any root-leaf path (in each branch). Therefore, the best non-adaptive strategy must correspond to a root-leaf path $P'$ in $\calT$ (although in general, non-adaptive strategy, such as \nonadaptivestrategy is randomized, a standard averaging argument shows that there must exist an optimal deterministic non-adaptive strategy). We compare the non-adaptive reward for $P'$ to that of $P$, the path in the original tree $\calT$ that corresponds to $P'$ under the transformation. We prove that for each coordinate $i$, the random variable $\max_{t\leq j\leq k_i} c_{i,j}R_{i,j}$ is dominated\footnote{A (non-negative) random variable $X$ is dominated by another random variable $Y$ if and only if for any $z\geq 0$, $\Pr[X\geq z]\leq \Pr[Y\geq z]$.} by the original variable $X_i$, where $t$ means that the path $P'$ leaves $e_{i,t}$ and goes to $T_t$. It is clear that $\max_{t\leq j\leq k_i} c_{i,j}R_{i,j}$ takes the value $c_{i,j}$ if and only if $R_{i,j}=1$ and $R_{i,j}$ for larger $j$ are not active. By \cref{lm:treeequal}, the probability is exactly $p_{i,j}$, the same as the random variable $X_i$, for $t\leq j\leq k_i$. For the remaining probability $\sum_{j=0}^{t-1}p_{i,j-1}$, the random variable $\max_{t\leq j\leq k_i} c_{i,j}R_{i,j}$ takes zero, while $X_i\geq 0$, hence the assertion of dominance holds.  

Therefore, since dominated (multidimensional) random variables have a smaller $f$ value for monotone norm $f$, we conclude that the best non-adaptive strategy in the new instance is at most that of the original instance. Therefore, the adaptivity gap of the worst case for Bernoulli setting is at least the adaptivity gap for the underlying instance. Hence, the adaptivity gap is at most $\gapfunction_{\XOS}(n\log 8r,r\log 8r)$.
\end{proof}

Finally, we prove Theorem~\ref{thm:genesym}.

\begin{proof}[Proof of \textnormal{Theorem~\ref{thm:genesym}}]
We now define an objective $g':\mathbb R_{\geq 0}^{U'}\to \mathbb{R}_{\geq 0}$ as follows: $$
g'((x_{ij})_{i\in U,j\in[k_i]}) = f\left(\mathsf{Top}_n(x_{ij})\right),
$$ 
where $\mathsf{Top}_n$ stands for the $n$ largest values (break tie arbitrarily) in $x_{ij}$. Clearly, $g'$ is a monotone symmetric norm. The same proof as above shows that the expected gain following $\calT'$ with objective $g'$ is the same as the original optimal reward. Therefore, the \emph{optimal} adaptive reward is at least the original.

For the non-adaptive strategy in the new instance, we use the norm $f'$ in the proof of \textnormal{Theorem~\ref{thm:geneXOS}}:
$$
f'\bigl((x_{ij})_{i\in U,\,j\in[k_i]}\bigr)
  \ =
  f\Bigl(\bigl(\max_{j\in[k_i]}x_{i,j}\bigr)_{i\in U}\Bigr).
$$
We consider $P$, the path in the original tree $\calT$ that corresponds to $P'$ under the transformation. The same proof shows that the non-adaptive reward for $P$ with norm $f$ (note that the random variables $R_{ij}$ take value $\{0,c_{i,j}\}$, so we change $f'$ correspondingly) is at least that for $P'$ with norm $f'$. We now prove that the non-adaptive reward for $P'$ with norm $f'$ is least half the same path with norm $g'$. Indeed, it suffices to prove for any input $x_{i,j}\in \{0,c_{i,j}\}$, $2f'(x_{ij})\geq g^{\prime}(x_{ij}) $. Among the $n$ largest elements in $x_{ij}$, we assume that for each $i$, $x_{i,t_i} = c_{i,t_i}$ is taken, where $t_i\in [0,k_i]$ is the largest possible (if no $x_{i,j}$ appears in the $n$ largest, or its value is zero, set $t_i = 0$). Then, by definition of $f'$, we have $f'(x_{ij})\geq f(x_{i,t_i}) = f(c_{i,t_i})$. On the other hand, we can partition $x_{i,j}$ appearing in $\mathsf{Top}_n(x_{ij})$ for each $i$ with respect to the distance $t_i-j$. Therefore, by symmetry, subadditivity and monotonicity of $f$ (treat $x_{i,t_i-j} = 0$ if $t_i-j\leq 0$), $$
g'(\mathsf{Top}_n(x_{ij}))\leq \sum_{j=0}^{\infty }f(x_{i,t_i-j})\leq  \sum_{j=0}^{\infty }f(c_{i,t_i-j}) \leq  \sum_{j=0}^{\infty }2^{-j}f(c_{i,t_i}) = 2f(c_{i,t_i}) \leq 2f'(x_{ij}).
$$
Therefore, the adaptivity gap of the worst case for Bernoulli setting is at least half the adaptivity gap for the underlying instance. Hence, the adaptivity gap for general random variables is at most $2\gapfunction_{\sym}(n\log 8n)$, using $r = \max\{2, \max\limits_{F\in \calF} |F|\}\leq n$ in \cref{general:assumption}.
\end{proof}
\end{appendices}
\newpage

\bibliography{main}

\end{document}